\documentclass[aps,prd,nofootinbib,twocolumn,superscriptaddress,preprintnumbers,balancelastpage,longbibliography]{revtex4-1}

\usepackage{placeins}
\usepackage{amsmath,amssymb,mathtools,bm}
\usepackage{graphicx, color, hepunits}
\usepackage[dvipsnames]{xcolor}
\usepackage{float}
\usepackage{filecontents}
\usepackage{multirow}
 \usepackage{hyperref} 
\hypersetup{
    colorlinks=true,       
    linkcolor=blue,        
    citecolor=blue,        
    filecolor=magenta,     
    urlcolor=blue          
}

\usepackage{url}
\usepackage[utf8]{inputenc}
\usepackage[english]{babel}
\usepackage{mathrsfs}
\let\vec\mathbf

\newlength{\dhatheight}

\newcommand{\mW}{m_{\scriptscriptstyle W}}

\newcommand{\es}[2] {\begin{equation} \label{#1} \begin{split} #2 \end{split} \end{equation}}
\usepackage{placeins}

\begin{document}

\title{CTA and SWGO can Discover Higgsino Dark Matter Annihilation}

\author{Nicholas L. Rodd}
\email{nrodd@lbl.gov}
\affiliation{Berkeley Center for Theoretical Physics, University of California, Berkeley, CA 94720, U.S.A.}
\affiliation{Theoretical Physics Group, Lawrence Berkeley National Laboratory, Berkeley, CA 94720, U.S.A.}

\author{Benjamin R. Safdi}
\email{brsafdi@berkeley.edu}
\affiliation{Berkeley Center for Theoretical Physics, University of California, Berkeley, CA 94720, U.S.A.}
\affiliation{Theoretical Physics Group, Lawrence Berkeley National Laboratory, Berkeley, CA 94720, U.S.A.}

\author{Weishuang Linda Xu}
\email{wlxu@lbl.gov}
\affiliation{Berkeley Center for Theoretical Physics, University of California, Berkeley, CA 94720, U.S.A.}
\affiliation{Theoretical Physics Group, Lawrence Berkeley National Laboratory, Berkeley, CA 94720, U.S.A.}

\date{\today}

\begin{abstract}
Thermal higgsino dark matter (DM), with a mass near 1.1 TeV, is one of the most well-motivated and untested DM candidates.
Leveraging recent hydrodynamic cosmological simulations that give DM density profiles in Milky Way analogue galaxies we show that the line-like gamma-ray signal predicted from higgsino annihilation in the Galactic Center could be detected at high significance with the upcoming Cherenkov Telescope Array (CTA) and Southern Wide-field Gamma-ray Observatory (SWGO) for all but the most pessimistic DM profiles. We perform the most sensitive search to-date for the line-like signal using 15 years of data from the Fermi Large Area Telescope, coming within an order one factor of the necessary sensitivity to detect the higgsino for some Milky Way analogue DM density profiles.  We show that H.E.S.S. has sub-leading sensitivity relative to Fermi for the higgsino at present. In contrast, we analyze H.E.S.S. inner Galaxy data for the thermal wino model with a mass near 2.8 TeV; we find no evidence for a DM signal and exclude the wino by over a factor of two in cross-section for all DM profiles considered.  In the process, we identify and attempt to correct what appears to be an inconsistency in previous H.E.S.S. inner Galaxy analyses for DM annihilation related to the analysis effective area, which may weaken the DM cross-section sensitivity claimed in those works by around an order of magnitude.
\end{abstract} 

\maketitle

\section{Introduction}

\begin{figure}[!t]
\begin{center}
\includegraphics[width=0.48\textwidth]{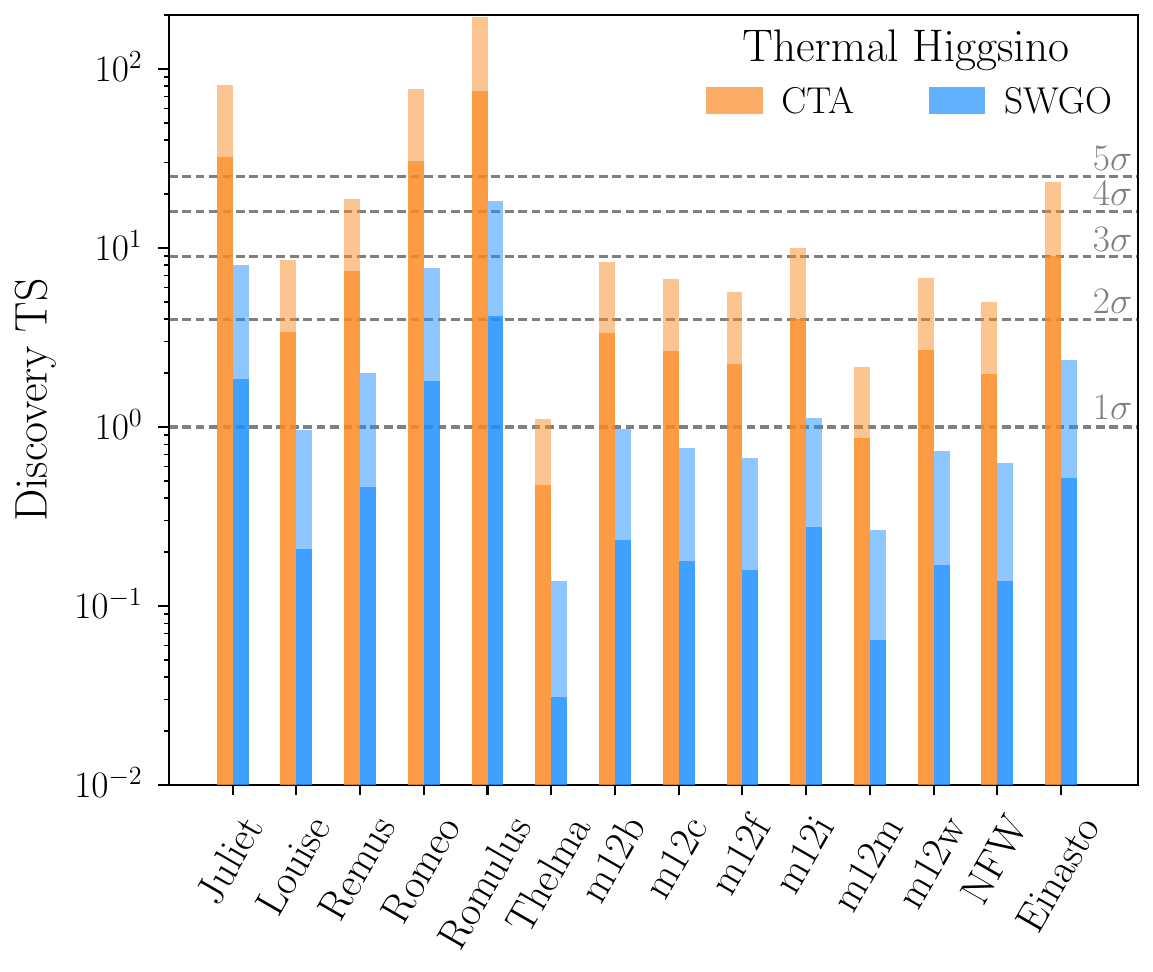}
\end{center}
\vspace{-0.7cm}
\caption{A projection of the expected discovery test statistics (TSs) in favor of the thermal higgsino annihilation signal for CTA (orange) and SWGO (blue).
The projected reach is shown for a wide range of DM profiles and for optimal analyses (light) and more realistic analyses that account for mismodeling uncertainties (dark), as discussed in this work.}
\label{fig:TS_proj_CTASWGO}
\end{figure}

Weakly interacting massive particle (WIMP) dark matter (DM) has been increasingly constrained in recent years from null results for DM scattering at large-scale direct detection experiments using liquid noble gases~\cite{PandaX-4T:2021bab,LZ:2022lsv}, searches for DM production at the large hadron collider~\cite{Boveia:2018yeb}, and gamma-ray telescope searches for DM annihilation~\cite{Fermi-LAT:2016uux,HESS:2022ygk}.  
For all this progress, it has been emphasized recently that in many ways the most canonical WIMP candidate -- the higgsino -- has remarkably yet to be definitively probed by any direct or indirect experiment~\cite{Rinchiuso:2020skh,Co:2021ion,Bottaro:2022one,Dessert:2022evk,Martin:2024pxx}.
In this work we demonstrate that this situation will soon change.
As shown in Fig.~\ref{fig:TS_proj_CTASWGO}, for a wide range of assumptions about the amount of DM in the inner Galaxy, the upcoming Cherenkov Telescope Array (CTA) and the Southern Wide-field Gamma-ray Observatory (SWGO) will be able to detect the thermal higgsino.

The higgsino is an example of minimal WIMP DM~\cite{Cirelli:2005uq}, whereby the DM is assumed to interact with the Standard Model (SM) through the electroweak force $SU(2)_L \times U(1)_Y$ and is put into a representation of the electroweak theory containing a neutral component that becomes the DM mass eigenstate after electroweak symmetry breaking. Scenarios where the SM is augmented only with a thermal bino ($SU(2)_L$ singlet) or wino ($SU(2)_L$ triplet) are largely ruled out by collider~\cite{ATLAS:2014jxt,CMS:2015flg} and indirect detection~\cite{Fan:2013faa,Cohen:2013ama} searches, respectively. Conversely, the thermal higgsino is an $SU(2)_L$ doublet fermionic DM candidate that is too heavy to be produced at existing colliders, with an expected mass around 1 TeV, too weakly interacting to scatter in direct detection experiments~\cite{Hisano:2011cs,Hill:2013hoa,Hill:2014yka,Hill:2014yxa}, and so far too weakly annihilating to give a decisive signal in gamma-ray searches. 
The higgsino is not, however, unreachable; existing studies have suggested that the forthcoming Cherenkov Telescope Array (CTA) South in Chile~\cite{CTAConsortium:2010umy} may be able to reach the thermal signal~\cite{Rinchiuso:2020skh}.  In this work we go beyond~\cite{Rinchiuso:2020skh} by projecting the sensitivity to higgsino DM annihilation accounting for realistic analysis procedures that are robust to systematic background mismodeling while maintaining sensitivity to the DM signal and also accounting for uncertainties in the DM density profile using recent hydrodynamic cosmological simulations.  
We validate our procedures using High Energy Stereoscopic System (H.E.S.S.) and Fermi Large Area Telescope (LAT) inner Galaxy data; we show that while H.E.S.S. and Fermi are at present unable to conclusively reach the higgsino, our procedure opens the path to a high significance discovery with near-term instruments such as CTA and the SWGO air shower array~\cite{Albert:2019afb}.\footnote{SWGO  has yet to finalize a site location but plans to be located in South America at a latitude between roughly $10^\circ$ and $30^\circ$ South.} 

The higgsino DM model has few free parameters, since the higgsino's interactions are fixed by its representation under the electroweak force. One free parameter is the mass of the $SU(2)_L$ doublet, which sets the relic abundance under a standard freeze-out cosmology. To produce the correct DM abundance the higgsino should have a mass $m_\chi = 1.08 \pm 0.02$ TeV~\cite{Bottaro:2022one}.
Below the scale of electroweak symmetry breaking, the higgsino is described by a charged Dirac fermion and a pair of neutral Majorana fermions.
Radiative corrections break the mass degeneracy between the charged and neutral components of the $SU(2)_L$ doublet; the charged Dirac fermion state acquires a mass approximately 355 MeV heavier than the neutral Dirac fermion~\cite{Thomas:1998wy}. In the absence of higher-dimensional operators or additional states, the higgsino is excluded by $Z$-exchange direct detection searches. On the other hand, higher-dimensional operators may split the neutral Dirac fermion into two Majorana mass states; if the neutral mass splitting $\Delta m$ is greater than around 200 keV the $Z$-exchange rate becomes inelastic and is kinematically shut off.
The dimension-five operators giving rise to this mass splitting may emerge from mixing of the higgsino with the bino or wino states in a supersymmetric ultraviolet (UV) completion, where the higgsino arises as the superpartner of the Higgs doublet in that theory. The same operators may too contribute additional mass splitting between the charged and neutral states.
More broadly, the higgsino is motivated in the context of split-spectrum supersymmetry models that have light ($\sim$TeV scale) gauginos and higgsinos to preserve gauge coupling unification but otherwise decouple the scalar superpartners to higher masses~\cite{Wells:2003tf,Giudice:2004tc,Arkani-Hamed:2004ymt,Hall:2011jd,Arvanitaki:2012ps,Arkani-Hamed:2012fhg}.  
In this case the requirement $\Delta m \gg 200 \, \, {\rm keV}$ is translated, generically, to the requirement that either the bino or wino have a mass less than $10^8$ GeV~\cite{Nagata:2014wma}.  The bino or wino admixture in principle opens up an avenue for direct detection through Higgs mediated spin-independent or $Z$ mediated spin-dependent interactions.
The implications of constraints on these processes for the higgsino purity was studied in Ref.~\cite{Martin:2024pxx}.
Broadly, current constraints only require the bino and wino to be heavier than $\sim 1.5-3$ TeV, with the exact value depending on the full set of model parameters.
 
While the mass splitting $\Delta m$ dramatically affects direct detection prospects, the DM annihilation signal is relatively unaffected and is determined through the same diagrams, with the exception of those involving coannihilations, that go into determining the DM relic abundance.  There are two distinct signatures of prompt higgsino DM annihilation: (i) a continuum of gamma-rays extending from the DM mass $m_\chi \simeq 1.1$ TeV to lower energies from the tree-level annihilations to $WW$ and $ZZ$ final states; and (ii) a narrow gamma-ray line at $m_\chi$ from the one-loop direct annihilation to $\gamma\gamma$. As the line is produced at one loop, higher body final states such as $W W \gamma$ contribute comparably and generate a sharp endpoint spectrum that peaks at $\mW^2/m_\chi^2 \simeq 0.5\%$ below the DM mass, with $\mW$ the $W$-boson mass, making it an indistinguishable contribution to the line for the energy resolution of any proposed instrument. An effective field theory calculation of this effect for the higgsino has been performed and is included in our analyses~\cite{Beneke:2019gtg}.

To-date the most sensitive search for higgsino DM annihilation made use of 14 years of data from the Fermi-LAT towards the Milky Way Galactic Center (GC)~\cite{Dessert:2022evk}.  That work focused on gamma-ray energies below roughly 100 GeV and the continuum component of the signal from $WW$ and $ZZ$ final states. Interestingly, Ref.~\cite{Dessert:2022evk} found a modest excess of gamma-rays above the expected continuum background, which could be interpreted as arising from higgsino DM annihilation, depending on the DM density profile.  As part of the present work we revisit the Fermi-LAT analysis of~\cite{Dessert:2022evk} but focus exclusively on the line-like signal at energies near $m_\chi$.  We show that while Fermi-LAT is not sensitive enough to see the expected line-like signal from higgsino annihilation, it is the most sensitive instrument at present to this spectral feature, surpassing the sensitivity of H.E.S.S..

With the above in mind, a key focus of our study is on investigating the sensitivity of the existing H.E.S.S. telescope to a higgsino signal. Even though H.E.S.S. is not the most competitive instrument at present, its similarities to the upcoming CTA, which as we show will be the most sensitive instrument when it comes online, presents an ideal testing ground for analysis strategies.
The H.E.S.S. telescope has been collecting data since 2004; while the collaboration has never performed a dedicated search for the higgsino spectral signature they have searched for the $W W$ and $\gamma\gamma$/$Z\gamma$ final states~\cite{HESS:2018cbt,HESS:2022ygk} (see also~\cite{Montanari:2022buj}). 
Using the information provided in \cite{HESS:2022ygk} we reconstruct their data set and perform a search for the line-like signal associated with higgsino annihilation; we find no evidence for any excess flux.

Most importantly, however, we show that in extracting and analyzing the data from the prior H.E.S.S. analysis in Ref.~\cite{HESS:2022ygk} that the work exhibits apparent inconsistencies, which put in question the validity of their results.  The central inconsistency in~\cite{HESS:2022ygk} is as follows.   Ref.~\cite{HESS:2022ygk} used an ON/OFF analysis procedure, whereby an ON ({\it i.e.}, signal) region is defined for a given observation within the field of view (FOV) towards the GC, while an OFF ({\it i.e.}, background) region is defined to have the same angular size as the ON region and to be the same distance from the beam center but in a direction away from the GC.
The motivation is that cosmic-ray-induced backgrounds should show up the same in the ON and OFF regions, so by modeling the ON-region data as the sum of the OFF region data and the DM-induced signal, one is able to account for the cosmic-ray-induced background in a data-driven manner.
Of course, any astrophysical backgrounds that are not spatially uniform, such as the diffuse emission from our own Galaxy (that has been well-characterized by the Fermi-LAT~\cite{Neronov:2019ncc}), should also appear in the OFF subtracted ON data set.  
However, we show that the apparent astrophysical residuals do not appear at the expected level unless the H.E.S.S. analysis effective area is smaller than that assumed in~\cite{HESS:2022ygk} by a factor of around $8$.  Correcting the effective area by the appropriate amount, we find striking evidence, for the first time, of Galactic diffuse emission over a broad region within the inner Galaxy in the H.E.S.S. data, with a spectral shape above roughly 500 GeV consistent with that measured by Fermi. (For a prior detection within the inner 200 pc, see Ref.~\cite{HESS:2017tce}.)

A key consequence of the correction described above, however, is that the limits on DM annihilation in~\cite{HESS:2022ygk} should also be weakened by around a factor of 8.  An alternative approach to understanding the issue, as we show, is that the effective area assumed in~\cite{HESS:2022ygk} implies, given the total reported photon counts, a cosmic-ray rejection efficiency of  $\gtrsim 99.9\%$, far beyond what H.E.S.S. has claimed to be achievable.

We also show that the analysis procedure performed in~\cite{HESS:2022ygk} for searching for DM annihilation is sub-optimal because (i) DM annihilation contributes to both the ON and OFF regions and is thus partially subtracted in the ON minus OFF procedure, and (ii) the use of the ON minus OFF procedure greatly limits the data volume.
On the other hand, Ref.~\cite{HESS:2022ygk} presented their spectral ON and OFF data, summed over their full region of interest (ROI) in the inner $3^\circ$ of the Galaxy, which we use to perform dedicated higgsino and wino analyses. 
We further use the ON spectral data from~\cite{HESS:2022ygk} to validate an un-subtracted analysis procedure, which we show is more sensitive than the ON/OFF procedure.
We perform a direct search for the higgsino signal, while modeling the background emission with parametric and non-parametric models to illustrate different approaches to how robust and sensitive analyses may be performed.
For the non-parametric approach we use Gaussian Process (GP) modeling, along the lines of the searches performed in~\cite{Frate:2017mai,Foster:2021ngm} in the context of bump hunts at colliders and in astrophysical X-ray data sets.  
We show that the GP and parametric, spectral-template-based approaches could be performed in future analyses of H.E.S.S. data and the data sets collected with upcoming instruments such as CTA and SWGO.

We search for evidence of thermal wino DM annihilation, with a mass near 2.8 TeV, in the un-subtracted H.E.S.S. data using the corrected effective area.  We find no evidence for wino DM annihilation and strongly exclude the wino model for all Milky Way analogue galaxy DM profiles considered. This result add to the previous but weaker results in~\cite{Fan:2013faa,Cohen:2013ama}, which used smaller H.E.S.S. data sets and also assumed only a sample of ad-hoc, parametric DM density profiles.

The remainder of this article proceeds as follows. We detail our approach for computing the thermal higgsino annihilation signal in Sec.~\ref{sec:theory}.
In Sec.~\ref{sec:detector} we outline our modeling of three of the ground-based detectors considered in this work --- H.E.S.S., CTA, and SWGO --- and we perform parametric estimates of the sensitivity of each instrument to a higgsino signal.
In Sec.~\ref{sec:fermi}, we present an analysis searching for the higgsino line-like signature in Fermi-LAT data.
In Sec.~\ref{sec:HESS_real} we analyze the H.E.S.S. inner Galaxy data that we extract from the information provided in \cite{HESS:2022ygk}.
We first demonstrate that the analysis in that work has likely overestimated their effective area, and after correcting this we perform an analysis for the higgsino and wino signals in the data.
We provide projected sensitivities for CTA and SWGO 
in Secs.~\ref{sec:CTA_projections} and~\ref{sec:SWGO_projections}, respectively, justifying that as illustrated in Fig.~\ref{fig:TS_proj_CTASWGO} they could detect thermal higgsino annihilation for a wide array of DM profile scenarios.  
We conclude in Sec.~\ref{sec:discussion}.

The appendices are dedicated to further cross-checks of H.E.S.S. results, with the first three focused on \cite{HESS:2022ygk}.
Appendix~\ref{app:HESS_detector} provides a more complete description of how we extract the data from~\cite{HESS:2022ygk} and independent checks that there is an issue with the results in that paper.
In App.~\ref{app:annuli} we show that adding spatial information to our H.E.S.S. analysis (which we neglect in the main text) has negligible impact.
As a final validation of our procedure, in App.~\ref{app:WW} we demonstrate that we are able to faithfully reproduce the primary science result of \cite{HESS:2022ygk} -- a limit on $\chi \chi \to W W$ -- only in the case where we adopt the effective area that we suspect has been over reported.
In Apps.~\ref{app:Line} and~\ref{app:dwarf} we explore whether this issue is present in other H.E.S.S. searches for DM annihilation, finding that there may be an issue with the search for a line-like signature in the inner Galaxy presented in~\cite{HESS:2018cbt}, but no apparent issue in the dwarf analysis of~\cite{HESS:2020zwn}.
Lastly, in App.~\ref{app:CTA_comp} we demonstrate that our CTA projections are in close agreement with those that appeared recently in \cite{Abe:2024cfj}.

\section{Gamma-ray flux from Higgsino Annihilation}
\label{sec:theory}

We begin by describing the indirect detection signal expected for higgsino DM.
As reviewed above, the higgsino is a well-motivated DM candidate.
In the UV, it could originate as the supersymmetric partner of the SM Higgs, specifically taking the form of two fermionic electroweak doublets that carry hypercharges $\pm 1/2$.
After electroweak symmetry breaking, the spectrum is reorganized into a single charged Dirac fermion and two neutral Majorana fermions split in mass by $\gtrsim 200~{\rm keV}$.
The lightest of the neutral states could then be the DM of our Universe.\footnote{In the minimal DM framework, stability of the lightest state is guaranteed as there are no further operators that induce $\chi$ decay. In the supersymmetric context the lightest higgsino state is a viable DM candidate if it is also the lightest supersymmetric particle and $R$-parity is conserved.}
Given its SM couplings, higgsino DM could emerge from a thermal freeze-out cosmology, with the correct relic abundance requiring $m_\chi = 1.08 \pm 0.02$ TeV~\cite{Bottaro:2022one}, and we adopt the central value of that range as the canonical mass for the thermal higgsino in this work.\footnote{Higher and lower higgsino DM masses are allowed if one considers non-standard cosmological histories or if the higgsino is only a sub-fraction of the DM. We do not consider these possibilities further in this work.}

Within this scenario, we wish to determine the number of counts that higgsino annihilations could produce in our telescope as a function of energy.
The flux incident on the detector depends on two factors, which we discuss in detail in the upcoming subsections.
The first of these is the differential cross section to produce a photon, $d\langle \sigma v \rangle_{\gamma}/dE$, where the subscript $\gamma$ denotes that we are considering the semi-inclusive process $\chi \chi \to \gamma + X$, with $X$ denoting any other final state particle. This contribution describes both the probability of higgsino annihilation and the spectrum of photons produced per annihilation.
The second factor is the $J$-factor: a measure of the DM annihilation flux controlled by the amount of DM in the location observed, or in detail an integral over the DM density squared along the line-of-sight, $J = \int ds\,\rho_{\rm DM}^2(s,\Omega)$ (see, {\it e.g.},~\cite{Lisanti:2017qoz}).
The incident flux can then be combined with the properties of the instrument to determine the differential number of observed counts in units of [cts/TeV/sr], as (see, {\it e.g.},~\cite{Safdi:2022xkm})
\begin{equation}
\frac{dN}{d E_{\rm r} d \Omega} = \frac{J t_{\rm exp}}{8 \pi m_\chi^2} \int dE_{\rm t}\, P(E_{\rm r} | E_{\rm t} ) A_{\rm eff} \frac{d \langle \sigma v \rangle_{\gamma}}{d E_{\rm t}}.
\label{eq:dN/dE}
\end{equation}
Here $t_{\rm exp}$ is the exposure time, $A_{\rm eff}$ is the energy-dependent effective area of the telescope, and lastly $E_{\rm t}$ and $E_{\rm r}$ are the true and reconstructed photon energies, mapped into each other by the energy response $P(E_{\rm r} | E_{\rm t} )$.\footnote{More precisely, $P(E_{\rm r}|E_{\rm t})$ represents the probability of recording a photon of energy $E_{\rm r}$, given the instrument response function and a true photon energy $E_{\rm t}$.} Our prescription for each of these quantities for H.E.S.S., CTA, and SWGO is presented in Sec.~\ref{sec:detector}, while for the Fermi-LAT we use the publicly-available analysis software \texttt{Fermitools}.\footnote{\url{https://fermi.gsfc.nasa.gov/ssc/data/analysis/documentation/}} 

\subsection{The spectrum of higgsino annihilations}

The higgsino annihilation photon spectrum receives contributions from several channels.
The annihilation cross section is dominated by $WW$ and $ZZ$ final states, which proceed at tree level and produce continuum photons, with cross sections of the rough size $\langle \sigma v\rangle_{W} \simeq 8 \times 10^{-27}$ and $\langle \sigma v\rangle_{Z} \simeq 5 \times 10^{-27}$ cm$^3$/s.
Hard photons with $E = m_{\chi}$ can be produced at loop level, and so the annihilation cross section to a line ($\gamma \gamma + \gamma Z/2$) is significantly smaller, around $\langle \sigma v\rangle_{\rm line} \simeq 1 \times 10^{-28}$ cm$^3$/s.
The neutral mass splitting does not have a large impact on these rates or the indirect detection signal generally. Formally it modifies the Sommerfeld enhancement discussed below, although adjusting it over the allowed range of $100~{\rm keV} \lesssim \Delta m \lesssim 1~{\rm GeV}$ varies the cross sections given above by ${\cal O}(10\%)$; throughout, we adopt a representative value of $\Delta m = 20~{\rm MeV}$ and 355 MeV for the chargino-neutralino splitting. The above cross sections are estimated purely using the tree level amplitudes with Sommerfeld enhancement but receive corrections from large logarithms, which we include in the spectrum as discussed below.

The three cross sections provided above are annihilation cross sections. What enters the calculation in \eqref{eq:dN/dE}, however, is the cross section to produce a photon.
The two are roughly related by,\footnote{For further discussion of the different possible cross sections that could be used to parameterize the annihilation rate and spectrum, see~\cite{Baumgart:2017nsr,Rinchiuso:2018ajn}.}
\begin{equation}
\frac{d\langle \sigma v \rangle_{\gamma}}{dE}
\!\simeq\! \langle \sigma v \rangle_{W} \frac{dN_{W}}{dE}
+ \langle \sigma v \rangle_{Z} \frac{dN_{Z}}{dE}
+ \langle \sigma v \rangle_{\rm line} \frac{dN_{\rm line}}{dE}.
\label{eq:2xsec}
\end{equation}
Here $dN/dE$ represent the differential numbers of photons produced per annihilation. For instance, $dN_{\rm line}/dE = 2 \delta(E-m_{\chi})$, whereas the equivalent for the $W$ and $Z$ describe the continuum of photons produced from these final states.
The result in~\eqref{eq:2xsec} is only approximate as there can be additional contributions to the spectrum. An important one that we include is that the line cross section can receive an ${\cal O}(1)$ enhancement from endpoint photons (three or higher body final states involving a photon).

To account for all of these contributions we use the code package \texttt{DM$\gamma$Spec}~\cite{Beneke:2022eci,Cirelli:2010xx}.\footnote{The continuum contribution in \texttt{DM$\gamma$Spec} is determined using results from \texttt{PPPC4DMID}~\cite{Cirelli:2010xx}.
As shown recently in Ref.~\cite{Jager:2023njk}, there are continuum contributions this approach does not fully capture that become particularly relevant for neutralino masses below a TeV.}
This package further accounts for the cross section corrections induced by Sommerfeld enhancement~\cite{Hisano:2003ec,Hisano:2004ds,Arkani-Hamed:2008hhe,Beneke:2014gja,Blum:2016nrz} and large logarithms, described below.
The Sommerfeld enhancement arises when the annihilating DM particles interact through a relatively light mediator particle, as in the case for TeV-scale higgsino DM interacting through electroweak bosons.
(Note that \texttt{DM$\gamma$Spec} computes the Sommerfeld enhancement using the NLO potentials determined in \cite{Beneke:2019qaa,Urban:2021cdu}.)
The emission of soft and collinear radiation generates Sudakov double logarithms of the form $\ln^2(4 m_\chi^2/\mW^2)$.
Further, when focusing on hard photons with $E \simeq m_{\chi}$, this places a kinematic restriction on the process $\chi \chi \to \gamma + X$ that generates additional large logarithms associated with the endpoint spectrum.
A reliable calculation of the DM cross section requires both sets of logarithms to be resummed, which can be done in the framework of effective field theory, as developed for TeV scale electroweak DM over the last decade~\cite{Baumgart:2014saa,Baumgart:2014vma,Bauer:2014ula,Ovanesyan:2014fwa,Baumgart:2015bpa,Ovanesyan:2016vkk,Baumgart:2017nsr,Beneke:2018ssm,Baumgart:2018yed,Beneke:2019vhz,Beneke:2022eci,Beneke:2022pij,Baumgart:2023pwn}.

\begin{figure}[!t]
\begin{center}
\includegraphics[width=0.49\textwidth]{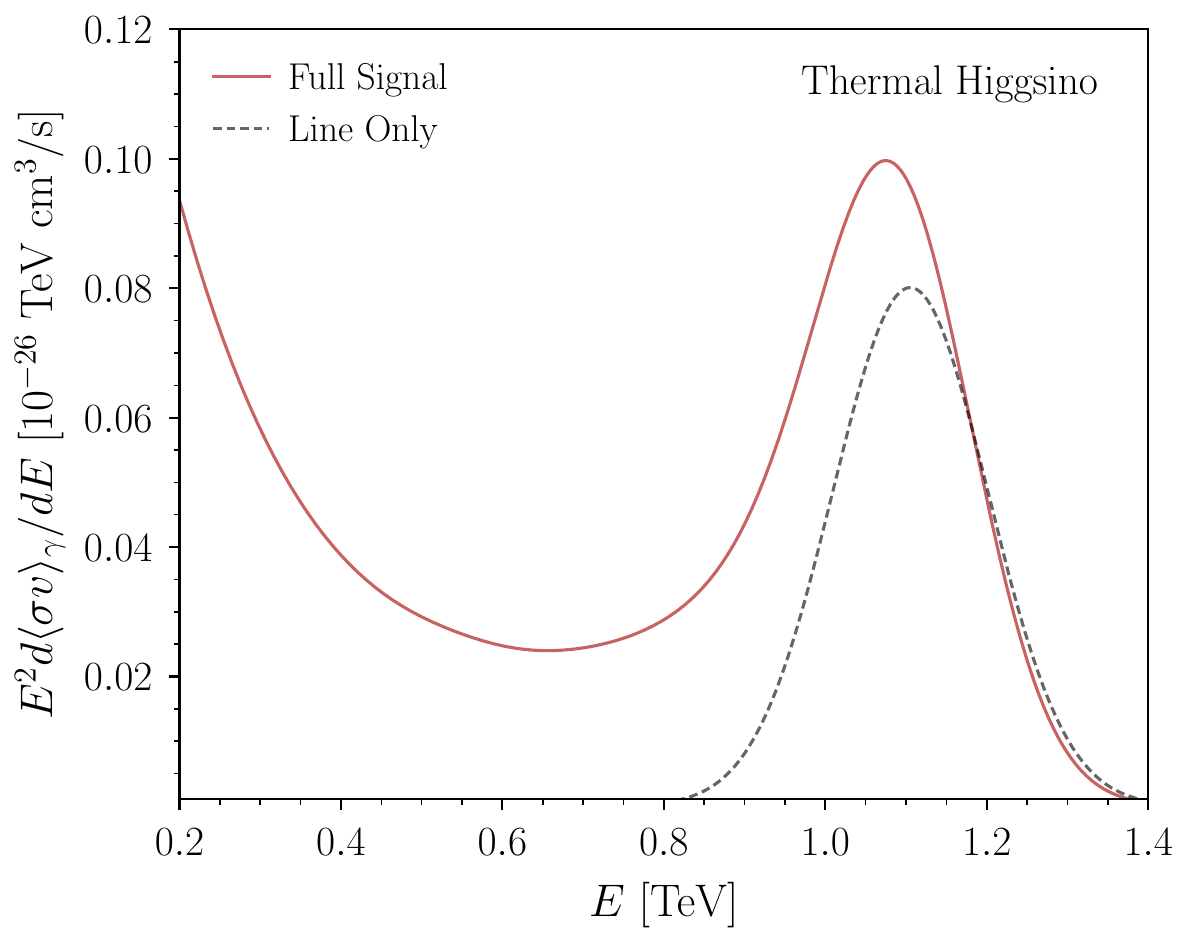}
\end{center}
\vspace{-0.7cm}
\caption{The differential higgsino annihilation cross section $d \langle \sigma v \rangle_{\gamma}/d E$ computed using \texttt{DM$\gamma$Spec} and assuming a thermal higgsino mass of $m_\chi = 1.08$ TeV.
We show the distribution after it has been convolved with a Gaussian model for the H.E.S.S.-like energy resolution.
We show the full signal, which is provided by \texttt{DM$\gamma$Spec}, and for contrast we also show what would be obtained for a pure line $\sim \delta(E-m_{\chi})$, normalized by the higgsino prediction for two-body final states involving a photon.
As seen, the full signal differs considerably from a line, which highlights that dedicated higgsino searches are necessary; searches for spectral lines in isolation may return incorrect results for the higgsino given the different shape and normalization of the signal.}
\label{fig:xsec}
\end{figure}

From $d\langle \sigma v \rangle_{\gamma}/dE$ we can determine the distribution of photons that arrives at the detector, which we then need to correct with the instrumental energy response.
For simplicity we model the energy responses of the H.E.S.S, CTA, and SWGO detectors in this work as Gaussians with an energy-dependent width, $\sigma_x E_t$; for the Fermi-LAT we use the full energy redistribution matrix provided in \texttt{Fermitools}.
In Fig.~\ref{fig:xsec} we show the differential cross section convolved with the instrument energy resolution, taking $\sigma_x = 0.1$, as roughly appropriate for H.E.S.S. (see Sec.~\ref{sec:detector}). 
We illustrate both the full spectrum including all relevant final-state photons, along with a comparison to what would be obtained if we simply had the line contribution arising from two-body final states that include a photon. We use this to demonstrate that accounting for the full spectrum is important in part because the peak is shifted to slightly lower energies, but even more so as there is considerable flux at lower energies.

The primary focus of our work is on the detectability of the largely un-probed thermal higgsino.
Nevertheless, our approach can be easily extended to other TeV electroweak DM candidates.  To demonstrate this point we also consider the thermal wino, which has a mass $\sim$2.8 TeV.
The wino is the neutral component of a Majorana triplet of fermions, which has similarly strong motivations to the higgsino, and its spectrum can also be computed with \texttt{DM$\gamma$Spec}.
Note that indirect detection searches already disfavor the wino, unless there is a considerable reduction in the DM density profile in the inner galaxy~\cite{Fan:2013faa,Cohen:2013ama,Rinchiuso:2020skh}. We return to the wino in Sec.~\ref{sec:HESS_real} by showing that the thermal wino is disfavored from our analysis of the H.E.S.S. inner Galaxy data for all DM profiles considered.

\subsection{DM density profile in the inner Galaxy}

A major complication when searching for DM annihilation in the inner Galaxy is that the DM density profile is expected to be strongly affected by baryons within the inner few degrees.
The Galactic bulge extends to roughly half a kpc from the GC, corresponding to around $4^\circ$ away from the GC in terms of the closest distance along the line of sight from Earth; within the bulge radius the bulge contributes around $10^9$ $M_\odot$ of baryonic matter~\cite{Bovy:2014vfa}, while the amount expected from the DM halo -- assuming a standard Navarro–Frenk–White (NFW)~\cite{Navarro:1995iw,Navarro:1996gj} profile --  is around $10^8$ $M_\odot$.  It is thus expected that DM density profiles motivated by DM-only $N$-body simulations, such as the NFW profile and the Einasto profile~\cite{Graham:2005xx}, are poor estimators of the DM density in the inner parts of Milky Way mass galaxies.  To more accurately estimate the DM density profile in the inner galaxy, and more importantly the inherent uncertainty therein, we use 12 Milky Way analogue galaxies from the Feedback In Realistic Environments (FIRE-2) zoom-in hydrodynamic cosmological simulations~\cite{Hopkins:2017ycn,2022MNRAS.513...55M}, which include baryon dynamics in addition to DM. Although there is a spread between simulations, generically baryonic feedback leads to an enhancement in the density of DM at the GC through adiabatic contraction, thereby boosting the expected DM-induced signal.

We follow~\cite{2022MNRAS.513...55M} and adjust the FIRE-2 DM density profiles such that the local DM density is $0.38$ GeV$/$cm$^3$ with a distance to the Sun of $r_\odot = 8.2$ kpc.\footnote{We note that differences in the mass enclosed with the solar radius between the 12 Milky Way analogue DM density profiles, after normalizing to the same local DM density, are at maximum around 20\%.  Given current uncertainties in the local DM density~\cite{deSalas:2020hbh} all of these DM profiles can likely thus be considered consistent with local and global measurements of the DM content of the Milky Way.} 
We compare the FIRE-2 profiles to the NFW profile~\cite{Navarro:1995iw,Navarro:1996gj},
\begin{equation}
\rho_{\rm NFW}(r) = {\rho_0 \over r/r_s (1 + r/r_s)^2} \,,
\end{equation}
and to the Einasto profile~\cite{Graham:2005xx,Retana-Montenegro:2012dbd,Navarro:2008kc},
\begin{equation}
\rho_{\rm Ein}(r) = \rho_s \exp \left[- {2 \over \alpha_s} \left( \left( {r \over r_s}\right)^{\alpha_s} - 1 \right) \right]\!,
\end{equation}
which are both motivated by DM-only cosmological simulations.  We normalize all DM density profiles to the same local DM density.
For the NFW profile we take the fiducial value for the scale radius $r_s = 15$ kpc, while for the Einasto profile we assume $r_s = 20$ kpc and $\alpha_s = 0.17$ (for more discussion, see~\cite{Safdi:2022xkm}).
\begin{figure}[!t]
\begin{center}
\includegraphics[width=0.49\textwidth]{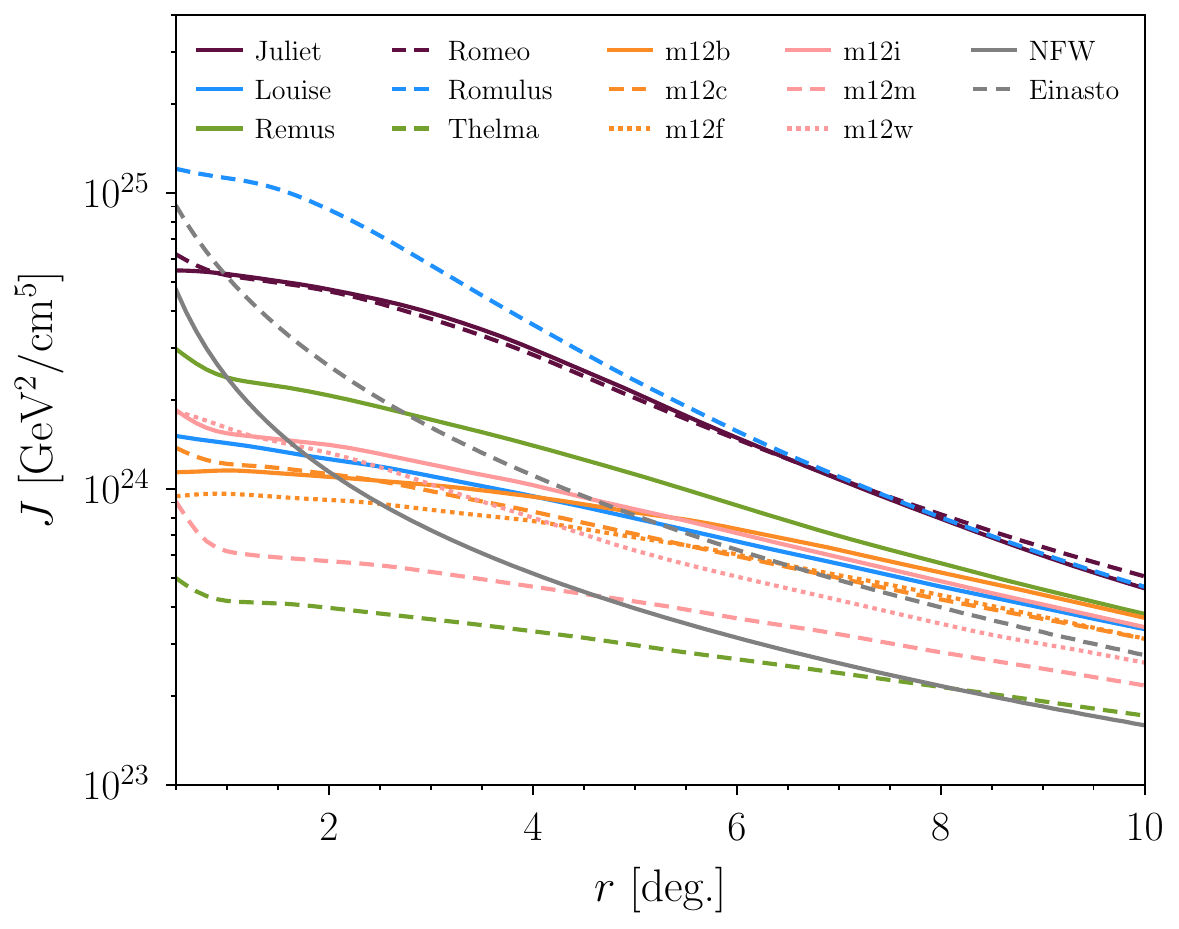}
\end{center}
\vspace{-0.7cm}
\caption{The $J$-factor profiles computed using the 12 FIRE-2 Milky Way analogue galaxies~\cite{Hopkins:2017ycn,2022MNRAS.513...55M} (also reproduced from~\cite{Dessert:2022evk}). 
We compare the FIRE-2 profiles to the NFW and Einasto profiles, all normalized to match the local DM density at solar system distances. The named FIRE-2 galaxies were evolved in pairs to mimic Milky Way-Andromeda dynamics.}
\label{fig:J}
\end{figure}
From the DM profiles, we then compute the $J$-factor as a function of the observed angle from the GC, using 
\begin{equation}
J \equiv \int ds\,\rho_{\rm DM}^2(s,\Omega),
\label{eq:J}
\end{equation}
where $s$ is the line-of-sight distance from Earth and $\Omega$ denotes the angular position on the sky.

The $J$-factor profiles in the inner parts of the Galaxy are illustrated in~\cite{Dessert:2022evk} and are also reproduced in Fig.~\ref{fig:J}. 
Note that there is around an order of magnitude spread in predicted profiles between different FIRE-2 Milky Way analogue galaxies.  Of the 12 FIRE-2 galaxies, the Romulus profile has the most similar baryonic features (thick disk, stellar bulge, etc.) as those in the Milky Way, and 6 of the 12 galaxies (including Romulus) were evolved in pairs to mimic the interactions of the Milky Way with M31.
For this reason, we use Romulus as a benchmark in many of our analyses.
The resolution in the FIRE-2 simulations is estimated at 2.75$^\circ$~\cite{2022MNRAS.513...55M}; however, we use the simulation output to make projections down to smaller radii, though the $J$-factors should be treated with caution at such small angular scales.

\section{H.E.S.S., CTA, and SWGO detector characterizations}
\label{sec:detector}

In this section we describe our parameterizations of the existing H.E.S.S. telescope and the future CTA and SWGO detectors. (As previously mentioned, for the Fermi-LAT we simply use the \texttt{Fermitools}.)  We are primarily interested in the performance of these instruments near 1 TeV.  Nevertheless, we also consider the instrument responses at higher and lower energies. The lower energies are important to capture the low-energy continuum photons produced by the higgsino annihilation, though the primary focus of this work is on the line-like signature.
The higher energies are irrelevant for the thermal higgsino, however, considering them allows us to search for additional DM candidates such as the $\sim$2.8 TeV thermal wino.

We caution the reader that (i) the modern H.E.S.S. data, instrument responses, and observation strategies are not publicly available, and (ii) the precise design configurations for CTA and SWGO have not been decided and/or publicized.  With regards to H.E.S.S. this implies that we are forced to rely on approximations, described below, based on small amounts of public data from older instrument configurations. These will differ from the true current instrument response and are certainly unable to capture variation on an observation-by-observation basis. In terms of CTA and SWGO, we assume the default configurations for these instruments, but in reality there will undoubtedly be differences between the final versions of these detectors  and the configurations assumed here. Conversely, we highlight through back-of-the-envelope estimates in this section and more careful calculations in subsequent sections how the higgsino sensitivity depends parametrically on the detector parameters, such that future detectors such as CTA and SWGO may be further optimized for higgsino detection.   

\subsection{H.E.S.S.}

\begin{figure*}[!htb]
\begin{center}
\includegraphics[width=0.49\textwidth]{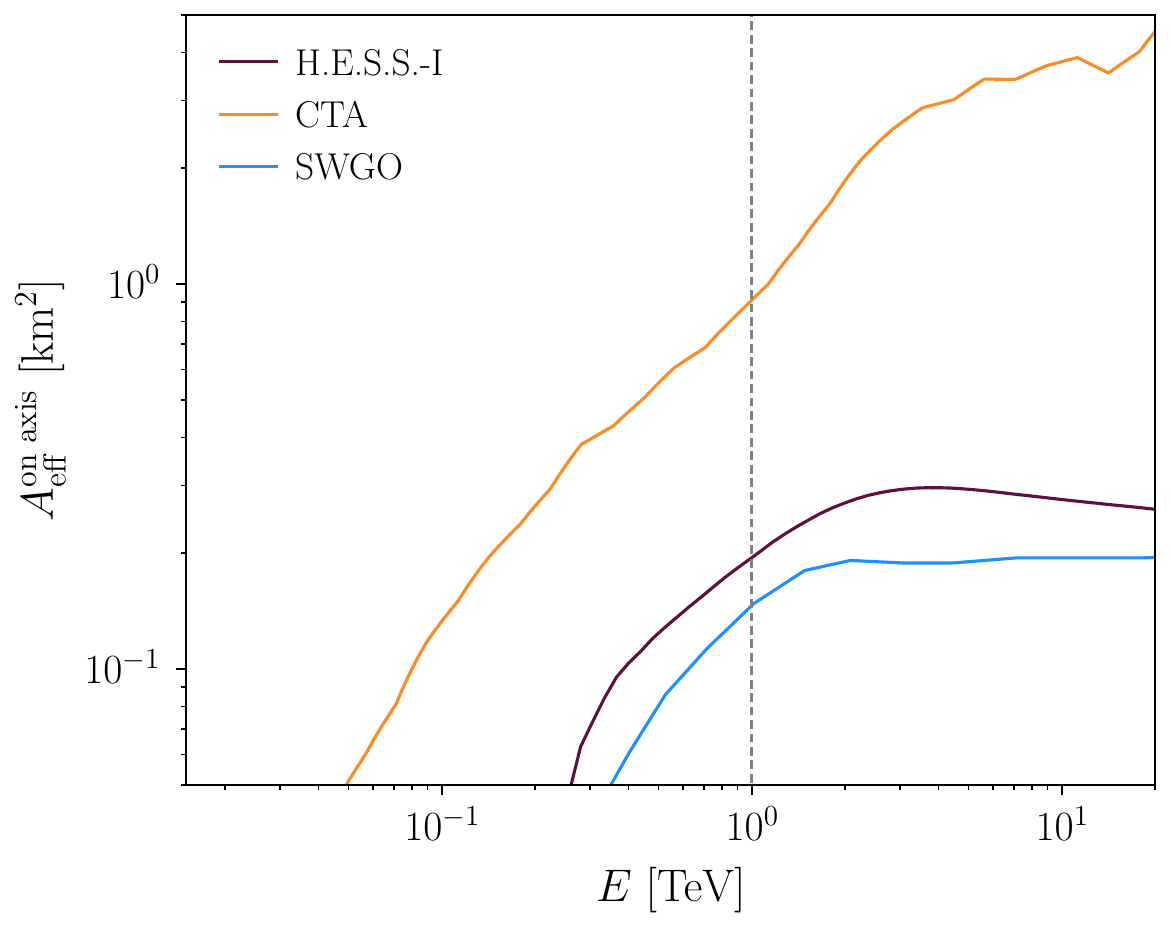}
\includegraphics[width=0.48\textwidth]{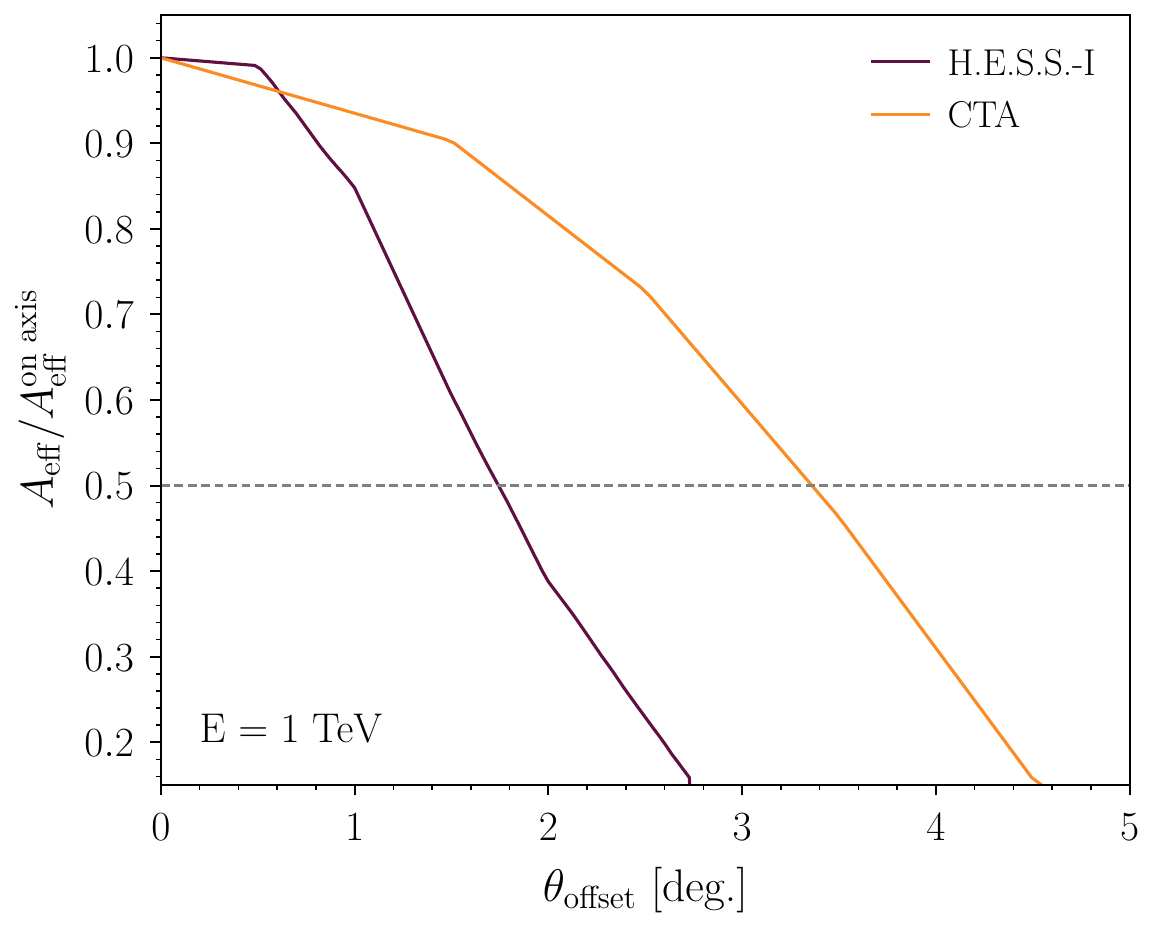}
\end{center}
\vspace{-0.7cm}
\caption{(Left panel) The on-axis effective area of H.E.S.S.-I as a function of gamma-ray energy, along with the projections for the on-axis effective areas of CTA and SWGO.
(Right panel) The fall-off of the H.E.S.S. effective area as a function of the off-set angle from the beam center $\theta_{\rm offset}$, with the horizontal curve showing where the effective area drops by a factor of two. The projected CTA effective area covers a larger field of view (FOV), as illustrated. We do not show SWGO because of its much larger FOV, on the order of 1 sr.}
\label{fig:Aeff}
\end{figure*}

\begin{figure*}[!htb]
\begin{center}
\includegraphics[width=0.485\textwidth]{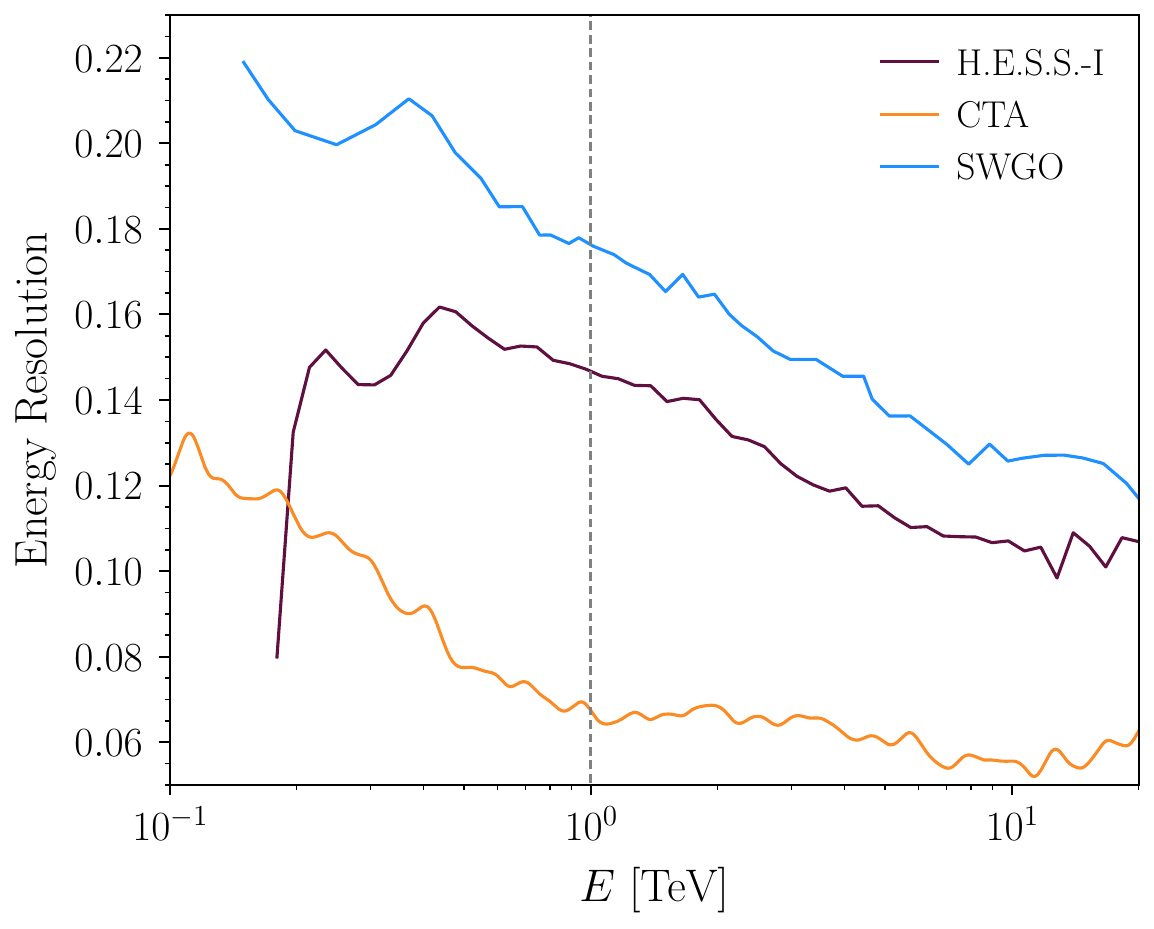}
\includegraphics[width=0.49\textwidth]{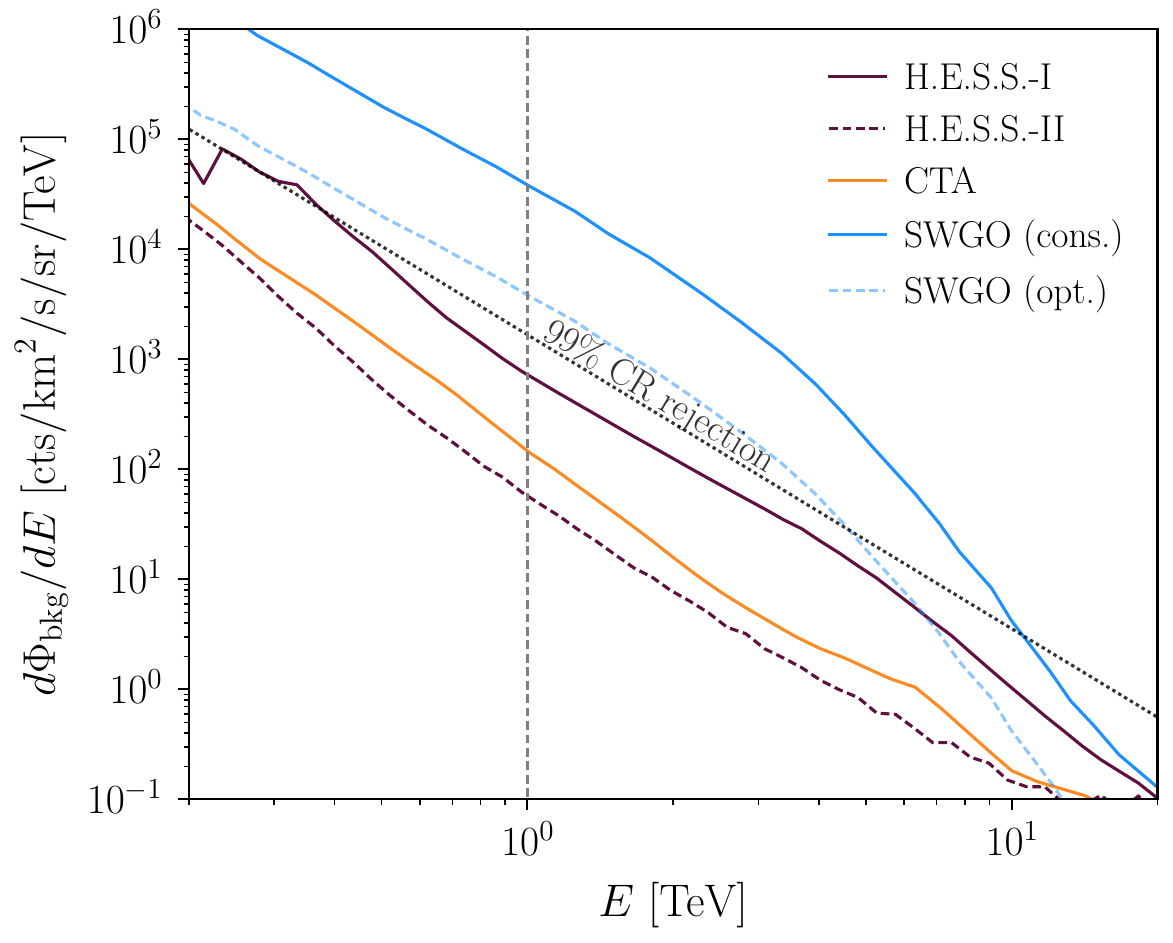}
\end{center}
\vspace{-0.7cm}
\caption{(Left panel) The energy resolution as a function of energy for H.E.S.S., CTA, and SWGO.
We model the energy resolution as a Gaussian distribution with width $\sigma_x E_{\rm t}$, with $E_{\rm t}$ the true incident energy of the photon and $\sigma_x$ the energy resolution.
(Right panel) The background rates at H.E.S.S., CTA, and SWGO from misidentified cosmic rays, which are predominantly high-energy protons.  The H.E.S.S.-I background rate is that inferred from their public data release~\cite{HESS:2018zix}, while that labeled H.E.S.S.-II is quoted as being currently achieved by the telescope in~\cite{HESS:2022ygk}. As we illustrate, however, the H.E.S.S.-II background rate corresponds to $\gtrsim99.9\%$ cosmic ray (CR) rejection efficiency and even surpasses the proposed CTA efficiency; we suggest in this work that the quoted background rate in~\cite{HESS:2022ygk} is incorrect and is due to an overestimate of their effective area.
The projected SWGO background rate, SWGO (cons.), is significantly higher than that of the IACTs and taken to be the background rate observed by {\it e.g.} HAWC, though improvements at the level of an order of magnitude may be achievable and are included in our projections and illustrated through SWGO (opt.).}
\label{fig:edispbkg}
\end{figure*}

H.E.S.S. has been collecting data since 2004 with four telescopes (H.E.S.S.-I), though in 2012 they added a larger, fifth telescope (H.E.S.S.-II upgrade).
In Fig.~\ref{fig:Aeff} we show the typical on-axis H.E.S.S.-I effective area (left panel) as a function of energy along with the falloff of the effective area with angle from the beam axis (right panel) at $E = 1$ TeV.
These results are taken from the H.E.S.S.-I public data release~\cite{HESS:2018zix}; we take the median response over 38 blank-sky observations, restricting to those at zenith angles lower than $40^\circ$ in attempt to match the criteria in~\cite{HESS:2022ygk}.
The equivalent H.E.S.S.-II results are not publicly available, although we expect them to be similar other than increased effective area from the additional H.E.S.S.-II telescope, primarily at $\sim$100 GeV~\cite{HESS:2015cyv}.
In Sec.~\ref{sec:HESS_real} and App.~\ref{app:HESS_detector} we outline a procedure for inferring the H.E.S.S.-II effective area from the results in Ref.~\cite{HESS:2022ygk}, and although there we demonstrate that the effective area used in that work appears to be anomalously large, we note that in general we expect the H.E.S.S.-I and II performances to be comparable.\footnote{In our reference to H.E.S.S.-I versus H.E.S.S.-II we also, for brevity, refer to the reconstruction and cosmic-ray-rejection algorithms used in the public data release~\cite{HESS:2018zix} as H.E.S.S.-I with the newer algorithms implemented in~\cite{HESS:2022ygk} as H.E.S.S.-II.  }
In particular, for the angular response, we assume throughout that the behavior of H.E.S.S.-II matches H.E.S.S.-I.
The energy resolution, which we define as the standard deviation $\delta E$ of the energy response over the true energy $E_t$, for H.E.S.S.-I as extracted from the H.E.S.S.-I public data release~\cite{HESS:2018zix} is illustrated in Fig.~\ref{fig:edispbkg} (left panel). We assume the energy response is a Gaussian with width $\sigma_x E_{\rm t}$ as discussed in Sec.~\ref{sec:theory}, which is then determined by the energy resolution, $\sigma_x$.
H.E.S.S.-I has $\sigma_x \simeq 0.15$ around $E_t \simeq 1$ TeV as shown in in Fig.~\ref{fig:edispbkg}.
An improved resolution of at least 10\% above 200 GeV was quoted for H.E.S.S.-II in \cite{HESS:2022ygk}, and we correspondingly adopt $\sigma_x = 0.1$ as a fixed value for H.E.S.S.-II throughout.

To project sensitivity to higgsino annihilation signals we need to know the expected background rates. The largest source of backgrounds arises from cosmic rays (CRs), primarily from high-energy protons that are misidentified as gamma-rays, even though the majority of cosmic ray proton events are vetoed. The residual background rate of fake gamma-rays from misidentified cosmic rays, which we refer to as $d \Phi_{\rm bkg} / d E$, is shown in Fig.~\ref{fig:edispbkg} (right panel).  We take as background rate (labeled H.E.S.S.-II) the observed fluxes reported in~\cite{HESS:2022ygk} averaged over their full analysis region, making the assumption that the data are dominated by $d \Phi_{\rm bkg} / d E$. We compare these rates to 1\% of the incident flux of hadronic cosmic rays (dominantly from protons and helium) on the atmosphere as determined by AMS-02~\cite{AMS:2015tnn}, assuming an energy-independent rejection rate. We also show the typical background seen by H.E.S.S.-I blank-sky observations, which achieves a CR rejection of approximately this order. 
This figure illustrates that \cite{HESS:2022ygk} has an exceedingly high cosmic-ray rejection efficiency, at the order of $\gtrsim 99.9\%$, although as discussed in Sec.~\ref{sec:HESS_real} we are concerned this flux may have been underestimated. We suspect that this under-reported flux is a direct consequence of the over-reported effective area, which in turn impacts both the science results of Ref.~\cite{HESS:2022ygk} and projections of H.E.S.S.'s future reach, as we describe.

\begin{figure}[!htb]
\begin{center}
\includegraphics[width=0.49\textwidth]{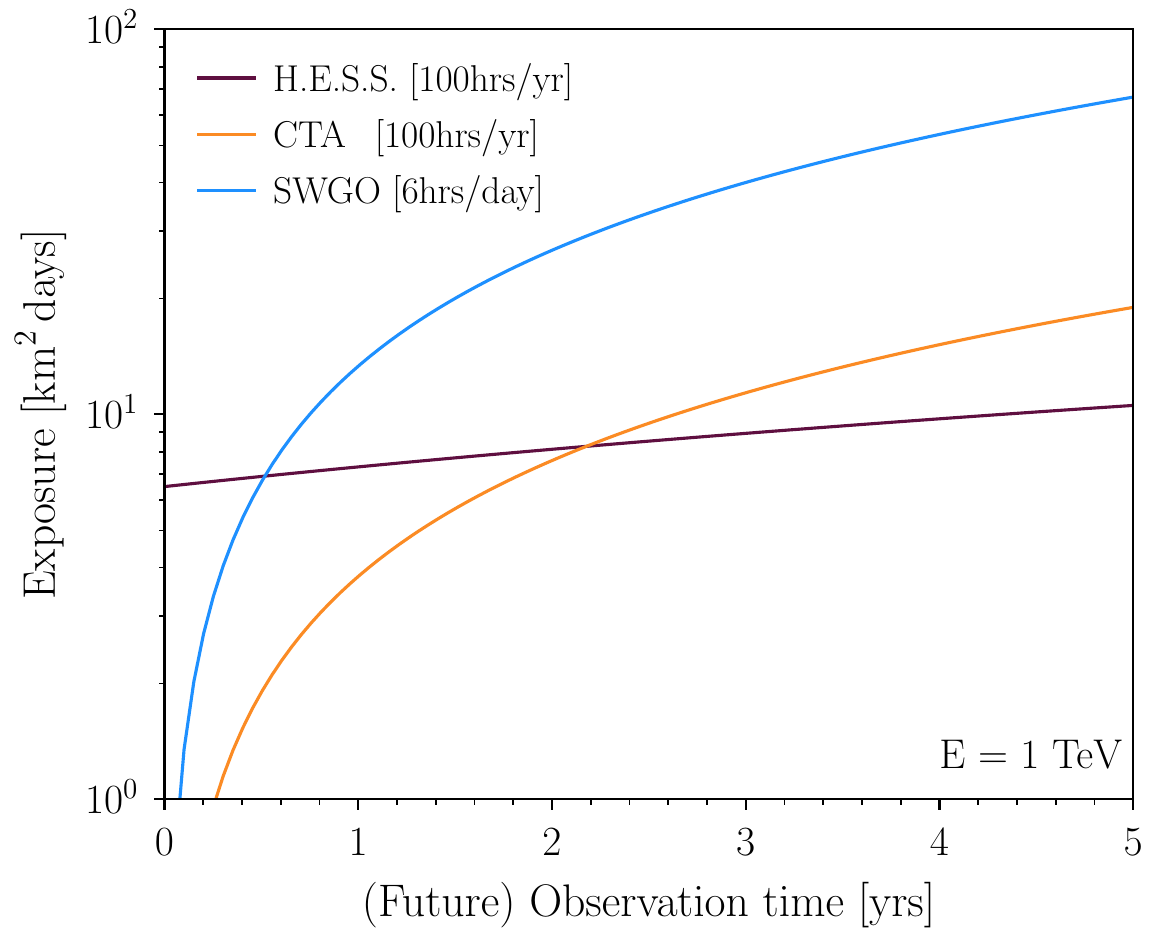}
\end{center}
\vspace{-0.7cm}
\caption{Estimate for the accumulated exposure (${\mathcal E} \equiv A_{\rm eff} \times t_{\rm exp}$, with $t_{\rm exp}$ the exposure time and $A_{\rm eff}$ the effective area), for H.E.S.S., CTA, and SWGO at the GC at $E = 1$ TeV.  Note that we include the data accumulated to-date at the GC for H.E.S.S. and we approximate all of the GC pointings to be at the same location, such that we use $A_{\rm eff}^{\operatorname{on-axis}}$ for the effective area, for illustrative purposes only.}
\label{fig:exposure}
\end{figure}

Lastly, it is useful -- especially in comparing to extensive air-shower arrays (EASs) such as the future SWGO -- to illustrate, roughly, the expected data collection rate. Imaging atmospheric Cherenkov telescopes (IACTs) such as H.E.S.S. and the future CTA collect data at a much slower rate than EASs, since EASs can collect data for multiple hours each day towards a typical point on the sky within sight of the detector. On the other hand, IACTs conventionally collect data under pristine and moonless night-time conditions, with the target overhead.
H.E.S.S. has, to-date, collected on the order of 800 hrs of data in the inner $\sim$5$^\circ$ of the GC, as we discuss in more detail in Sec.~\ref{sec:HESS_real}.
Let us assume that around 100 hrs of data could be collected at the GC per year going into the future; that value is motivated by the fact that H.E.S.S.-II collected around 550 hrs of data at the GC over roughly 6 years~\cite{HESS:2022ygk} (although up to 150 hours per year may be achievable~\cite{Montanari:2022buj}). Then, the predicted H.E.S.S. accumulated GC exposure -- going into the future -- is illustrated in Fig.~\ref{fig:exposure}.  For illustrative purposes, we show the exposure ${\mathcal E} \equiv A_{\rm eff} \times t_{\rm exp}$, where $t_{\rm exp}$ is the exposure time, and we take $A_{\rm eff}$ to be the on-axis effective area as shown in Fig.~\ref{fig:Aeff}; we also account for the data collected to-date.

\subsection{CTA}

The upcoming CTA South observatory will have many similarities with H.E.S.S. but with a larger effective area that extends to lower energy photons, a wider FOV, improved energy resolution, and improved cosmic-ray rejection efficiency~\cite{CTAConsortium:2017dvg}. Here, we assume the proposed CTA Observatory ``Alpha Configuration" performance~\cite{CTA_performance}. The projected on-axis effective area for this configuration is illustrated in the left panel of Fig.~\ref{fig:Aeff}, with the right panel showing the fall-off of the effective area with the offset angle $\theta_{\rm offset}$ from the beam axis. Note that the effective area is around a factor of five higher than that of H.E.S.S. at 1 TeV and also the effective area has non-trivial support around twice as far away from the beam center relative to H.E.S.S..

The CTA energy resolution is expected to be better than that of H.E.S.S., as shown in Fig.~\ref{fig:edispbkg} using data provided in~\cite{CTA_performance}. We use this projected energy resolution, assuming Gaussian energy dispersion as for H.E.S.S., when making CTA projections. (Note that for CTA we vary $\sigma_x$ with energy according to Fig.~\ref{fig:edispbkg}.) The CTA South Alpha Configuration projected background rate as provided in~\cite{CTA_performance} is shown in Fig.~\ref{fig:edispbkg}, where it is seen to be better than that achieved by H.E.S.S.-I~\cite{HESS:2022ygk}.

Lastly, as illustrated in Fig.~\ref{fig:exposure} we assume that CTA acquires exposure at the GC at a comparable rate to H.E.S.S., such that around 500 hours of data could be collected within the first five years. Thus, after $\sim$2 years the CTA integrated exposure at the GC should surpass that of H.E.S.S..  On the other hand, this figure underestimates the improvement of CTA relative to H.E.S.S. because, as we discuss more in Sec.~\ref{sec:boe}, the larger FOV of CTA relative to H.E.S.S. also improves the sensitivity to higgsino annihilation.

\subsection{SWGO}

The proposed SWGO EAS~\cite{Albert:2019afb} is fundamentally different from the IACTs like H.E.S.S. and CTA, with a number of important implications for higgsino annihilation searches.  IACTs use telescopes to search for the Cherenkov radiation produced in the atmosphere by particle cascades initiated by incident high-energy gamma-rays; on the other hand, EASs search directly for the particles in the shower that make it to the Earth's surface. SWGO will be similar in concept to the high-altitude water Cherenkov observatory (HAWC)~\cite{historical:2023opo}, which is located in Mexico, but with a larger effective area and a location in the Southern Hemisphere.  In HAWC, the shower is detected by imaging the charged particles passing through water tanks that detect the resulting Cherenkov radiation of the transiting particles.

Unlike IACTs, EASs are able to continuously take data, even in the middle of the day. The effective area of EASs such as HAWC are comparable to that of IACTs such as H.E.S.S. and moreover EASs have large FOVs, on the order of 1 sr. On the other hand, EASs tend to have significantly worse energy resolution and background rates.  We illustrate the proposed ``Straw Man" SWGO performance~\cite{Viana:2019ucn} in Fig.~\ref{fig:Aeff}, Fig.~\ref{fig:edispbkg}, and Fig.~\ref{fig:exposure} for the effective area, energy resolution, background rate, and integrated exposure in the inner Galaxy, respectively.  The effective area of SWGO is planned to be comparable to that of H.E.S.S., while the energy resolution is around a factor of two worse. Note that we do not illustrate the fall off of $A_{\rm eff}$ with angle from the beam center in Fig.~\ref{fig:Aeff} (right panel), since SWGO has a large FOV $\sim$1 sr and the effective area would thus be completely flat over the illustrated angular scales.  Despite the smaller effective area, the SWGO integrated exposure rapidly surpasses that of H.E.S.S. as illustrated in Fig.~\ref{fig:exposure} due to the fact that it may observe the GC for $\sim$6 hrs per day~\cite{Albert:2019afb}, instead of the 100 hrs per year in the case of H.E.S.S./CTA.
Note that in Fig.~\ref{fig:edispbkg} we show two SWGO projected background rates: conservative and optimistic. The conservative background rate is the same observed with HAWC~\cite{Viana:2019ucn}, without accounting for any improvements in going to SWGO. On the other hand, the collaboration has conjectured an order of magnitude improvement in the background rejection around 1 TeV~\cite{BarresdeAlmeida:2022lgy}; the background rate labeled `optimistic' thus assumes precisely a factor of 10 improvement relative to the `conservative' rate.

\subsection{Astrophysical diffuse emission}
\label{sec:astro}

The ground-based gamma-ray telescopes that we focus on in this work -- CTA, SWGO, and H.E.S.S. -- are dominated by cosmic-ray-induced backgrounds, which are approximately isotropic on Earth.   
However, sub-leading to these backgrounds are astrophysical backgrounds, primarily from Galactic diffuse emission at $E \sim 1$ TeV.
The diffuse emission can arise from several sources, but includes cosmic-ray protons interacting with gas throughout the Milky Way to generate neutral pions and thereby gamma-rays.
Here, we describe our modeling of the diffuse emission. 

Our starting point is the data set and Galactic diffuse models obtained by the Fermi-LAT Collaboration. 
The sky map collected by Fermi extends in energy up to $\sim$2 TeV and thus has accumulated data near the GC at energies $\sim$1 TeV, which we may use to estimate the Galactic diffuse emission.
In particular, we use 813 weeks of Fermi data with photons in the \texttt{SOURCE} event class, and we further select the top three quartiles of data as ranked by the energy resolution; our description of this data set is described in Sec.~\ref{sec:fermi}. In this section we bin the data in 40 logarithmically-spaced energy bins between 2 GeV and 2 TeV, though we only illustrate the data above $\sim$500 GeV.  We additionally make use of the \texttt{gll\_iem\_v07} (\texttt{p8r3}) Galactic diffuse model reprocessed for this data set (publicly available \href{https://fermi.gsfc.nasa.gov/ssc/data/access/lat/BackgroundModels.html}{here}).

\begin{figure}[!t]
\begin{center}
\includegraphics[width=0.49\textwidth]{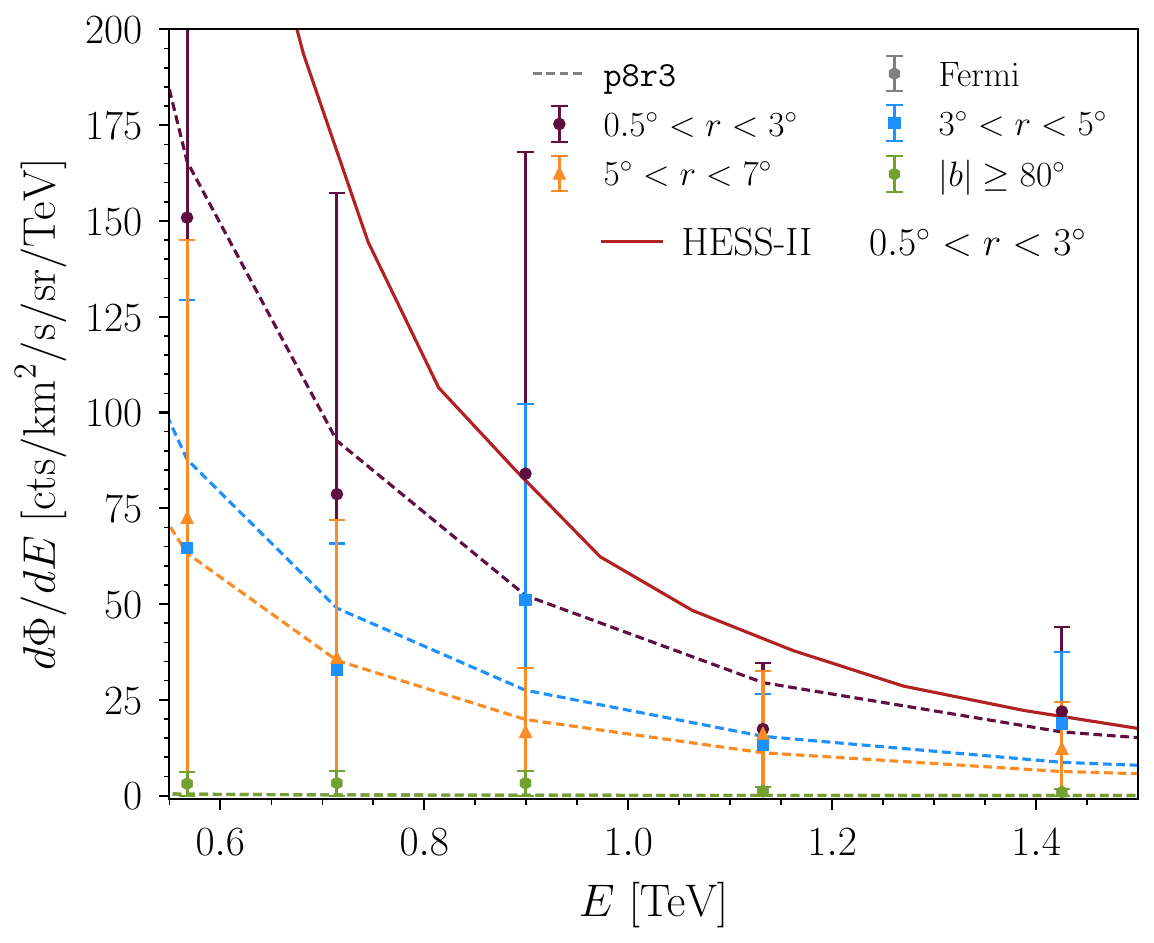}
\end{center}
\vspace{-0.7cm}
\caption{The Galactic diffuse emission as measured by Fermi using 729 weeks of data (data points and associated 1$\sigma$ error bars from counting statistics) for three different annuli around the GC, as indicated, with the Galactic plane masked at $|b| \leq 0.3^\circ$ in each case. While the data points show the binned Fermi data (see text for details), the smooth dashed curves illustrate the Fermi \texttt{p8r3} Galactic diffuse model (see Fig.~\ref{fig:fermi_diffuse_map}), which agrees with the Fermi data up to the statistical uncertainties shown. Also shown is the observed flux around the north and south Celestial poles, showing that cosmic-rays and extragalactic photons make a negligible contribution to the data. 
In addition, in red we show the H.E.S.S.-II reported flux in the inner annulus from~\cite{HESS:2022ygk}.
From this it is clear that an analysis with this level of flux sensitivity must account for the Galactic diffuse emission (cf. Fig.~\ref{fig:edispbkg}).
}
\label{fig:diffuse}
\end{figure}

In Fig.~\ref{fig:diffuse} we show the binned Fermi data in Galactocentric rings of indicated width.  We present the data in units of [cts/km$^2$/s/sr/TeV], where we divide the observed counts in each annulus by the average exposure in that annulus, the angular size of the region, and the width of the energy bin.  Note that we mask the Galactic plane at $|b| \leq 0.3^\circ$ to match the Galactic plane mask in the H.E.S.S. analysis~\cite{HESS:2022ygk}. In addition to the Fermi data, with 1$\sigma$ statistical error bars from finite photon statistics, we also show the prediction from the \texttt{p8r3} Galactic diffuse model in each of these energy bins. Finally, as a control we show the observed flux in regions far away from the Galactic plane. The negligible flux seen around the north and south Celestial poles indicates that the Fermi data around the GC is primarily Galactic photons rather than mismodeled cosmic rays or extragalactic emission, both of which would be isotropic.

There are a number of features in Fig.~\ref{fig:diffuse} that are worth commenting on. First, there is a clear trend where the annuli closer to the GC have a higher flux than those further from the GC. This is in notable contrast to the isotropic cosmic-ray-induced background. Secondly, within the statistical uncertainties the observed Fermi data matches the Fermi \texttt{p8r3} model, which is not surprising considering that the model was fit to the data by the Fermi Collaboration.
Based on this observation, for the rest of this work we use the Fermi \texttt{p8r3} model when making projections for Galactic diffuse emission.
Third, we show the H.E.S.S.-II inner Galaxy stacked ON data from~\cite{HESS:2022ygk}, which has a region similar to the $0.5^\circ < r < 3^\circ$ region.
As mentioned previously and discuss further below, we suspect that the H.E.S.S. flux -- shown in Fig.~\ref{fig:diffuse} -- was underestimate. One reason is that, as seen in Fig.~\ref{fig:diffuse}, given the Fermi measurements one would conclude that the H.E.S.S. stacked ON data should mostly arise from Galactic diffuse emission, which does not appear to be the case when looking at the stacked ON minus OFF data. A plausible explanation is that H.E.S.S. overestimated their effective area. Correcting the H.E.S.S. effective area we find, as discussed more, than the ON minus OFF H.E.S.S. data shows residual emission with a spectral shape consistent with the Galactic diffuse emission illustrated in Fig.~\ref{fig:diffuse}.

\begin{figure}[!t]
\begin{center}
\includegraphics[width=0.49\textwidth]{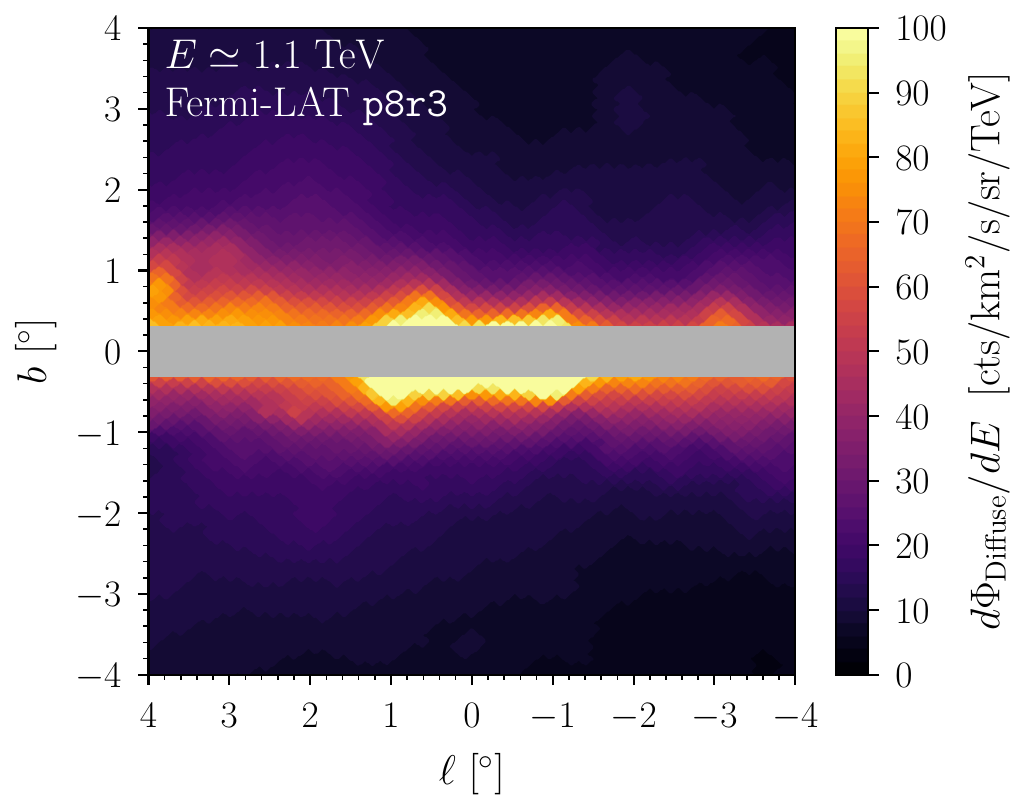}
\end{center}
\vspace{-0.7cm}
\caption{As in Fig.~\ref{fig:diffuse} we illustrate the Galactic diffuse emission with the Fermi \texttt{p8r3} model, but here we fix the energy $E \simeq 1.1$ TeV and illustrate the spatial distribution of the astrophysical diffuse emission in the vicinity of the GC, with $|b| \leq 0.3^\circ$ masked.  Comparing this with the right of Fig.~\ref{fig:edispbkg} we see that at $E \sim 1$ TeV and extremely close to the GC the Galactic diffuse emission can be comparable to the cosmic-ray-induced emission for observatories such as H.E.S.S. and CTA.}
\label{fig:fermi_diffuse_map}
\end{figure}

In Fig.~\ref{fig:fermi_diffuse_map} we show the \texttt{p8r3} diffuse model at $E \simeq 1.1$ TeV in the vicinity of the Galactic plane, with $|b| \leq 0.3^\circ$ masked, and binned using HEALPix~\cite{Gorski:2004by,Zonca2019} with {\tt nside}=512.
One important point visible in that figure is that the diffuse emission extends well away from the Galactic plane and is highly anisotropic, meaning that it is not fully subtracted in ON minus OFF analysis procedures, as we discuss further in this work.

\subsection{Back-of-the-envelope sensitivity estimates}
\label{sec:boe}

Before performing detailed analysis projections it is useful to roughly estimate the sensitivities of the different instruments we consider in this work: H.E.S.S., CTA, SWGO, and Fermi. This is especially relevant given that our assumed instrumental parameters are only approximate for the ground-based observatories, so the estimates below allow us to understand the parametric sensitivity on general grounds.

We consider H.E.S.S. first. Suppose that we have $t_{\rm exp} \sim 500$ hrs of observation time distributed equally across an ROI centered at the GC of radius  $r_{\rm ROI} = 5^\circ$. At $E = 1.08$ TeV we take the on-axis effective area $A_{\rm eff}^{\operatorname{on-axis}} \sim 0.2$ km$^2$.  Averaged over the radius $r_{\rm FOV} \sim 1.5^\circ$, which we define to be the angular scale away from the beam axis where the effective area drops by a factor of two, the averaged effective area is $A_{\rm eff}^{\rm FOV} \simeq 0.7 \cdot A_{\rm eff}^{\operatorname{on-axis}}$, assuming a normally distributed fall-off of the beam with angle from the beam center.  This implies that any given point within our $5^\circ$ region of interest (ROI) has an exposure ${\mathcal E}$ (exposure time times effective area) of
\begin{equation}\begin{aligned}
{\mathcal E} &\simeq 0.7 \cdot \left( {r_{\rm FOV} \over r_{\rm ROI}} \right)^2  t_{\rm exp} \cdot A_{\rm eff}^{\operatorname{on-axis}}  \\
&\simeq 0.7 \cdot \left({1.5^\circ \over 5^\circ} \right)^2 \cdot 500 \, \, {\rm hrs} \cdot 0.2 \, \, {\rm km}^2  \\
&\sim 5\, \, {\rm km}^2 \cdot {\rm hrs}.
\end{aligned}\end{equation}
Combining this with \eqref{eq:dN/dE} we can determine the expected number of counts higgsino annihilations could generate within the ROI, which we denote $N_{\rm sig}$. For this simple estimate we take the average $J$-factor in the ROI to be $J \sim 5 \times 10^{24}$ GeV$^2$/cm$^5$. 
We ignore the dependence of this quantity on distance from the GC, and we focus just on the photons from the line for which $\langle \sigma v\rangle_{\gamma} \simeq 2 \times 10^{-28}$ cm$^3$/s.
In detail,
\begin{equation}\begin{aligned}
N_{\rm sig} &\simeq \Delta \Omega \cdot  {\mathcal E} \cdot {J \over 8 \pi m_\chi^2} \langle \sigma v \rangle_{\gamma} \\
&\sim 2 \cdot 10^2 \left( {r_{\rm FOV} \over 1.5^\circ}\right)^2 \left( {\rm t_{\rm exp} \over 500\, {\rm hrs}} \right) \left( {A_{\rm eff}^{\operatorname{on-axis}} \over 0.2\, {\rm km}^2} \right) {\rm cts},
\end{aligned}\end{equation}
where we take $\Delta \Omega \simeq  \pi r_{\rm ROI}^2$ as the ROI size. 

The signal counts all appear within, roughly, $\delta E \simeq \sigma_x E \sim 100\,$GeV of $E \simeq m_\chi = 1.08$ TeV, smeared from a line by the detector energy resolution.
We can also estimate the number of background counts in this same region. 
If we take the H.E.S.S.-I background rate, then from Fig.~\ref{fig:edispbkg} we arrive at $d \Phi_{\rm bkg} / dE \sim 600$ cts/km$^2$/s/sr/TeV at $E = 1.08$ TeV.
At that same energy, the observed H.E.S.S.-II flux in~\cite{HESS:2022ygk} is roughly a factor of thirteen smaller, and we comment on how that would impact the higgsino sensitivity below.
Combining these results, the number of background counts within the ROI and within $\delta E$ of $m_{\chi}$ is expected to be 
\begin{align}
N_{\rm bkg} &\simeq {d \Phi_{\rm bkg} \over dE} \cdot \Delta \Omega \cdot {\mathcal E} \cdot \delta E \\
&\sim 4\cdot 10^4 \left( {r_{\rm FOV} \over 1.5^\circ}\right)^2\! \left( {\rm t_{\rm exp} \over 500 \, {\rm hrs}} \right)
\!\left( {\sigma_x \over 0.1}\right)\!\left( {A_{\rm eff}^{\operatorname{on-axis}} \over 0.2 \, {\rm km}^2} \right) {\rm cts}. \nonumber
\end{align}
(For simplicity, we ignore possible differences between the gamma-ray and cosmic-ray effective areas.)

If we have a perfect model for the expected background counts, which is a point that we return to later on, then we are limited by the statistical noise in the number of background counts, with the standard deviation being $\sqrt{N_{\rm bkg}}$. In particular, the expected discovery test statistic (TS) or $\Delta \chi^2$ in favor of the higgsino model over the null hypothesis scales as 
\begin{align}
\Delta \chi^2 &\sim {N_{\rm sig}^2 \over N_{\rm bkg}}
\label{eq:Delta_chi2_estimate} \\
&\sim {\cal O}(1) \, \left({t_{\rm exp} \over 500 \, {\rm hrs}} \right)  \left( {A_{\rm eff}^{\operatorname{on-axis}} \over 0.2 \, {\rm km}^2 } \right) \left( {r_{\rm FOV} \over 1.5^\circ} \right)^2 \left( {0.1 \over \sigma_x} \right)\!. \nonumber
\end{align}
This result tells us that, depending on the precise $J$-factor profile and how well the real analysis strategy can reach this sensitivity estimate, we should expect at best H.E.S.S. has a marginal sensitivity to the higgsino.
If we instead adopted the flux observed by H.E.S.S.-II~\cite{HESS:2022ygk}, we would find a significantly improved sensitivity of $\Delta \chi^2 \sim 13$, which would imply that H.E.S.S. is already sensitive to the signal.
However, as already mentioned, we discuss in Sec.~\ref{sec:HESS_real} that the H.E.S.S.-II flux in~\cite{HESS:2022ygk} is likely considerably underestimated.

Looking to the future, we can similarly use~\eqref{eq:Delta_chi2_estimate} to estimate the expected reach of CTA. Referring to Fig.~\ref{fig:Aeff}, we take for CTA the approximate values $r_{\rm FOV} \sim 3.5^\circ$ and $A_{\rm eff}^{\operatorname{on-axis}} \sim {\rm km}^2$.
Further, from Fig.~\ref{fig:edispbkg} we take $\sigma_x \sim 0.06$ and a background flux a factor of 5 smaller than H.E.S.S.-I at $E = 1.08$ TeV.
With these values, we estimate the discovery TS in favor of the higgsino model for CTA as
\begin{equation}
\Delta \chi^2_{\rm CTA} \sim {\cal O}(200)\, \left( {\rm t_{\rm exp} \over 500 \, {\rm hrs}} \right)\!.
\end{equation}
Note that a 5$\sigma$ discovery corresponds to $\Delta \chi^2 \sim 25$, which means that CTA should be well suited to discover thermal higgsino DM through the line spectrum with around five years of data (see also~\cite{Rinchiuso:2020skh}).\footnote{Comparing the CTA versus H.E.S.S. estimates, we can infer that under the null hypothesis and for comparable observation times we would expect CTA to set limits around ten times stronger, in terms of cross-section, than H.E.S.S.. On the other hand, the recent CTA projections for $\chi \chi \to \gamma\gamma$ in~\cite{Abe:2024cfj} (which we reproduce in App.~\ref{app:CTA_comp}) are within $\sim$50\% of the H.E.S.S. expected limits from their analysis in~\cite{HESS:2018cbt}.  This is an additional indirect piece of evidence that the H.E.S.S. sensitivity from their recent GC analyses for DM annihilation may be overestimated.}

We may also apply the rough estimates above to SWGO, although we emphasize that the FOV of SWGO, which is roughly $1 \, \, {\rm sr}$, is much larger than any ROI that would be used for an analysis at the GC. Thus, in applying~\eqref{eq:Delta_chi2_estimate} we take $r_{\rm FOV} \to r_{\rm ROI} \sim 5^\circ$, although we caution that we expect the estimate in this case to be far less accurate for SWGO, given that it will collect data over a much larger FOV relative to H.E.S.S. and CTA.  This means, for example, that data beyond $5^\circ$ could also be incorporated into the analysis. 
Referring to Fig.~\ref{fig:Aeff} the SWGO effective area at $1.08$ TeV is $A_{\rm eff} \simeq 0.15$ km$^2$, while $\sigma_x \simeq 0.17$ (see Fig.~\ref{fig:edispbkg}).
Perhaps the most significant difference, however, between SWGO and H.E.S.S. or CTA is the exposure time. As discussed previously, we can assume that SWGO collects roughly 6 hours of data each day. Thus, after 5 years we expect $t_{\rm exp} \sim 10^4$ hrs.
At $E = 1.08$ TeV the HAWC background rate is approximately a factor of $300$ larger than what CTA is projected to achieve.
As previously discussed, the SWGO background rejection may improve relative to HAWC, and we consider the possibility that it improves by an order of magnitude following~\cite{BarresdeAlmeida:2022lgy}. 
We then project the discovery $\Delta \chi^2_{\rm SWGO}$ of
\begin{equation}
\Delta \chi^2_{\rm SWGO} \sim {\mathcal O}\left(2.5 - 25\right) \times  \left( {t_{\rm exp} \over 10^4 \, \, {\rm hrs}} \right)\!,
\end{equation}
with the higher (lower) number corresponding to the more efficient (HAWC-level) background rejection efficiency.

Lastly, we consider the sensitivity of the Fermi-LAT to the line-like higgsino signature.
In the inner Galaxy, Fermi has an exposure on the order of ${\mathcal E} \sim 6 \times 10^{-3}$ km$^2 \cdot$hr.
This can be roughly understood from the detector being $\sim$m$^2$ in size, taking data for roughly 15 years, and seeing around 1/5 of the sky at once.
In the inner 5$^\circ$ we thus expect, using the $J$-factor from the previous estimates, around $N_{\rm sig} \approx 0.2$ signal counts.
To estimate the background, we note that around a TeV, the Fermi-LAT energy resolution is $\sigma_x \simeq 0.1$.
Further, unlike for the ground-based detectors discussed above, Fermi is dominated by astrophysical gammma-ray diffuse emission backgrounds at $\sim$1 TeV.
Referring to Fig.~\ref{fig:diffuse} we estimate $d \Phi / dE \sim 30$ cts/km$^2$/s/sr/TeV as the diffuse background flux in the inner $5^\circ$ at 1.08 TeV.
Accordingly, for Fermi we expect $N_{\rm bkg} \sim 2$.
Since the number of counts is not large, we cannot estimate $\Delta \chi^2$ by $N_{\rm sig}^2 / N_{\rm bkg}$.
Nevertheless, as $N_{\rm sig} \ll 1$ it is clear that Fermi does not have the necessary sensitivity to reach the thermal higgsino signal. On the other hand, as we show in the following section incorporating data beyond the inner $5^\circ$ dramatically improves the sensitivity of Fermi to the higgsino signal, making it the most competitive instrument to-date for the line-like signal (though still not sensitive enough to probe the expected higgsino cross-section).
Moreover, Fermi does have the necessary sensitivity at lower energies to probe the low-energy tail of the tree-level annihilation products~\cite{Dessert:2022evk}. 
The above estimates motivate the detailed analyses presented in the remainder of this work.

\section{Search for a higgsino line signature in Fermi-LAT data}
\label{sec:fermi}

We now begin our consideration of how to confront the thermal higgsino hypothesis with  data, beginning with observations from the Fermi-LAT $\gamma$-ray space telescope.
Reference~\cite{Dessert:2022evk} searched for the higgsino annihilation signal in Fermi data, but that work focused on the low-energy continuum emission from the $W^+W^-$ and $ZZ$ tree-level final states.  In this work we restrict to energies above $\sim$500 GeV and focus exclusively on the line-like signal near threshold (see Fig.~\ref{fig:xsec}), which is also the signal we focus on for H.E.S.S., SWGO, and CTA.
As the Fermi data is completely publicly available, we can analyze it without assumptions about the instrument response.

\begin{figure}[!t]
\begin{center}
\includegraphics[width=0.49\textwidth]{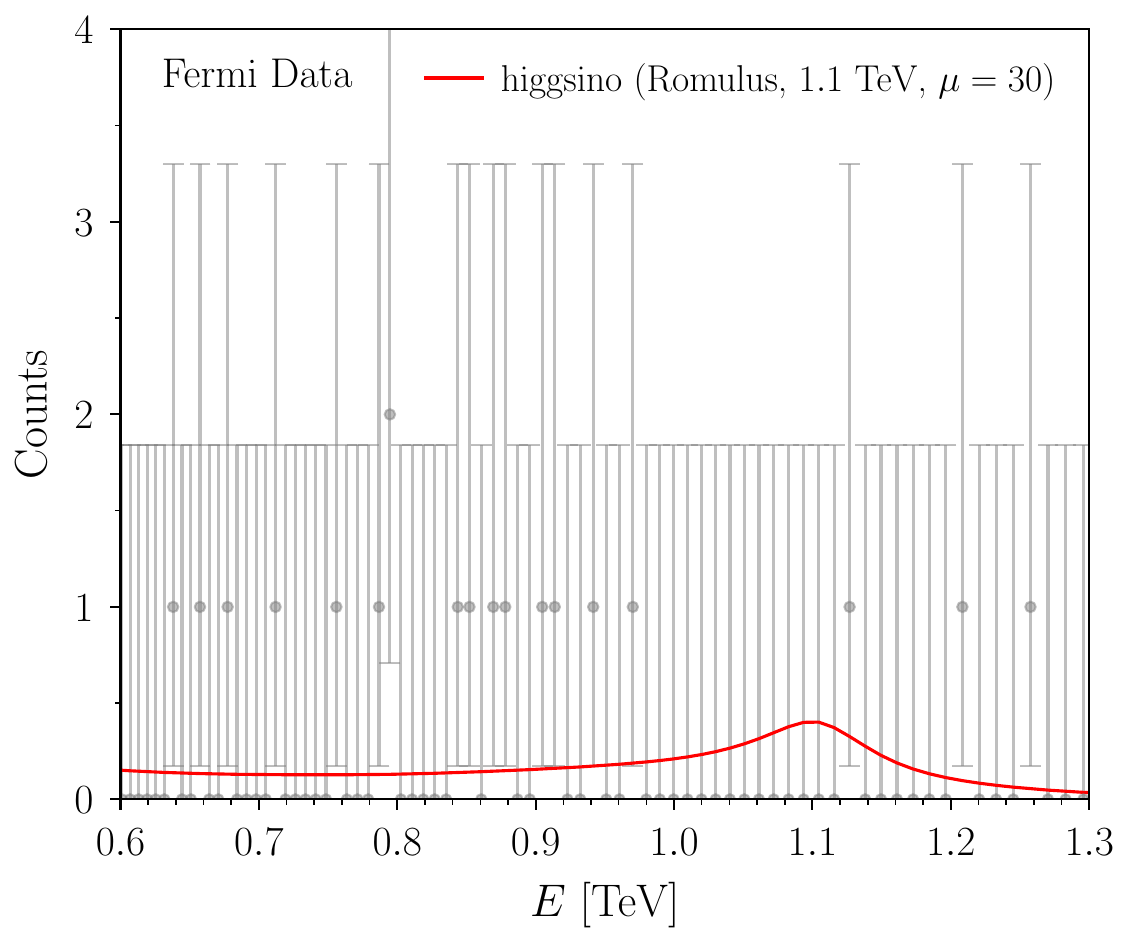}
\end{center}
\vspace{-0.7cm}
\caption{An illustration of the Fermi data used in this work to search for evidence of the gamma-ray line from higgsino annihilation. We show the observed counts as a function of energy within a representative region defined to be the inner $5^\circ$ around the GC with a plane mask described in the main text.
We show the predicted counts from higgsino annihilation for $m_\chi = 1.1$ TeV and assuming the Romulus DM profile, with $\mu = 30$ (corresponding to the expected signal multiplied by a factor of 30, for illustration).  Note that our actual analysis analyzes the data independently for each \texttt{edisp} quartile and each annulus, out to $30^\circ$ from the GC, and then combines the results using a joint likelihood (see text for details).}
\label{fig:fermi_Data}
\end{figure}

Our analysis follows almost identically the procedure used in~\cite{Foster:2022nva}, which searched for DM annihilation to gamma-ray lines in 14 years of gamma-ray data.  One important difference between~\cite{Foster:2022nva} and this work, however, is that while~\cite{Foster:2022nva} only included the monochromatic $\gamma\gamma$ final state, in this work we search for the unique spectral signature associated with higgsino annihilation and illustrated in Fig.~\ref{fig:xsec}.  We also include an additional $\sim$two years of data relative to~\cite{Foster:2022nva}, and we perform a more accurate statistical interpretation of the results.

We reduce the Fermi data using the same set of quality cuts described in~\cite{Foster:2022nva}. 
Specifically, we use the {\tt SOURCE} class photon classification and the top three quartiles of data as ranked by energy dispersion (\texttt{edisp}).  We analyze the data in each \texttt{edisp} quartile separately and then join the results using a joint likelihood.
We reduce the data using the \texttt{Fermitools}.
While Ref.~\cite{Foster:2022nva} incorporated 729 weeks of Pass 8 Fermi data, we use 813 weeks of data, including data up to 23 February 2024.
(See Ref.~\cite{Foster:2022nva} for a description of the additional quality cuts.)
We bin the data into 531 energy bins, logarithmically spaced, between 10 GeV and 2 TeV, though only the data above $\sim$500 GeV are incorporated into our analysis.
In Fig.~\ref{fig:fermi_Data} we illustrate the Fermi-LAT data summed over the inner 5$^\circ$ and summed over all three \texttt{edisp} quartiles, around the GC in the vicinity of 1 TeV.
We follow the masking procedure used in~\cite{Foster:2022nva} and leave the GC unmasked  but mask latitudes $|b| \leq 1.5^\circ$ for longitudes $|\ell| \geq 3^\circ$ in order to reduce emission from bright point sources along the Galactic plane. 
In Fig.~\ref{fig:fermi_Data} we overlay the predicted higgsino annihilation signal assuming the Romulus DM profile, with a rescaled normalization for illustrative purposes.

In our analysis, we follow~\cite{Foster:2022nva} and bin the data in each \texttt{edisp} quartile into annuli, subject to the plane mask described above, of radial width $1^\circ$ each.  The innermost annulus is in fact a disk extending from $0^\circ$ to $1^\circ$ from the GC, while the outermost annulus extends from $29^\circ$ to $30^\circ$; in total, there are 30 annuli. For a given DM mass $m_\chi$, we use a floating energy window analysis that includes $k_{\rm max} = 25$ energy bins above and below the energy bin that includes $m_\chi$. The data are described using a Poisson likelihood where the mean signal model prediction is the sum, in each energy bin, of the higgsino model (with model parameter for the cross-section, as described below) and the background model.  The background model is described by a power-law with free nuisance parameters for the spectral index and overall normalization, independent for each $m_\chi$ point and for each \texttt{edisp} quartile.

More precisely, for a given mass point $m_\chi$, we compute the Poisson likelihood
\es{eq:LL_fermi}{
p({\bf d} | {\mathcal M}, {\bm \theta}) = &\prod_{i=1}^3 \prod_{j = 1}^{30}\prod_{k = {\bar k} - k_{\rm max}}^{k = {\bar k} + k_{\rm max}} \\
&\times{\mu_{ijk}({\bm \theta})^{N_{ijk}} e^{-\mu_{ijk}({\bm \theta})} \over N_{ijk}!},
}
where ${\bf d}$ denotes the data set consisting of the set of counts $\{N_{ijk}\}$ in \texttt{edisp} quartile $i$, annulus $j$, and energy bin $k$. The quantity $\bar k$ denotes the energy bin that contains $m_\chi$.  Above, $\mu_{ijk}({\bm \theta})$ is the mean model prediction in the appropriate detector bin for the model parameters ${\bm \theta}$.  The nuisance parameter vector consists of $3 \times 30 \times 2$ parameters, at fixed $m_\chi$. In each \text{edisp} quartile and annulus the background model is given by taking the power-law in flux $A_{ij} E_k^{\alpha_{ij}}$, with model parameters $A_{ij}$ and $\alpha_{ij}$ and across energies $E_k$, and forward modeling this power-law through the appropriate instrument response to obtain predicted detector counts. 

While the higgsino model has a well-defined and calculable annihilation cross section, it is useful to consider -- at fixed higgsino mass $m_\chi$ -- the modified higgsino model where the annihilation cross section is treated as a free parameter. Note that higher (lower) cross sections relative to the fiducial cross section simply re-scale the expected flux upwards (downwards).  
In this case, at fixed $m_\chi$ the signal model has one free model parameter $\langle \sigma v\rangle$ controlling the normalization of the signal. Towards this end, we introduce the $\mu$ parameter, which re-scales the annihilation cross section relative to the expected higgsino cross section $\langle \sigma v \rangle_{\rm higgsino}$ computed for the full higgsino model:\footnote{The $\mu$ parameter should not be confused by the mean expected number of counts $\mu_{ijk}$.} 
\begin{equation}
\langle \sigma v \rangle \equiv \mu \langle \sigma v \rangle_{\rm higgsino}.
\label{eq:mu}
\end{equation}
The $\mu$ parameter may be positive or negative, even though only positive values of $\mu$ are physical.
(A preference for a negative $\mu$ could arise in the real data from a downward fluctuation in the expected background.)
We make use of the $\mu$ parameter when referring to the annihilation cross section at fixed DM mass $m_\chi$ throughout the rest of this work.

Our parameter of interest at fixed $m_\chi$ is $\mu$, but we additionally have 120 nuisance parameters that we must profile over. (Note that we have, at fixed $m_\chi$, 4500 data points in the analysis.) 
We denote the vector of nuisance parameters by ${\bm \theta}_{\rm nuis}$, such that the full model parameter vector is ${\bm \theta}  = \{ \mu , {\bm \theta}_{\rm nuis} \}$.  It is then useful to compute the profile likelihood 
\es{eq:prof_LL}{
\lambda( \mu ) \equiv { p({\bm d}|{\mathcal M}, \{ \mu , \hat {\hat {\bm \theta}}_{\rm nuis}\}) \over p({\bm d}|{\mathcal M}, {\hat {\bm \theta}}) },
}
with ${\hat {\bm \theta}}$ denoting the parameter space vector that maximizes the likelihood and $\hat {\hat {\bm \theta}}_{\rm nuis}$ denoting the nuisance parameter vector that maximizes the likelihood at fixed $\mu$.
The one-sided discovery test statistic (TS) is defined as
\begin{equation}
q \equiv \left\{ 
\begin{array}{cl}
-2 \ln \lambda(0) & {\hat{\mu}} > 0, \\
0 & {\hat{\mu}} < 0,
\end{array}
\right.
\end{equation}
with ${\hat{\mu}}$ denoting the best-fit rescaling of the annihilation cross section. Assuming Wilks' theorem the Z-score ({\it i.e.}, the number of ``$\sigma$" quantifying how likely it is to observe such a value of $q$ under the null hypothesis) is given by $Z = \sqrt{q}$.
See {\it e.g.}~\cite{Safdi:2022xkm} for a review and a summary of how to use the profile likelihood to compute upper limits.

\begin{figure}[!t]
\begin{center}
\includegraphics[width=0.49\textwidth]{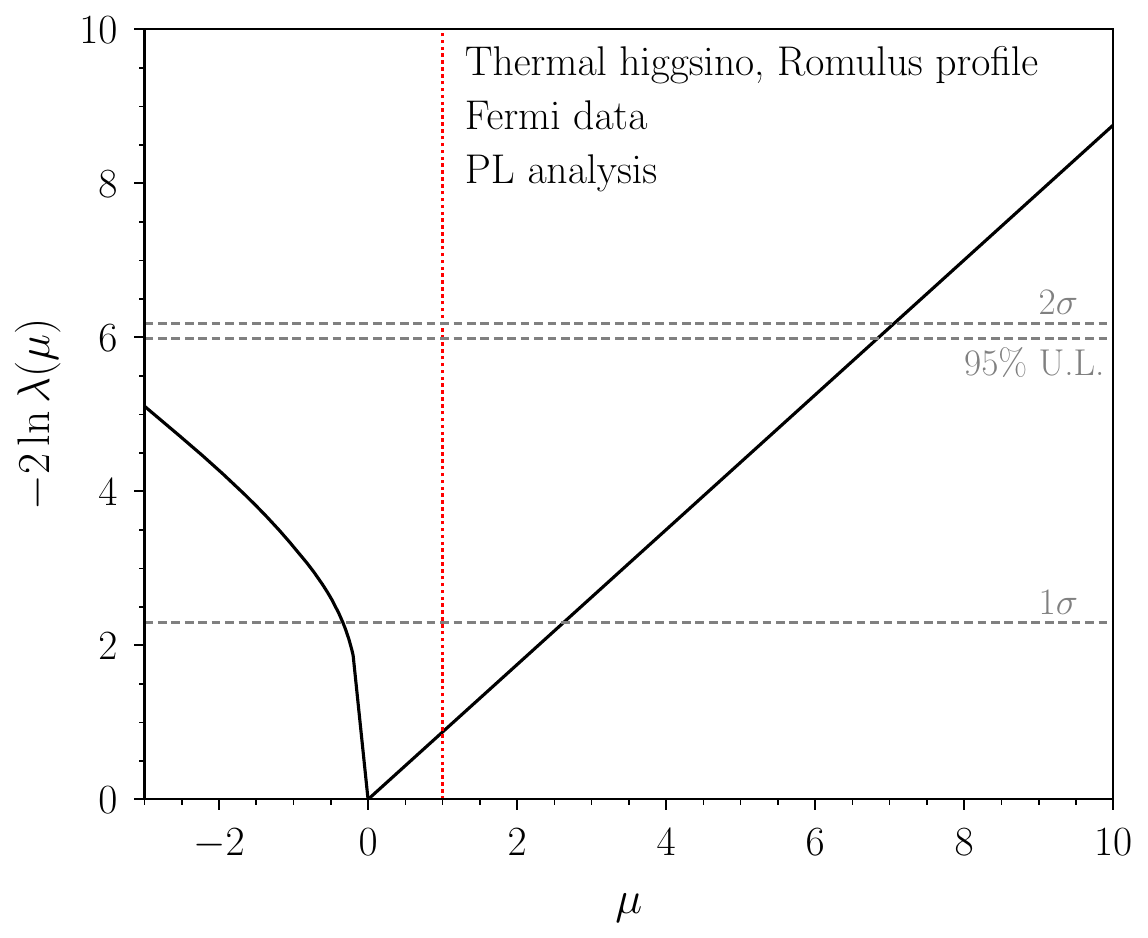}
\end{center}
\vspace{-0.7cm}
\caption{The profile likelihood in~\eqref{eq:prof_LL} as a function of the annihilation cross-section parameter $\mu$ for a 1.08 TeV higgsino signal from the Fermi analysis for a narrow spectral line at energies near $m_\chi$.
In particular, we show the statistic $-2 \ln \lambda(\mu)$; this statistic has a minimum at $\hat \mu = 0$, meaning that there is no preference for the signal model over the null hypothesis. Indeed, the likelihood behaves as that expected for a zero-count Poisson distribution in the vicinity of $\mu \approx 0$ in that it rises steeply for negative values of $\mu$ and is approximately linear at positive $\mu >0$. This is consistent with the low number of counts observed by Fermi in the inner Galaxy at energies near a TeV (see Fig.~\ref{fig:fermi_Data}). We approximate the profile likelihood thresholds for 1$\sigma$/2$\sigma$ containment intervals and the threshold for the 95\% one-sided upper limit as those appropriate for the Poisson distribution with zero counts, as indicated, though we note that this is not exact but a conservative choice.}
\label{fig:profile_LL_ill}
\end{figure}

\begin{figure}[!t]
\begin{center}
\includegraphics[width=0.49\textwidth]{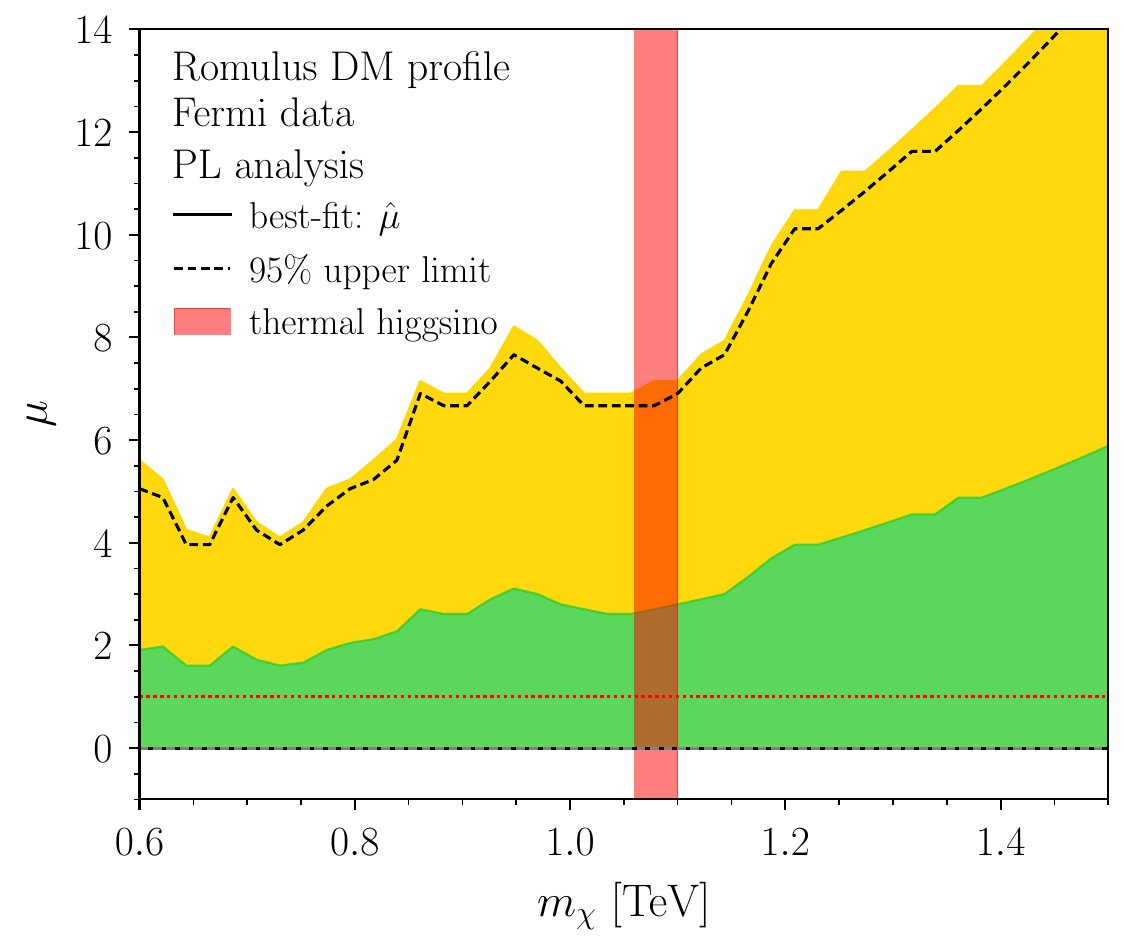}
\end{center}
\vspace{-0.7cm}
\caption{Using the Fermi data we compute the best-fit value for the rescaled annihilation cross section, at fixed mass $m_\chi$, for the higgsino model; we find $\hat \mu = 0$ for all $m_\chi$ given the low counts.  We also show the 1$\sigma$ and 2$\sigma$ containment intervals for $\mu$ at fixed $m_\chi$ in green and gold, respectively. (To compute these intervals we approximate the likelihood as that of a zero-count Poisson distribution; see text for details.) We also illustrate our 95\% upper limit.
The expected thermal higgsino model has a mass within the shaded vertical band and $\mu = 1$, which is also indicated with a dotted horizontal red line.  
(Note that at each $m_\chi$ we assume the higgsino is 100\% of the DM.)
This figure assumes the Romulus DM profile and focuses on the narrow line-like feature predicted for higgsino annihilations (see Fig.~\ref{fig:fermi_Data}).
Fermi has insufficient sensitivity at present to probe the line-like signal from the higgsino model, though as shown in Ref.~\cite{Dessert:2022evk} Fermi does have the necessary sensitivity to probe the higgsino model through the low-energy continuum. }
\label{fig:fermi_bf}
\end{figure}

While Wilks' theorem may be applied to good approximation in the discussions of H.E.S.S., CTA, and SWGO, we find that given the small number of counts that Fermi observes at energies $\sim$1 TeV (see Fig.~\ref{fig:fermi_Data}), the resulting profile likelihood (as given in~\eqref{eq:prof_LL}) behaves more like that expected for a zero-count Poisson distribution than expected in the large-counts regime. In particular, note that if the likelihood is given by a single Poisson distribution with mean $\nu$ and no counts are observed, then the best-fit signal parameter is $\hat \nu =0$ and the profile likelihood is given simply by $-2 \ln \lambda(\nu) = 2\nu$, for $\nu > 0$. (The profile likelihood is formally divergent for $\nu < 0$, meaning that such values should not be considered in this case.)

In Fig.~\ref{fig:profile_LL_ill} we show the profile likelihood $\lambda(\mu)$ for the signal parameter for $m_\chi = 1.08$ TeV.  The profile likelihood is nearly linear for $\mu > 0$, while is rises steeply for $\mu < 0$. The shape of this profile likelihood is consistent with the near zero-count Poisson distribution expected given the limited counts Fermi observes at TeV energies. We note that this was also true for the line analysis in~\cite{Foster:2022nva}. However, while~\cite{Foster:2022nva} assumed Wilks' theorem to compute upper limits, given the profile likelihoods, in this work we use a more conservative assumption that the distribution of TSs follows that of the zero-count Poisson distribution when computing upper limits and confidence intervals. (More accurately, one should compute these intervals through a Monte Carlo procedure, but given the clear lack of a signal we instead make the stated approximation.) For a zero-count Poisson distribution with mean predicted counts $\nu$, the best-fit is $\hat \nu = 0$ and the 1$\sigma$ (2$\sigma$) containment interval is given by all values of $\nu >0$ for which $-2\ln \lambda(\nu) \lesssim 2.30$ ($6.18$), while the one-sided 95\% upper limit is given by the value $\nu_{\rm 95}$ for which $-2\ln \lambda(\nu_{\rm 95}) \approx 5.99$.  We apply these profile likelihood thresholds to the ensemble of profile likelihoods we have from the Fermi analysis to compute confidence intervals and upper limits. (Using instead the Wilks' limit profile likelihood thresholds would lead to more aggressive confidence intervals and upper limits.)

As an aside, when projecting sensitivity to a putative DM annihilation signal, it is useful to work with the so-called Asimov data set~\cite{Cowan:2010js}.
For example, the mean expected discovery TS for a given annihilation cross section parameter $\mu$  may be found by computing the TS with the data set given by the mean expectation under the signal hypothesis, with no statistical noise added to the data (see~\cite{Cowan:2010js} for details on how to compute the containment intervals for the expected discovery TSs and upper limits).
We apply the Asimov procedure throughout this work when projecting sensitivity to the higgsino signal.

In Fig.~\ref{fig:fermi_bf} we show the best-fit annihilation cross section (parameterized by $\mu$) as a function of the higgsino mass within the vicinity of the thermal mass, whose range is indicated, from the analysis of the Fermi data assuming the Romulus DM profile.  As expected given the low numbers counts, the best-fit cross-section is identically zero for all masses shown. We also illustrate the 1$\sigma$/2$\sigma$ confidence intervals on $\mu$ at fixed DM mass $m_\chi$ in addition to the 95\% one-sided upper limit on $\mu$.  Note that at each $m_\chi$ we assume the higgsino is 100\% of the DM.  In the vicinity of the thermal higgsino mass ($m_\chi \sim 1.08$ TeV) the sensitivity to the line-like signal is around a factor of two too weak to probe the expected higgsino signal, for this DM profile.  The analysis finds no evidence for a higgsino signal. 

\begin{figure}[!t]
\begin{center}
\includegraphics[width=0.49\textwidth]{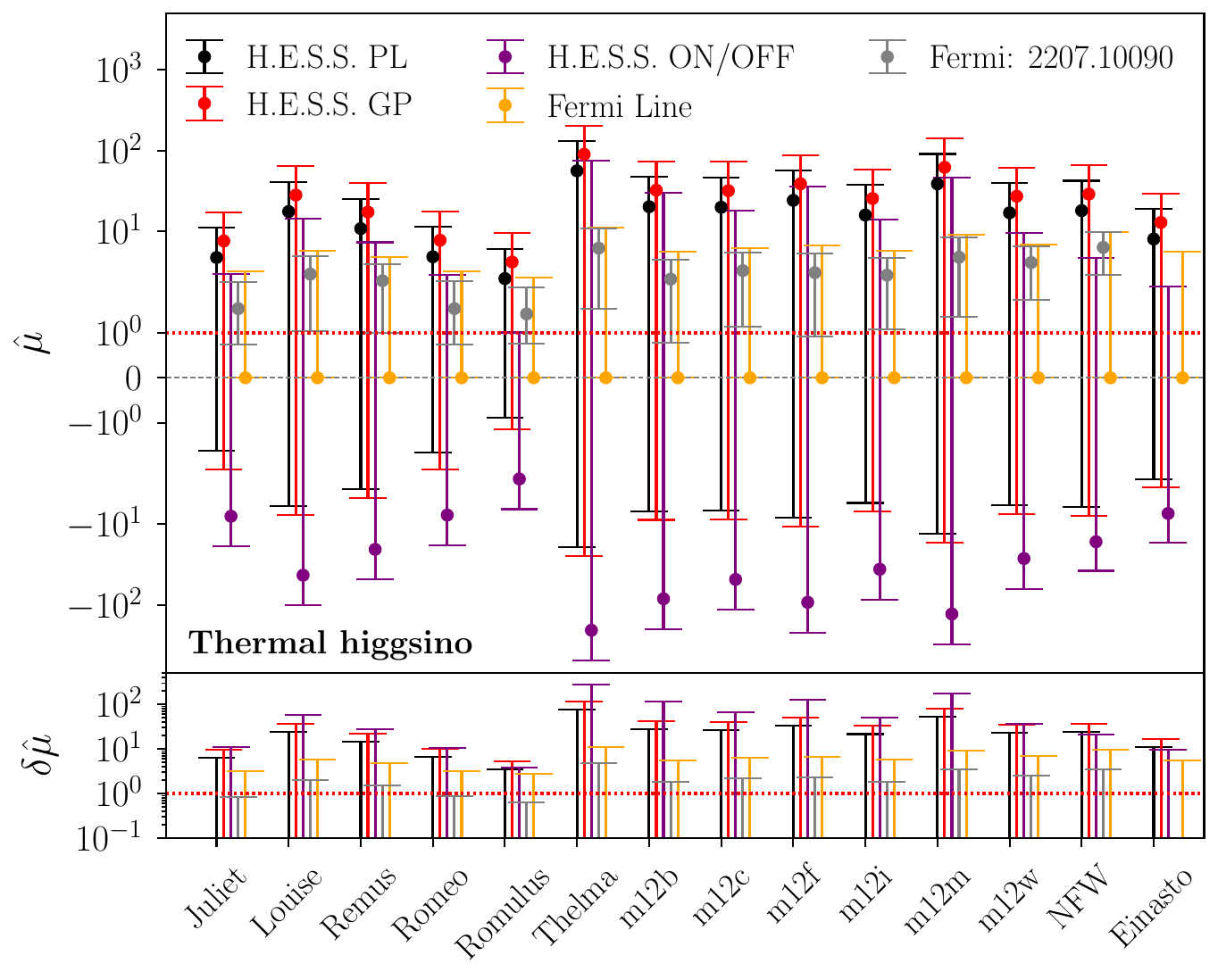}
\end{center}
\vspace{-0.7cm}
\caption{(Top panel) The best-fit cross section re-scaling parameter ($\hat \mu$), and 1$\sigma$ error bars, at $m_\chi = 1.08$ TeV for the analyses of the Fermi data and H.E.S.S. inner Galaxy data, assuming different DM profiles as indicated on the horizontal axis.  For Fermi we show the search for the line-like signal at the endpoint (this work) and the search for the lower-energy continuum (Ref.~\cite{Dessert:2022evk}), which is more sensitive but also subject to more systematic uncertainties from continuum mismodeling.  For H.E.S.S. we compare the results from this work for the power-law and GP analyses. 
The expected higgsino cross section is indicated by the horizontal, red line. (Bottom panel) We show the 1$\sigma$ error bars only for each of the analyses illustrated in the top panel. Those whose error bars are below the red band are able to probe the higgsino model at more than 1$\sigma$, assuming the indicated DM profile.
}
\label{fig:HESS_mu_data}
\end{figure}

In Fig.~\ref{fig:HESS_mu_data} we show the best-fit cross section recovered from this analysis at $m_\chi = 1.08$ TeV, along with the 1$\sigma$ uncertainties on the recovered best-fit, for the different DM profiles considered. 
We compare these results to that found in~\cite{Dessert:2022evk}, which performed a search for the continuum higgsino signal in Fermi. While Ref.~\cite{Dessert:2022evk} was able to achieve the necessary sensitivity to probe the expected higgsino cross section ($\mu = 1$) for some of the DM profiles considered, we are not able to reach this sensitivity through the line search (though we come close for {\it e.g.} Romulus). To emphasize this point in the bottom panel we compare the 1$\sigma$ uncertainties only, relative to the horizontal line that indicates the sensitivity needed to probe the higgsino at 1$\sigma$.
The figure also shows a point we demonstrate in the next section: H.E.S.S. data provides sub-leading sensitivity to the higgsino line signal.

\section{Analysis of public H.E.S.S. data}
\label{sec:HESS_real}

In this section we investigate and analyze the H.E.S.S. data presented in~\cite{HESS:2022ygk}.
We have two goals in this section.
First, we demonstrate that previous H.E.S.S. results, including the search in~\cite{HESS:2022ygk} for annihilation to continuum final states and possibly also~\cite{HESS:2018cbt} for annihilation to narrow gamma-ray lines, appear to have overstated their upper limits by around an order of magnitude.
In particular, we provide evidence that these works overestimated their analysis effective area, leading to artificially strong claimed sensitivity.
In Apps.~\ref{app:Line} and \ref{app:WW}, we show the impact of these changes on the specific DM analyses performed in those works.\footnote{To further test how widespread this issue may be, in App.~\ref{app:dwarf} we provide a cross-check on the search in \cite{HESS:2020zwn} for a DM signal in dwarf galaxies, and we find no evidence that the results stated there were overestimated.}

Our focus here, however, is on the higgsino.
Accordingly, after these corrections, we turn to our second goal which is to show that at masses $m_\chi \sim {\rm TeV}$ H.E.S.S. has strong but slightly subleading sensitivity to the line-like signal associated with higgsino annihilation, with the Fermi-LAT analysis described in the previous section slightly more sensitive. Neither instrument, however, is able to probe the expected higgsino annihilation cross section for the DM profiles considered.  In contrast, we show in the next section -- using analysis frameworks based on those we implement here on the H.E.S.S. data -- that the upcoming CTA will have sufficient sensitivity to probe the endpoint spectrum of higgsino annihilation.
Nevertheless, the H.E.S.S. data presented in~\cite{HESS:2022ygk} allows us to test analysis frameworks that could be applied in the future to {\it e.g.} upcoming data from CTA. We thus analyze the data in the context of the higgsino model for different analysis strategies. We also derive wino limits and show they rule out the thermal wino for all DM profiles considered. 

\subsection{Anchoring the H.E.S.S. effective area with astrophysical residuals}
\label{sec:HESS_ON_OFF}

We begin by considering the ON/OFF analysis strategy performed in~\cite{HESS:2022ygk}, which focused on continuum annihilation channels such as $\chi \chi \to W W$. We show first that using the analysis effective area presented in that work we are able to reproduce the claimed upper limits on the annihilation cross section but also that the astrophysical residuals would surpass the observed residuals by over an order of magnitude, pointing to an issue with the assumed analysis effective area.

\begin{figure}[!t]
\begin{center}
\includegraphics[width=0.49\textwidth]{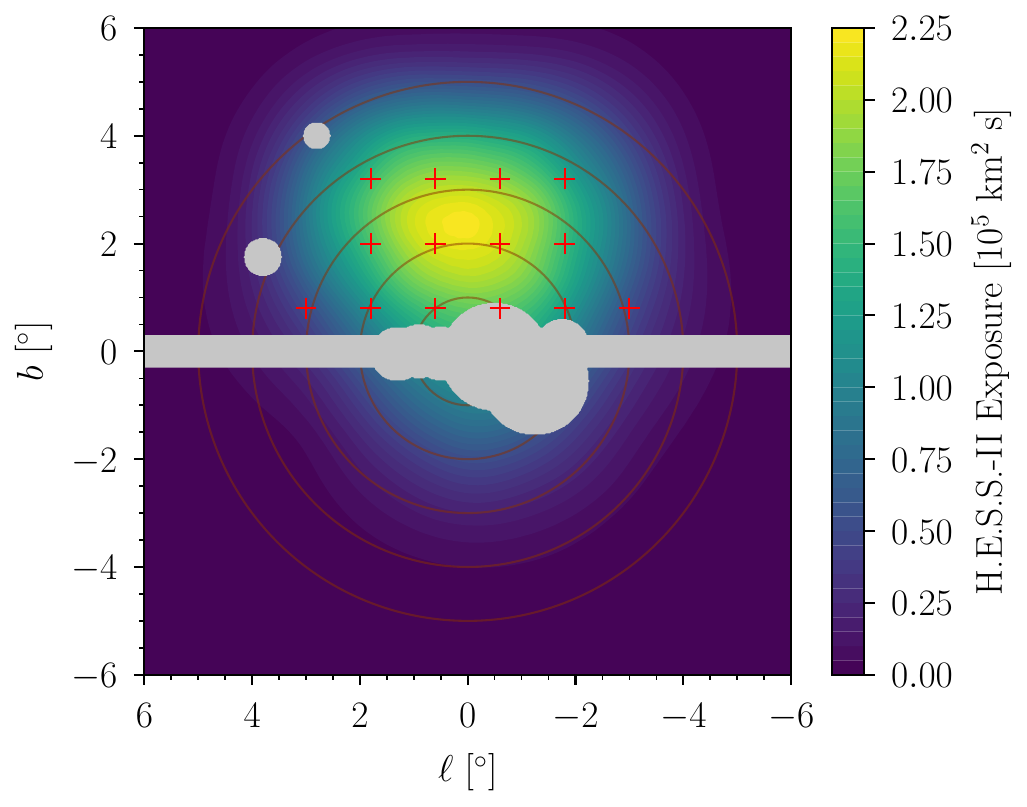}
\end{center}
\vspace{-0.7cm}
\caption{The H.E.S.S.-II exposure map, predominantly from the H.E.S.S. Inner Galaxy Survey (IGS), used in the $WW$ analysis in~\cite{HESS:2022ygk}. In our analysis of the public data in Sec.~\ref{sec:HESS_real} we do not use any spatial information, studying the data only as a function of energy not position.
However, overlaid here we show the five concentric annuli that we use as our spatial bins for our projected CTA analysis in Sec.~\ref{sec:CTA_projections}: five $1^\circ$ bins from 0-5$^\circ$.
}
\label{fig:HESS_II_exposure}
\end{figure}

First, we summarize the analysis performed in~\cite{HESS:2022ygk}.
That work predominantly used data collected between 2016 and 2020 as part of the H.E.S.S. inner Galaxy survey (IGS), though they also included data collected between 2014 and 2015 towards Sgr A*.  In Fig.~\ref{fig:HESS_II_exposure} we reproduce the exposure map presented in that analysis.
Ref.~\cite{HESS:2022ygk} specifies that 546 hrs of exposure time are used in that analysis, but while they provide the exposure map and pointing locations for the IGS, they do not specify the exposure times or acceptances on a pointing-by-pointing basis. Furthermore, the pointing locations for the Sgr A* observations, as well as the split of observing time between the 2014-2015 and 2016-2020 runs, are not public information.  The above information is necessary for a fully faithful reproduction of the analysis strategy described in this section.
In App.~\ref{app:HESS_detector} we describe our procedure for attempting to reproduce as accurately as possible the H.E.S.S. observations and appropriate aspects of the instrument response functions, which we use throughout this section.
 
The published H.E.S.S. analysis strategy has two key components, both of which reduce background but at the expense of reduced signal flux: (i) an ON/OFF background-subtraction procedure is used to account for the cosmic-ray-induced background using data collected during the same pointing; and (ii) masking of regions of high astrophysical emission to suppress astrophysical backgrounds. However, as we demonstrate, for the flux sensitivities quoted in \cite{HESS:2022ygk}, point (ii) is insufficient because of Galactic diffuse emission extending beyond the masks.
This can already be seen in Fig.~\ref{fig:diffuse}.
In more detail, the masks adopted in~\cite{HESS:2022ygk} can be seen as the gray regions in Fig.~\ref{fig:HESS_II_exposure}---a region which should be compared to the expected shape of the diffuse emission shown in Fig.~\ref{fig:fermi_diffuse_map}.
The Galactic plane is masked at $\pm 0.3^\circ$ along with the so-called PeVatron~\cite{HESS:2016pst} and high-energy gamma-ray point sources~\cite{HESS:2018pbp}.
The analysis in~\cite{HESS:2022ygk} assumes that these masks effectively eliminate astrophysical emission, at least to the necessary level of precision such that any further astrophysical emission need not be measured. We revisit this point shortly, however, and argue that this is not correct.

Let us describe in more detail the ON/OFF background subtraction procedure, though we refer to the original source~\cite{HESS:2022ygk} for more details. For a given pointing one defines an ON ROI and an OFF ROI within the FOV, defined symmetrically around the beam axis; {\it i.e.}, the OFF ROI is the reflection of the ON ROI about the beam center.
As these data sets are collected at the same time, and therefore under identical observational conditions, the assumption behind this method is that for a given pointing the expected cosmic-ray induced misidentified gamma-ray flux is comparable between the ON and OFF data; the OFF region provides a measure of the cosmic-ray background in the ON region.\footnote{Note that if a masked region is excluded in the ON region, its complement must also be excluded in the OFF region, and vice versa.}
Given this, in each energy bin Ref.~\cite{HESS:2022ygk} assigns a nuisance parameter that controls the background flux simultaneously in the ON and OFF regions. Profiling over this nuisance parameter in the large-counts limit, where the Poisson distribution may be treated in its Gaussian approximation, is equivalent to simply subtracting the OFF counts from the ON counts while increasing the error associated with the observed counts in the standard way appropriate for the subtraction of two normally-distributed random variables.

\begin{figure}[!t]
\begin{center}
\includegraphics[width=0.49\textwidth]{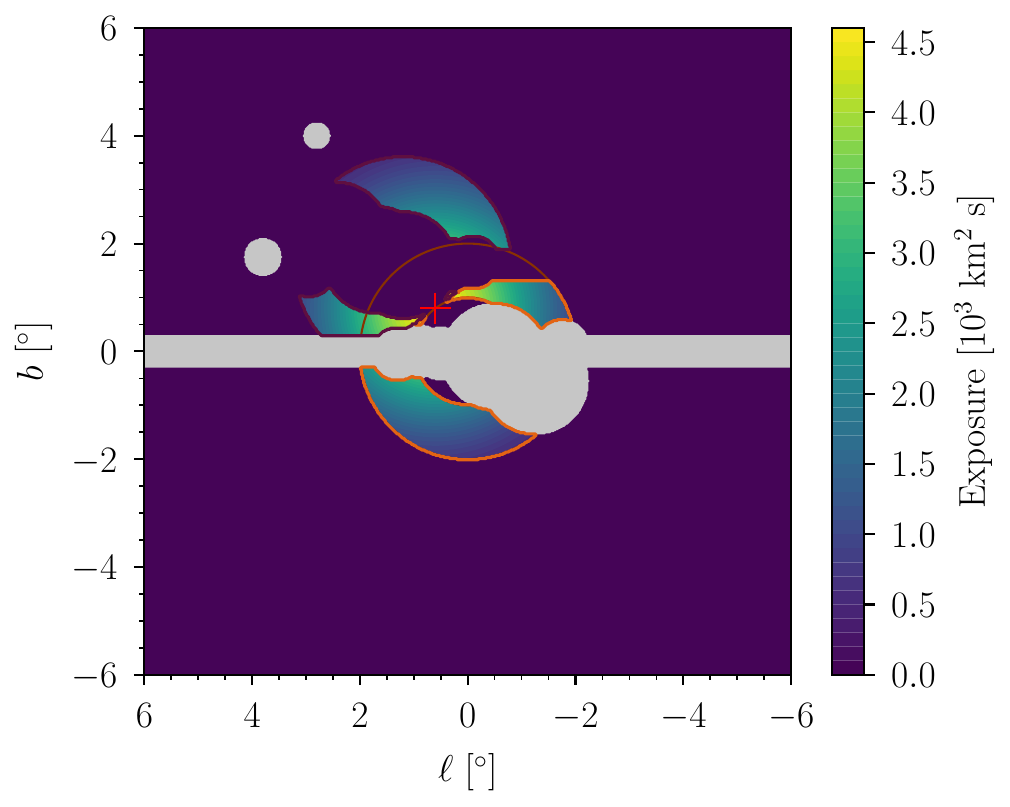}
\end{center}
\vspace{-0.7cm}
\caption{An example of ON (outlined in orange) and OFF (outlined in purple) regions symmetric around the pointing position centered at $\ell = 0.6^\circ$ and $b =0.8^\circ$, shown as a red plus.
We show the contribution to an annulus between $1-2^\circ$ from the GC, and with regions where the OFF region would have larger $J$-factor than the ON region omitted.
Further, the regions symmetric to masked areas have been removed from both ON and OFF regions to ensure identical exposure.}
\label{fig:mask_ring_1}
\end{figure}

In Ref.~\cite{HESS:2022ygk}, the data set was spatially divided into 25 concentric annuli extending from inner radius of $0.5^\circ$ to $2.9^\circ$, each of width $0.1^\circ$.
In this section, we do not incorporate any spatial information; unless stated otherwise, we take as ROI only a single annulus from $0.5^\circ$ to $3^\circ$,  covering the full set of ON regions in~\cite{HESS:2022ygk} and subtending a solid angle of $6.4\times 10^{-3}$ sr after masking.  This corresponds to the flux data made available in the publication, which is stacked over all 25 ON and OFF annuli.\footnote{We are able to estimate, though less reliably, the data in 8 sub-regions, and in App.~\ref{app:annuli} we analyze the data jointly over these sub-regions, finding consistent results with those presented in this section.} 

When available, spatial information can provide important information in distinguishing a DM signal from astrophysical backgrounds, and so in Fig.~\ref{fig:HESS_II_exposure} we show the five spatial bins we employ in the H.E.S.S. full-data projections discussed in the final part of this section.
These bins are each $1^\circ$ wide, and extend out to $5^\circ$. The first annulus, which we refer to as annulus 0, extends from $0^\circ$ to $1^\circ$ while the last, referred to as annulus 4, extends from $4^\circ$ to $5^\circ$. Our projections for CTA, discussed in the next section, use annuli of similar width, extending slightly further away from the GC.  

In Fig.~\ref{fig:mask_ring_1} we show an example pointing location and its contribution to annulus 1 (in our coarser annuli scheme, since it is easier to visualize than in the finer scheme of Ref.~\cite{HESS:2022ygk}).
We also indicate the complementary OFF region used to estimate the background flux for the ON contribution to that annulus. Note that this pointing location contributes to multiple annuli, though we just show in that figure the contribution to annulus 1.

We may then construct the stacked background-subtracted data in each annulus over the ensemble of H.E.S.S. pointings. We may also compute the model prediction for the expected number of signal counts in the ON region in each energy bin as a function the annihilation cross section $\langle \sigma v \rangle$, for a fixed DM mass $m_\chi$ and set of annihilation channels, by convolving with the instrument response. Note, however, in doing so it is crucial to account for the fact that signal counts are also produced in the OFF region, though at a smaller rate due to the fact that, by construction, the $J$-factors are smaller in the OFF region than in the ON region.

\begin{figure}[!t]
\begin{center}
\includegraphics[width=0.49\textwidth]{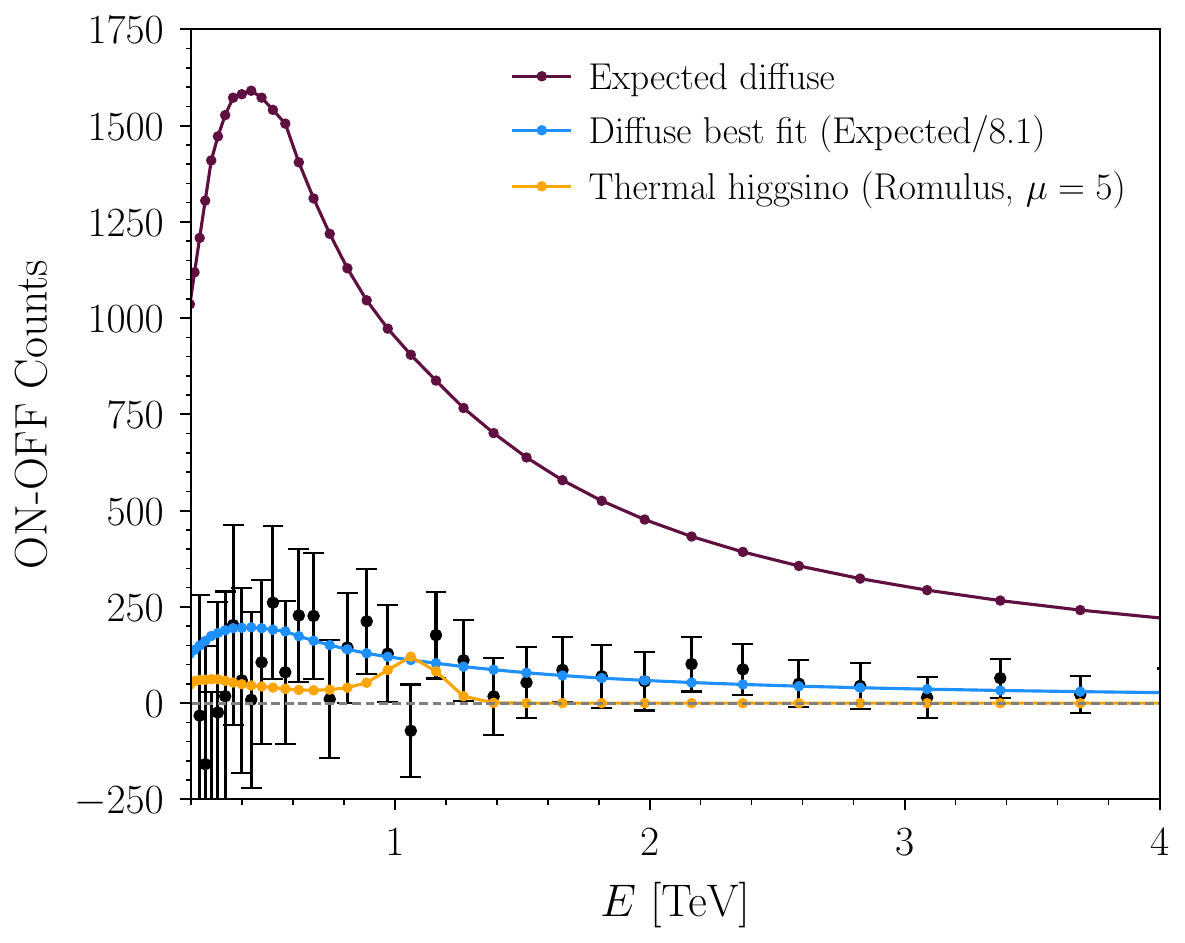}
\end{center}
\vspace{-0.7cm}
\caption{The difference between the ON and OFF counts from the H.E.S.S. DM annihilation search in the inner Galaxy in~\cite{HESS:2022ygk} stacked over all spatial bins from $0.5^\circ$ to $3^\circ$.
The residual counts appear strongly in tension with the expected residual emission from anisotropic astrophysical diffuse emission, which is indicated in purple, and estimated as described in the text.
The data prefers the presence of diffuse emission, but with an amplitude a factor of roughly 8.1 smaller than expected, as shown in blue.
This suggests the analysis effective area in \cite{HESS:2022ygk} has been overestimated by a similar factor.
Correcting the effective area by such a factor we find that H.E.S.S. is not sensitive to the thermal higgsino: the predicted signal scaled up by a factor of five ($\mu=5$) for the Romulus DM profile is shown in orange.
}
\label{fig:HESSonoffDiff}
\end{figure}

In order to determine the sensitivity of the H.E.S.S.-II data set to a higgsino annihilation signal, we also need a model for the effective area used in the analysis of \cite{HESS:2022ygk}.
We can extract their analysis effective area directly, by taking the ratio of Figs.~5 and 6 of the Supplemental Material in \cite{HESS:2022ygk}; see App.~\ref{app:HESS_detector} for details.
In order to cross check that the effective area we extract in this manner matches that which was used in their work, in App.~\ref{app:WW} we show that using the extracted analysis effective area from this procedure we are able to precisely reproduce the DM annihilation limits to $W^+W^-$ final states. This provides confidence in our general understanding of the analysis in~\cite{HESS:2022ygk}.

Nevertheless, as we now outline, we are concerned that \cite{HESS:2022ygk} overestimated their analysis effective area.
In particular, we identify three important issues.
First, the effective area in \cite{HESS:2022ygk} appears to be (numerically) too large, given the physical dimensions of H.E.S.S..  The analysis effective area is the effective area averaged over the ROI, accounting for {\it e.g.} the fall-off of the effective area from the beam center (see the right panel of Fig.~\ref{fig:Aeff}).  
Given our reconstructed pointing locations and exposure times, described in App.~\ref{app:HESS_detector}, we are then able to reconstruct the on-axis effective area from the analysis effective area.
This effective area approaches nearly 1 km$^2$ at high energies, which is around a factor of 6 higher than the expected physical on-axis effective area of H.E.S.S..  For example, we illustrate the on-axis effective area of H.E.S.S.-I in Fig.~\ref{fig:Aeff}, as extracted from the public data release, which is expected to roughly match H.E.S.S.-II at high energies but is much lower than our inferred on-axis effective area from~\cite{HESS:2022ygk}.  (Note that some observation-to-observation scatter in the effective area is expected, though not that at the level of a factor of 6.)

The second issue with the analysis effective area used in~\cite{HESS:2022ygk} is that it vastly over predicts the astrophysical residual emission in the ON minus OFF data versus the observed counts.  We illustrate this point in Fig.~\ref{fig:HESSonoffDiff}.  In that figure we show the ON minus OFF data over the full ROI as a function of energy. We then use the inferred pointing locations and the analysis effective area used in~\cite{HESS:2022ygk} to forward model the Galactic diffuse emission (computed using the Fermi-LAT diffuse model) into the predicted, mean residual ON minus OFF counts. This residual is illustrated in Fig.~\ref{fig:HESSonoffDiff} and is seen to overproduce the observed residual counts by approximately a factor of 8.
(In this figure and throughout our analysis of the H.E.S.S.-II data set we follow the energy binning of~\cite{HESS:2022ygk}: 67 logarithmically-spaced energy bins between $\sim$$0.172$ TeV and $\sim$$66.218$ TeV.)

Finally, as has been advanced in previous sections, the analysis effective area extracted from~\cite{HESS:2022ygk}, in conjunction with the total reported photon counts in their Tab.~I and the known total exposure time of 546h,  directly implies an unrealistically high efficiency of cosmic-ray rejection, $\gtrsim 99.9\%$.  This marks an observed flux roughly a factor of 13 lower than that seen by H.E.S.S.-I, although certainly improvements in reconstruction algorithms, more aggressive cuts, etc.~may explain some portion of this discrepancy. 

The first two of these issues above point to the analysis effective area being overestimated by a factor between 6 and 8. In particular, we stress that Fermi has measured the Galactic diffuse emission at a TeV, and since Fermi and H.E.S.S. observe the same sky that emission must also be observed, with the same normalization, by H.E.S.S..  We thus use the ON minus OFF residual counts to determine, in a data-driven way, the analysis effective area of H.E.S.S..  To do so, we use the spectral shape of the analysis effective area as provided by H.E.S.S. in~\cite{HESS:2022ygk} but we float the overall normalization of the effective area to determine the best-fit of the astrophysical diffuse emission to the residual ON minus OFF data.  In particular, restricting to energies $0.5 \, \,{\rm TeV} < E < 2 \, \, {\rm TeV}$, which is within the energy range of the Fermi-LAT, we find that the best-fit effective area is a factor of $\sim$$8.1$ times smaller than the analysis effective area used in~\cite{HESS:2022ygk}. The best-fit diffuse model prediction, after fixing the effective area normalization at its best-fit value, is illustrated in Fig.~\ref{fig:HESSonoffDiff}.  In the energy range $0.5 \, \,{\rm TeV} < E < 2 \, \, {\rm TeV}$ we find 3.4$\sigma$ evidence in favor of the diffuse model in the residual emission; that evidence grows to 5.0$\sigma$ when extending to 10 TeV, through we stress that the Fermi-LAT diffuse model is not trustworthy at such high energies as it has not been calibrated to data above $\sim$2 TeV.  Note, however, that even though the fit is only performed up to 2 TeV, the diffuse model prediction in Fig.~\ref{fig:HESSonoffDiff} continues to accurately describe the data at higher energies. In contrast, the diffuse model prediction over-predicts the residual emission at low energies, where the effective area changes rapidly with energy.

Our detection of Galactic diffuse emission in the ON minus OFF residuals is consistent with the claim made in Ref.~\cite{HESS:2022ygk} that above $\sim$0.48 TeV the data show an approximately 5$\sigma$ significance excess in the stacked ON minus OFF data.  We identify this excess with the expected residual astrophysical diffuse emission, as measured by {\it e.g.} the Fermi-LAT, after correcting the effective area. 

Moving forward, we fix the analysis effective area to give the correct (Fermi-LAT-level) astrophysical diffuse emission normalization in the $0.5 \, \,{\rm TeV} < E < 2 \, \, {\rm TeV}$ energy range, thereby rescaling down the value used in \cite{HESS:2022ygk} by a factor of $\sim$8.1.
Our inferred on-axis effective area is now in-line with that given in the H.E.S.S. public data release, solving the issue that the analysis effective area used in~\cite{HESS:2022ygk} appears much too large.  We caution, however, that our inference of the H.E.S.S. effective area is uncertain at the 10's of percent level for a number of reasons. First, it relies on the Fermi diffuse model, which as shown in {\it e.g.} Fig.~\ref{fig:diffuse} is only measured to ${\mathcal O}(10\%)$ accuracy at energies around a TeV. Secondly, we rely on our inference of the pointing locations and exposure times of H.E.S.S. in the IGS (see App.~\ref{app:HESS_detector}), which are uncertain at a similar level of accuracy.  Finally, we acknowledge there is an intrinsic variation to the acceptance on an observation-by-observation basis, and we only hope to reconstruct here the representative average quantity that connects fluxes to total observed counts.

The science implications of the reduced effective area versus that assumed in~\cite{HESS:2022ygk} are significant. In App.~\ref{app:WW} we show that using our more accurate effective area reduces the upper limits on the annihilation cross sections to $W^+W^-$ from the H.E.S.S. IGS data by almost an order of magnitude.  In App.~\ref{app:Line} we use the H.E.S.S. inner Galaxy data to search for DM annihilation to monochromatic photon final states ($\gamma\gamma$). This search is analogous to that performed in~\cite{HESS:2018cbt}, though we incorporate more data in the inner Galaxy than used in~\cite{HESS:2018cbt}. The search $\chi \chi \to \gamma\gamma$ is similar to but slightly different than that of higgsino annihilation because of the lack of end-point corrections.  Strikingly, however, our resulting upper limits are significantly weaker than those presented in~\cite{HESS:2018cbt}, calling into question the validity of the upper limits presented in~\cite{HESS:2018cbt} as well.
The issue does not appear to impact all H.E.S.S. analyses, however.
In App.~\ref{app:dwarf} we revisit some of the searches for DM annihilation in dwarf galaxies performed by H.E.S.S. (such as Reticulum II) and estimate a sensitivity in good agreement with those presented in that work.

\subsection{ON/OFF higgsino analysis of the H.E.S.S. inner Galaxy data}
\label{sec:hess_ON_OFF}

Using the updated effective area, we now perform an analysis of the ON minus OFF data presented in~\cite{HESS:2022ygk} and illustrated in Fig.~\ref{fig:HESSonoffDiff} for evidence of higgsino DM annihilation. To emphasize the caveats described above, one should keep in mind that our effective area -- and hence the final results -- are uncertain at the 10's of percent level.  Also note that a true re-analysis of H.E.S.S.-II IGS data by the collaboration should achieve a slightly higher sensitivity than we do, as they can leverage the annulus-by-annulus spatial information that we lack (though see App.~\ref{app:annuli}). 

We perform our analysis by computing the likelihood 
\begin{equation}
p({\bf d} | {\mathcal M}, {\bm \theta}) = \prod_{k} {\mathcal N}(N_{k}, \sigma_{k}^2 | \mu_{k} = \mu({\bm \theta})),
\label{eq:LL_gauss}
\end{equation}
where $k$ is an index over energy bins, and the likelihood in each bin is given by,
\begin{equation}
{\mathcal N}(N,\sigma^2|\mu) = {1 \over \sqrt{2 \pi} \sigma} e^{-{(N - \mu)^2 \over 2 \sigma^2}},
\end{equation}
which is the normal distribution with variance $\sigma^2$, observed value $N$, and expected value $\mu$.
(The use of a Gaussian rather than Poisson likelihood is justified by the large number of counts per bin.)
The OFF-subtracted ON data in each  energy bin is denoted $N_{k}$, with variance $\sigma_{k}^2$ arising from Poisson counting statistics in the ON and OFF regions; the full data vector is then denoted ${\bm d} = \{ N_{k}, \sigma_{k}^2 \}$.
The DM annihilation model ${\mathcal M}$ has parameter vector ${\bm \theta}$ and predicts $\mu_{k} = \mu_k({\bm \theta})$ counts in energy bin $k$, accounting self-consistently for the amount of signal that is lost during the OFF subtraction procedure.

Our model parameter vector ${\bm \theta}$ includes the signal parameter of interest $\mu$, at fixed $m_\chi$, in addition to an overall normalization nuisance parameter $A_{\rm bkg}$ for the Galactic diffuse emission, whose spectral template is determined from the Fermi diffuse model.  By definition of how we determine the effective area normalization, the best-fit value of $A_{\rm bkg}$ under the null hypothesis ($\mu = 0$) and including data between $0.5 \, \, {\rm TeV} < E < 2 \,\, {\rm TeV}$ is $A_{\rm bkg} = 1$. However, we profile over $A_{\rm bkg}$ when searching for evidence of a DM signal. Note that in Fig.~\ref{fig:HESSonoffDiff} we illustrate a higgsino signal assuming the Romulus DM profile in the ON minus OFF data with an enhanced amplitude of $\mu=5$.

\begin{figure}[!t]
\begin{center}
\includegraphics[width=0.49\textwidth]{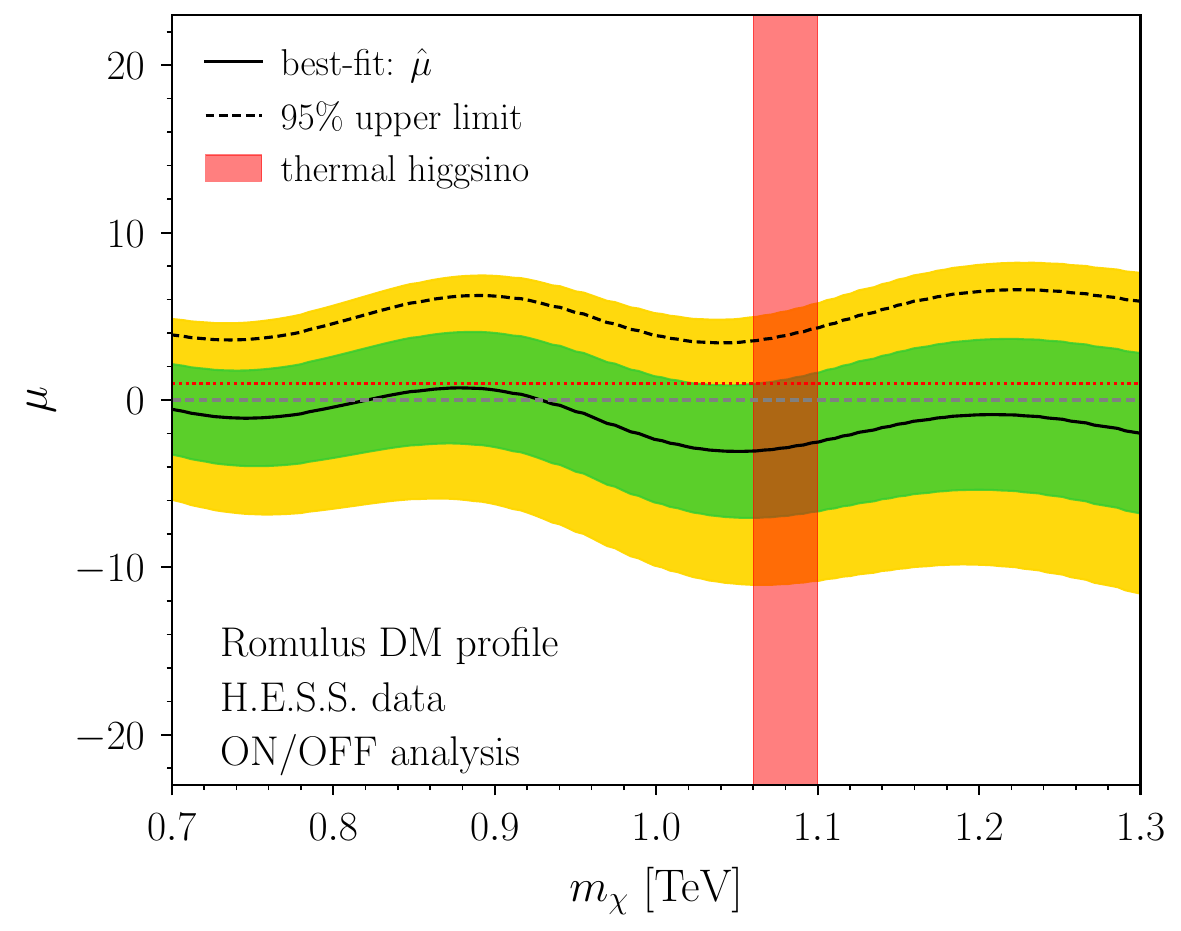}
\end{center}
\vspace{-0.7cm}
\caption{As in Fig.~\ref{fig:fermi_bf} but for the analysis of H.E.S.S. ON minus OFF data, with our corrected H.E.S.S. effective area. We find no evidence for higgsino DM annihilation in this analysis and also no evidence for mismodeling after accounting for emission associated with the Fermi diffuse model. }
\label{fig:HESS_bf_ON_OFF}
\end{figure}

In our analysis of the ON minus OFF data we incorporate all data with $0.6 \, \, {\rm TeV} < E < 4 \, \, {\rm TeV}$.  We do not use a floating energy window in this analysis because the Fermi diffuse model is able to accurately describe the residuals of the ON minus OFF data to the necessary precision over the full energy range considered. We then search for evidence of higgsino DM annihilation for $0.7 \, \, {\rm TeV} < m_\chi < 1.3 \, \, {\rm TeV}$, assuming the higgsino is 100\% of the DM at each mass.  In Fig.~\ref{fig:HESS_bf_ON_OFF} we illustrate the best-fit $\mu$ values and uncertainties at fixed $m_\chi$, for each value of $m_\chi$, assuming the Romulus DM profile.  We find no evidence in favor of the higgsino model, but also the sensitivity of this search is a factor of a few too weak to probe the higgsino model for this DM profile. In particular, our recovered $\mu$ parameter is consistent both with the null hypothesis $\mu = 0$ and the higgisino model prediction $\mu = 1$ at 1$\sigma$ significance for $m_\chi$ in the expected mass range for thermal higgsino DM, as indicated.  
To directly compare the sensitivity of the ON minus OFF H.E.S.S. search to the Fermi searches we overlay the best-fit cross-sections at $m_\chi = 1.08$ TeV for the different DM profiles considered in Fig.~\ref{fig:HESS_mu_data}.  Note that the H.E.S.S. sensitivities are still a factor of a few worse than the sensitivity achieved in~\cite{Dessert:2022evk} from the continuum search in Fermi-LAT data and are also slightly worse than those achieved in the previous section using Fermi-LAT data to search for the endpoint feature. In addition, the strength of the ON/OFF analysis is sensitively dependent on the assumed underlying DM profile, performing much worse on cored profiles such as Thelma and m12m compared to the cuspier NFW and Einasto.
In App.~\ref{app:annuli} we show that comparable but slightly improved sensitivity is achieved by using the ON minus OFF data broken up by annuli.

\subsection{Direct analysis of the ON H.E.S.S data}\label{sec:hess_on_only}

As described already, H.E.S.S. utilizes the ON minus OFF data set in order to account for the cosmic-ray-induced backgrounds.  While that approach has the advantage of removing the dominant background, it has the disadvantages of: (i) reducing the data volume that can be incorporated into the analysis, since the data must be divided into ON and complementary OFF regions; and (ii) subtracting a signal component from the data, since the DM annihilation signal appears in both the ON and the OFF data.  An alternate approach to analyzing the H.E.S.S. data is to avoid the separation in ON and OFF regions and to simply analyze the full collected data directly, modeling instead of subtracting the cosmic-ray-induced backgrounds.  Unfortunately, given the data released in~\cite{HESS:2022ygk} we only have access to the subsets of data in the independent ON and OFF regions over the full ROI.\footnote{In App.~\ref{app:annuli} we present an analysis of the data broken up into sub-annuli, though the data reconstruction is less reliable.}  In particular, more data should be available than that presented in~\cite{HESS:2022ygk} if one does not use ON and OFF separation. We explore the question of what could be gained by including these data in the analysis later in this section.

For now, we analyze the ON data over the full ROI, as presented in~\cite{HESS:2022ygk}, for evidence of the higgsino model, without performing OFF subtraction.  
To implement this analysis, however, we  need a spectral model for the cosmic-ray-induced emission that resides in the ON region.
With access to the full archival H.E.S.S. data, one could develop a spectral template for the cosmic-ray-induced background by using observations taken away from the GC with the same reconstruction algorithms and similar observing conditions as for the ON data set of interest.
(See {\it e.g.}~\cite{HESS:2018cbt}.)
Without access to these data, we instead implement two different data-driven approaches to modeling the cosmic-ray-induced background, and also the astrophysical background, one parametric and the other non-parametric, both described in detail below.
The approaches we employ are both less sensitive and robust relative to the ideal analysis, but they illustrate classes of analysis frameworks that may be more accurately performed with access to the full H.E.S.S. data archive and instrument response models.  These analyses also provide frameworks that may be implemented with the future CTA and SWGO data.

\subsubsection{Approach 1: Parametric background modeling}

As a first approach, we model the combination of astrophysical and cosmic-ray-induced backgrounds counts by a folded power-law, such that in a given energy bin $i$ the total expected number of counts from the background model is given by $\mu_k = A_{\rm eff}^k A E_k^{-n} + C$, such that we now have three nuisance parameters ${\bm \theta}_{\rm nuis} = \{ A,n,C\}$ that are profiled over in the analysis.
This model is undoubtedly simplistic and could be improved in a straightforward way by using more accurate spectral templates for the background model components, as we discuss further in the following section.
Without such improvements, we would not expect a single power law to provide an accurate description of the data, given the large number of counts.
Nevertheless, if we restrict our analysis to increasingly narrow energy ranges, the power-law model is expected to provide an increasingly good fit, and this is exactly the approach we adopt.
While narrow energy windows help mitigate mismodeling, this comes at the cost of a less constrained background model and heightened degeneracy between the signal and background components; thus narrower energy ranges are more robust but less sensitive than wider energy ranges (see {\it e.g.} Refs.~\cite{Dessert:2018qih,Dessert:2023fen} for a discussion).

\begin{figure*}[!tb]
\begin{center}
\includegraphics[width=0.49\textwidth]{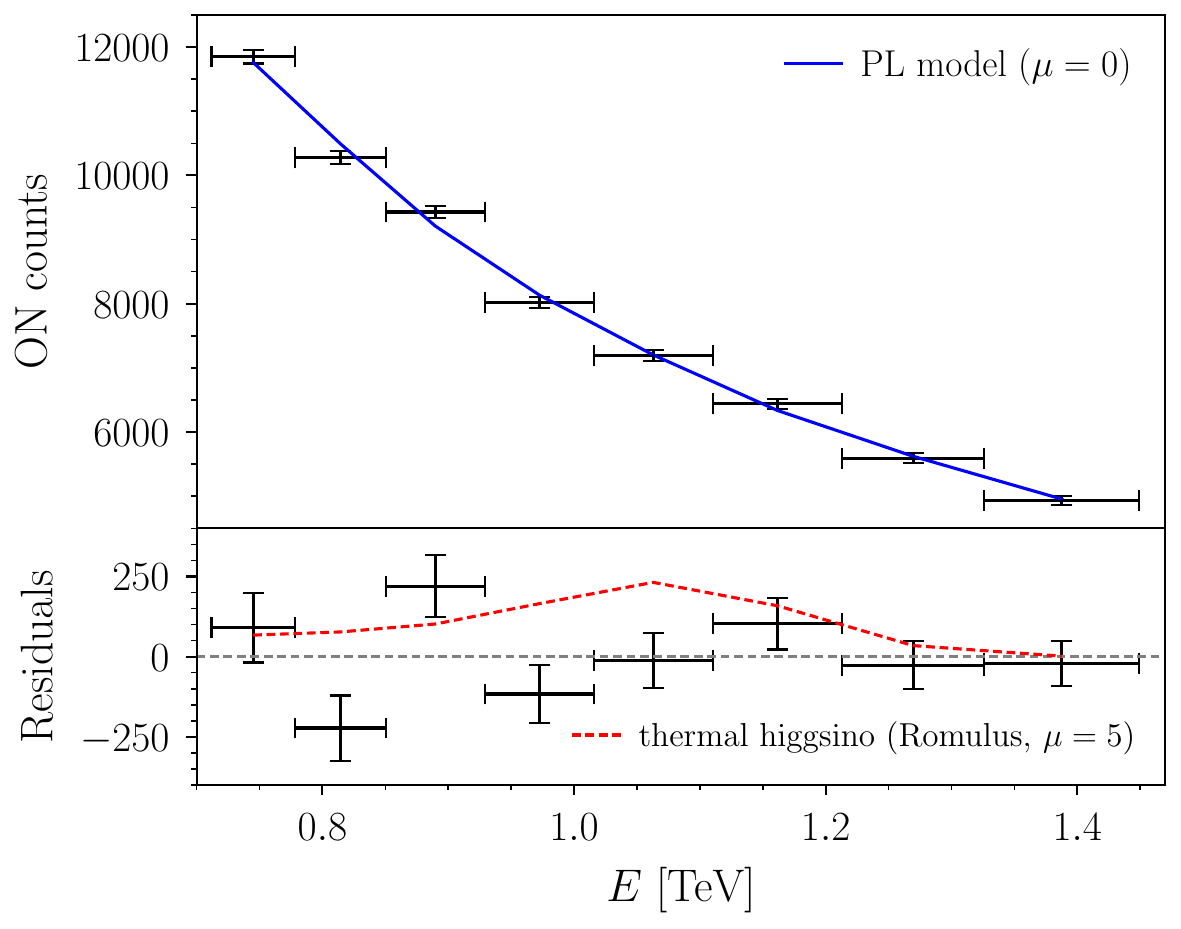}
\includegraphics[width=0.49\textwidth]{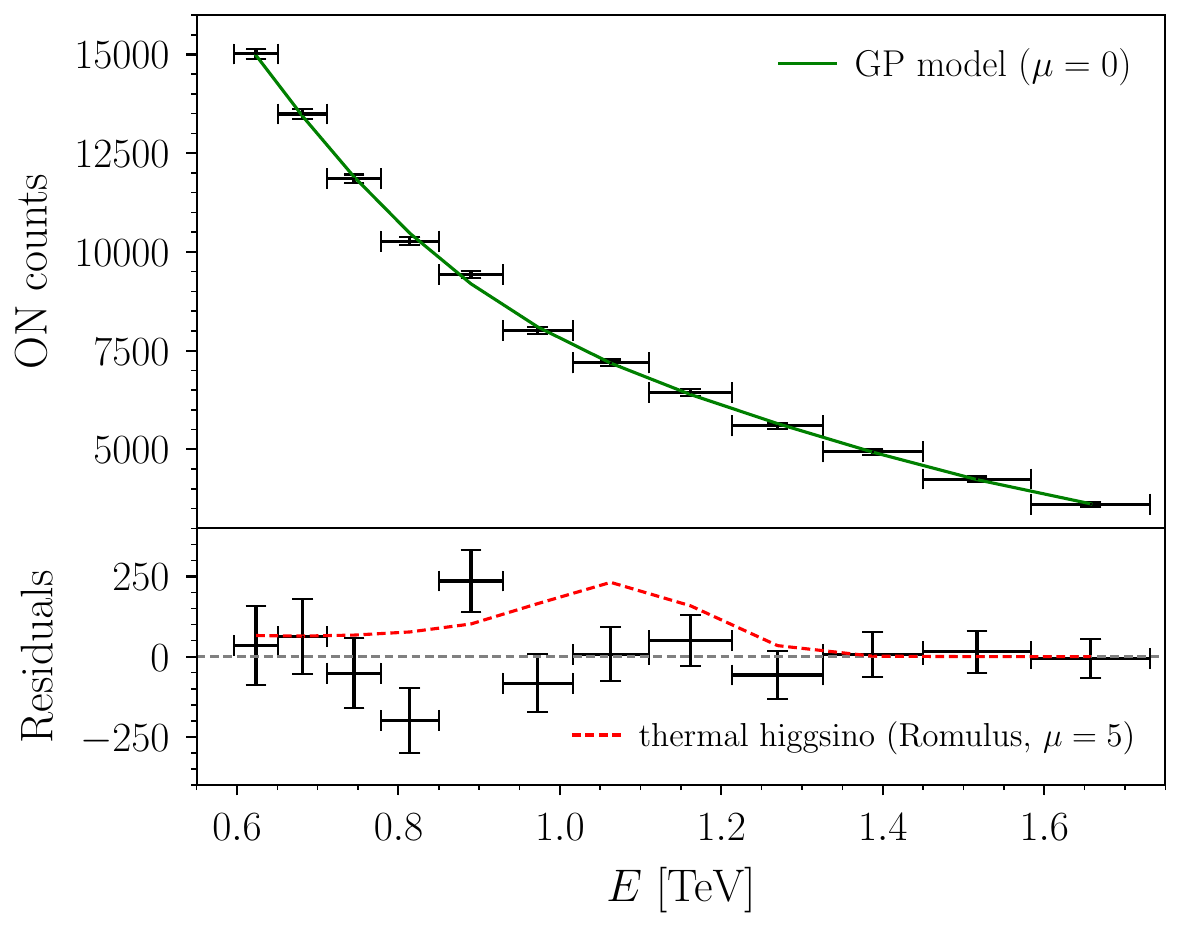}
\end{center}
\vspace{-0.7cm}
\caption{(Left panel) The ON counts provided by H.E.S.S. for their inner Galaxy analysis summed over the ON ROI (see App.~\ref{app:HESS_detector} for our reconstruction of these counts).  We model the data under the null hypothesis using a power-law model plus a constant off-set.  We use a floating energy window approach with six energy bins centered on the DM mass; this figure illustrates the null fit for a $m_\chi = 1.08$ TeV higgsino.
The bottom panel shows the residuals from the best-fit null model, indicating no clear signs of mismodeling.
We also overlay the expected spectral signal for a thermal higgsino rescaled up by a factor of 5 ($\mu=5$) for the Romulus DM profile.
(Right panel) As in the left panel, but for our alternate analysis framework using GP modeling.
This analysis also uses a sliding-energy window, and here we again illustrate the fit for $m_\chi = 1.08$ TeV.
}
\label{fig:HESS_null_signal_fits}
\end{figure*}

For the present analysis we choose a conservative sliding-energy window that, for a given DM mass $m_\chi$, contains 8 total energy bins; the energy bin containing the DM mass $m_\chi$, three energy bins directly above this bin, and then four energy bins directly below this bin.\footnote{Recall the data are binned logarithmically in energy into sixty-seven bins with the endpoints set to $0.172$ TeV and $66.218$ TeV.}
See Fig.~\ref{fig:HESS_null_signal_fits} (left panel) for an illustration of the data for the thermal higgsino ($m_\chi = 1.08$ TeV). 
We verify that small modifications to this window size do not qualitatively affect the final results. 
Taken together, we employ the sliding-window method to search for higgsino DM with masses between 0.7 and 1.3 TeV, as in the analyses presented in the previous sub-section using the ON minus OFF data.  

\subsubsection{Approach 2: non-parametric GP modeling}

Our second approach is non-parametric and thus less reliant on the assumptions inherent in the ad-hoc power-law background model.
In particular, we use GP modeling to describe the background emission non-parametrically.
This approach is justified here since we do not have alternate methods for constructing a cosmic-ray-induced background spectral template. 

To implement this analysis, we follow the procedure developed in {\it e.g.}~\cite{Frate:2017mai,Foster:2021ngm} for using GP models to describe background emission whose spectral shape is unknown for the purpose of searching for narrow spectral signatures.
The basic intuition is that when searching for narrow spectral features arising from new physics, the precise form of the background emission is unimportant so long as the background spectrum only varies over energy scales much larger than the spectral width of the signal, which for narrow DM lines is roughly the detector energy resolution.
With enhanced flexibility compared to the power law, the method can also partly account for the expected systematic fluctuations in the cosmic-ray-induced background.

We model the data under the null hypothesis as a combination of a zero-mean GP model and a constant off-set, which is treated as a nuisance parameter.
The zero-mean GP model has a Gaussian kernel \mbox{$K(E, \tilde E) = A_{\rm GP} \exp\big[ - (E - \tilde E)^2 / (2 \sigma_E^2) \big]$}.
This kernel describes the fluctuations in the mean model prediction, with amplitude $A_{\rm GP}$ and correlation length $\sigma_E$.
We treat $A_{\rm GP}$ as a nuisance parameter, while we fix $\sigma_E = 0.6 \times m_\chi$, for a given DM mass $m_\chi$, so that the GP model fluctuates over energy scales much wider than the expected signal.
We implement the GP fit using \texttt{george}~\cite{2015ITPAM..38..252A}. As in \cite{Frate:2017mai} we use the marginal likelihood determined from the GP fit in order to perform a frequentist profile likelihood analysis to constrain the signal parameter of interest.

As in the power-law analysis, we use a sliding-window energy range.
For a given DM mass $m_\chi$, we include the energy bin containing $m_\chi$, five energy bins above, and then six energy bins below.
We use more energy bins than in the power-law analysis because the GP analysis is less sensitive to mismodeling over wide energy ranges than for parametric models (see, {\it e.g.}, the discussion in~\cite{Foster:2021ngm}) and because the wider energy window helps with convergence.
We again search for DM with masses between 0.7 and 1.3 TeV.

\begin{figure}[!htb]
\begin{center}
\includegraphics[width=0.49\textwidth]{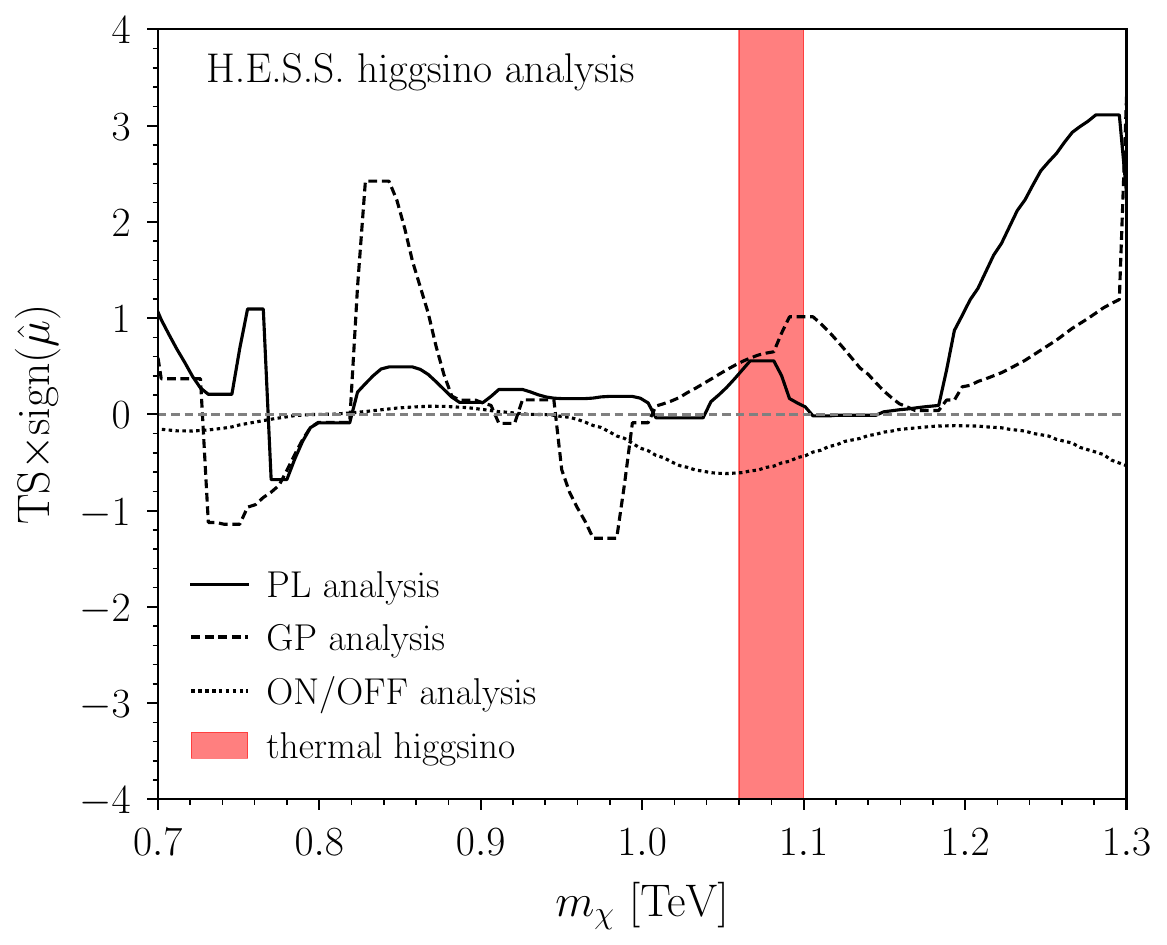}
\end{center}
\vspace{-0.7cm}
\caption{The discovery test statistics (TSs) for the three H.E.S.S. higgsino analyses we consider, with the thermal higgsino mass indicated.
None of the three analyses find evidence in favor of the higgsino model over the mass range considered.  Note that the jagged nature of the power-law (PL) and GP analysis TSs is physical and due to finite-binning effects in the sliding window approach.}
\label{fig:HESS_TS}
\end{figure}

\begin{figure*}[!htb]
\begin{center}
\includegraphics[width=0.49\textwidth]{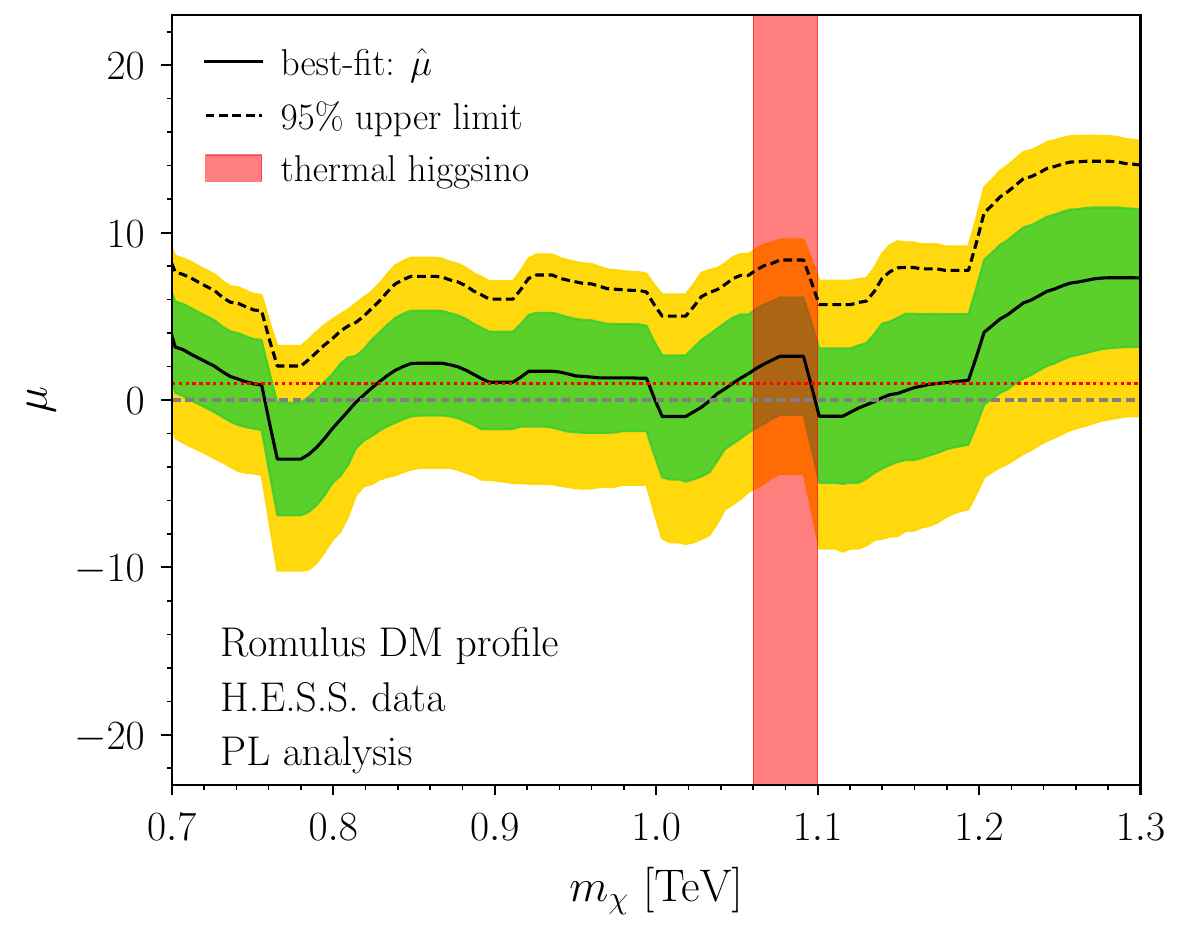}
\includegraphics[width=0.49\textwidth]{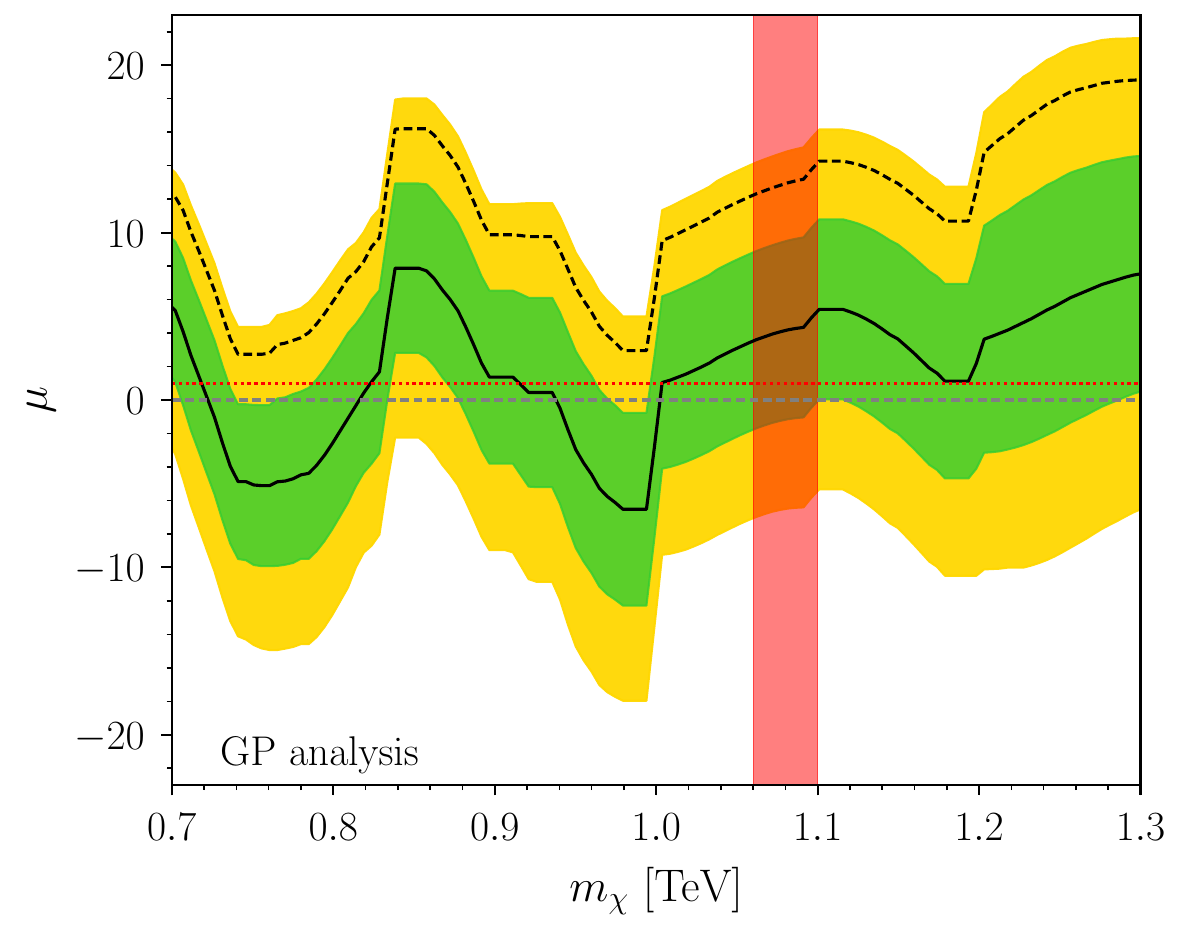}
\end{center}
\vspace{-0.7cm}
\caption{(Left panel) The recovered, best-fit cross section (illustrated through the re-scaling parameter $\mu$) for a power-law analysis of the stacked, ON H.E.S.S. data, as a function of the higgsino mass $m_\chi$. (Note that at each mass we assume 100\% DM.)
The green and gold intervals show the 1$\sigma$ and 2$\sigma$ recovered uncertainties on the best-fit cross section, respectively.  We also indicate the 95\% one-sided upper limit; $\mu$ values above the upper limit are excluded, assuming the Romulus DM profile. (Right panel) As in the left panel but for the GP analysis approach. The GP analysis framework is less sensitive than the power-law analysis because of increased degeneracy between the GP model and the signal.}
\label{fig:HESS_bf_data}
\end{figure*}

\subsubsection{Data analysis results}

In Fig.~\ref{fig:HESS_null_signal_fits} (left panel) we illustrate the fit of the null model ($\mu =0$) to the ON data for our first approach where the background is modeled by a power law.
The null model shows no significant evidence for mismodeling, as illustrated in the bottom sub-panel where the residual counts (data minus model) are shown.
The residual plot also overlays the expected signal spectral template for a thermal higgsino, assuming the Romulus DM profile, but with an enhanced signal strength $\mu = 5$.
We stress that the signal model has not been included in the analysis shown in the top panel.
Including the signal model, we find no evidence for a preference for the signal model over the null hypothesis for higgsino masses between 0.7 and 1.3 TeV, as illustrated in Fig.~\ref{fig:HESS_TS}, which shows the discovery TS times the sign of the best-fit signal parameter as a function of $m_\chi$.
The best-fit recovered cross section as a function of mass is shown in Fig.~\ref{fig:HESS_bf_data} (left panel).
We illustrate the best-fit value of $\mu$ at fixed $m_\chi$, along with the 1$\sigma$ and 2$\sigma$ containment intervals for the best-fit cross section, at fixed $m_\chi$.
The normalization of this figure assumes the Romulus DM profile, though we note that since there is a single ROI our results scale trivially with the appropriately-averaged $J$-factors for different DM profiles.

The right panel of Fig.~\ref{fig:HESS_null_signal_fits} shows the best-fit null model for the GP analysis.
The residuals, shown in the bottom panel, roughly agree between the power-law analysis and the GP analysis.
The discovery TS, while different in detail, largely follows the same trend as in the power-law analysis, as seen in Fig.~\ref{fig:HESS_TS}.
On the other hand, the GP analysis is slightly less sensitive than the power-law analysis, as seen by comparing the right panel of Fig.~\ref{fig:HESS_bf_data}, which shows the best-fit cross section for the GP analysis, to the left panel for the power-law analysis. This is because of the partial degeneracy between the signal model and the GP model.  

As mentioned above, going between different DM profiles amounts to a simple rescaling of the $\mu$ values, since there is a single ROI in these analyses.  In Fig.~\ref{fig:HESS_mu_data} we show the recovered signal strength for the thermal higgsino, along the associated 1$\sigma$ uncertainties, for both the power-law and GP analyses and for all the DM profiles considered in this work.  We see that the power-law analysis of the ON data is the most sensitive analysis performed with H.E.S.S. data in this work, with the GP-based analysis having sensitivity more comparable to that of the ON minus OFF data for many of the DM profiles considered. On the other hand, the GP analysis is robust in that it does not rely on ad hoc model assumptions for the background model.

\subsection{Wino analysis}

The wino DM candidate is closely related to the higgsino in that it is also an example of minimal DM~\cite{Cirelli:2005uq} that may emerge in the context of supersymmetric completions of the SM.  The wino is in the adjoint representation of $SU(2)_L$ and neutral under hypercharge; in the context of supersymmetry the wino is the superpartner of the $SU(2)_L$ gauge bosons. The wino multiplet has a neutral Majorana component along with a chargino, which acquires a slightly larger mass than the neutral fermion by electroweak corrections. 
To achieve the correct relic abundance in the standard cosmology the wino should have a mass of $2.84\pm 0.06$ TeV~\cite{Beneke:2020vff}; we refer to this DM candidate as the thermal wino.
(In Ref.~\cite{Beneke:2020vff} the prediction for the thermal wino mass is stated as 2.842 TeV with an error of several percent.)

The thermal wino has already been excluded for a variety of parametric DM profiles using H.E.S.S. data~\cite{Fan:2013faa,Cohen:2013ama}.  These works made use of the H.E.S.S. analysis in Ref.~\cite{HESS:2013rld}, which analyzes inner Galaxy H.E.S.S. data without OFF subtraction for evidence of DM annihilation to narrow gamma-ray lines. 
In more detail, Ref.~\cite{HESS:2013rld} analyzes the data using a phenomenological background model plus a narrow spectral line, in a similar spirit to the analyses performed here.
Following this, Refs.~\cite{Fan:2013faa,Cohen:2013ama} are able to exclude the thermal wino because the wino annihilation cross section is strongly affected by Sommerfeld enhancement at low velocities, which enhances the annihilation cross section relative to the higher-velocity cross section that sets the relic abundance in the early Universe.
In particular, unlike for the higgsino, the thermal wino mass falls near a Sommerfeld resonance, which results from the presence of a nearly zero-energy bound state appearing in the spectrum and provides a significant enhancement to the annihilation cross section (for further discussion see \cite{Blum:2016nrz}).
For example, Ref.~\cite{Fan:2013faa} finds that the thermal wino is excluded for the class of DM profiles that follow the NFW profile for radii larger than a given core radius, with the density profile constant within the core radius, for all cores smaller than $\sim$0.4 kpc (see also \cite{Baumgart:2017nsr}).
On the other hand, this class of DM profiles is somewhat ad-hoc, given that we do not expect sharp cores along these lines from hydrodynamic simulations. 

Here, we analyze the H.E.S.S. IGS ON data for evidence of the wino model using the 12 Milky Way analogue halos in addition to the NFW and Einasto profiles. 
As discussed in Sec.~\ref{sec:theory}, we also use \texttt{DM$\gamma$Spec} to compute the wino spectrum, which incorporates the Sommerfeld effect using the NLO potentials of \cite{Beneke:2019qaa,Urban:2021cdu}.
As we show below, we are able to rule out the thermal wino across all of the DM profiles considered, though we emphasize again the caveat that the data must be treated with caution given the various assumptions used to derive these results.

\begin{figure}[!t]
\begin{center}
\includegraphics[width=0.49\textwidth]{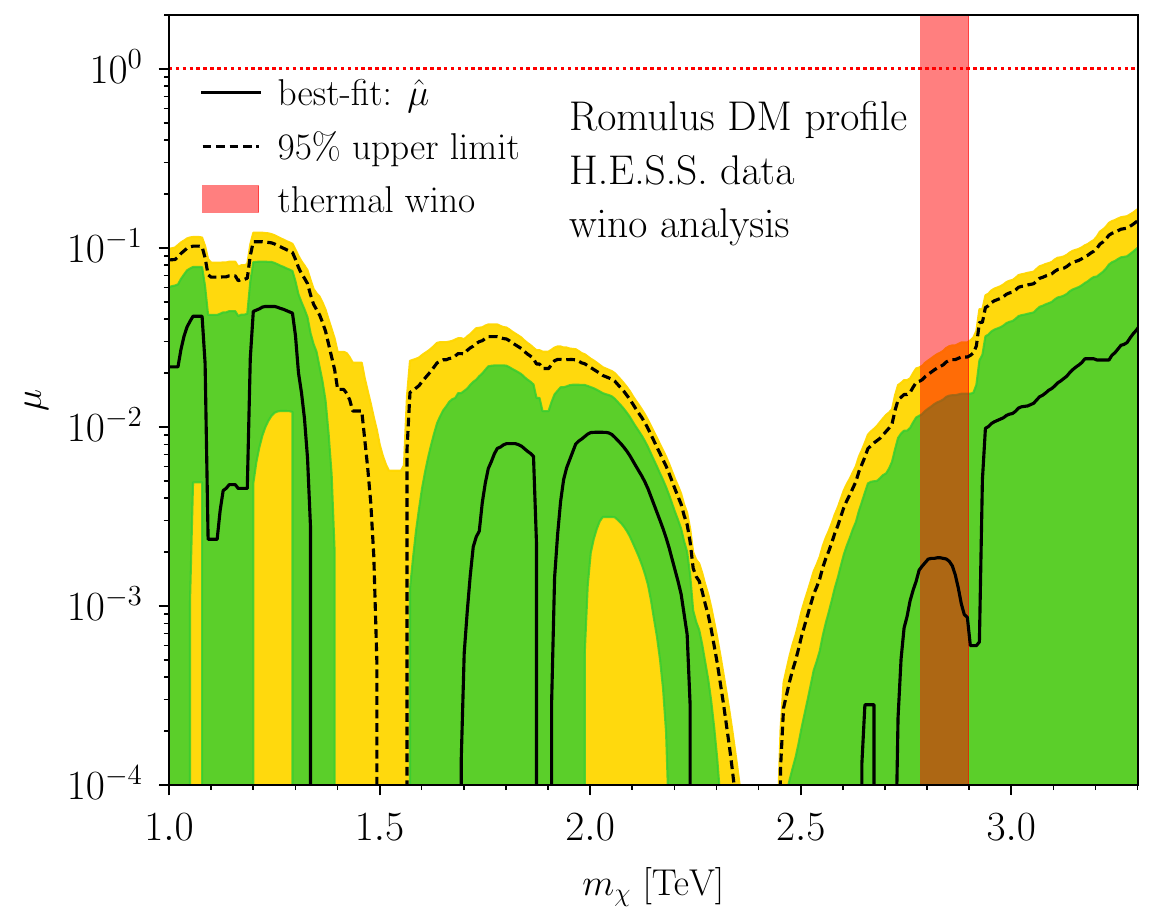}
\end{center}
\vspace{-0.7cm}
\caption{As in Fig.~\ref{fig:HESS_bf_data} (left panel), but for the wino model. The wino should have a mass $m_\chi \simeq 2.84$ TeV to produce the correct DM abundance in the standard cosmology, as indicated. Note that the wino annihilation today is strongly affected by Sommerfeld enhancement, leading to a sharp peak in the annihilation cross section for masses $m_\chi \sim 2.4$ TeV. For this reason we display the $y$-axis with a log-scale, even though some of the best-fit cross sections are negative and hence not shown.  Assuming the Romulus DM profile, the thermal wino model is excluded by almost a factor of fifty in the annihilation cross section.}
\label{fig:HESS_bf_wino}
\end{figure}

We repeat the power-law analysis procedure described previously for higgsino model instead using the wino model, with the $\mu$ parameter rescaling the wino cross section relative to to the expected annihilation cross section for an $SU(2)_L$ triplet with zero hypercharge.\footnote{Given the similarities between the GP and power-law analyses, we only show the power-law results; however, we verify that the conclusions of this section also hold if one performs a GP analysis instead.}  In Fig.~\ref{fig:HESS_bf_wino} we illustrate the best-fit cross section as a function of the wino mass $m_\chi$, with the 1$\sigma$/2$\sigma$ error bars on the recovered cross section at fixed $m_\chi$ shown in green/gold. The one-sided 95\% upper-limit is also indicated. This figure assumes the Romulus DM profile.  The expected wino cross section at the thermal mass is excluded by almost a factor of fifty. Note that we illustrate this figure with a log-scale for the $y$-axis, even though at some masses the best-fit $\mu$ values are negative (and hence not shown) because of the range of scales in the cross section that occur as we pass through a Sommerfeld resonance at $m_{\chi} \sim 2.4$ TeV.

\begin{figure}[!t]
\begin{center}
\includegraphics[width=0.49\textwidth]{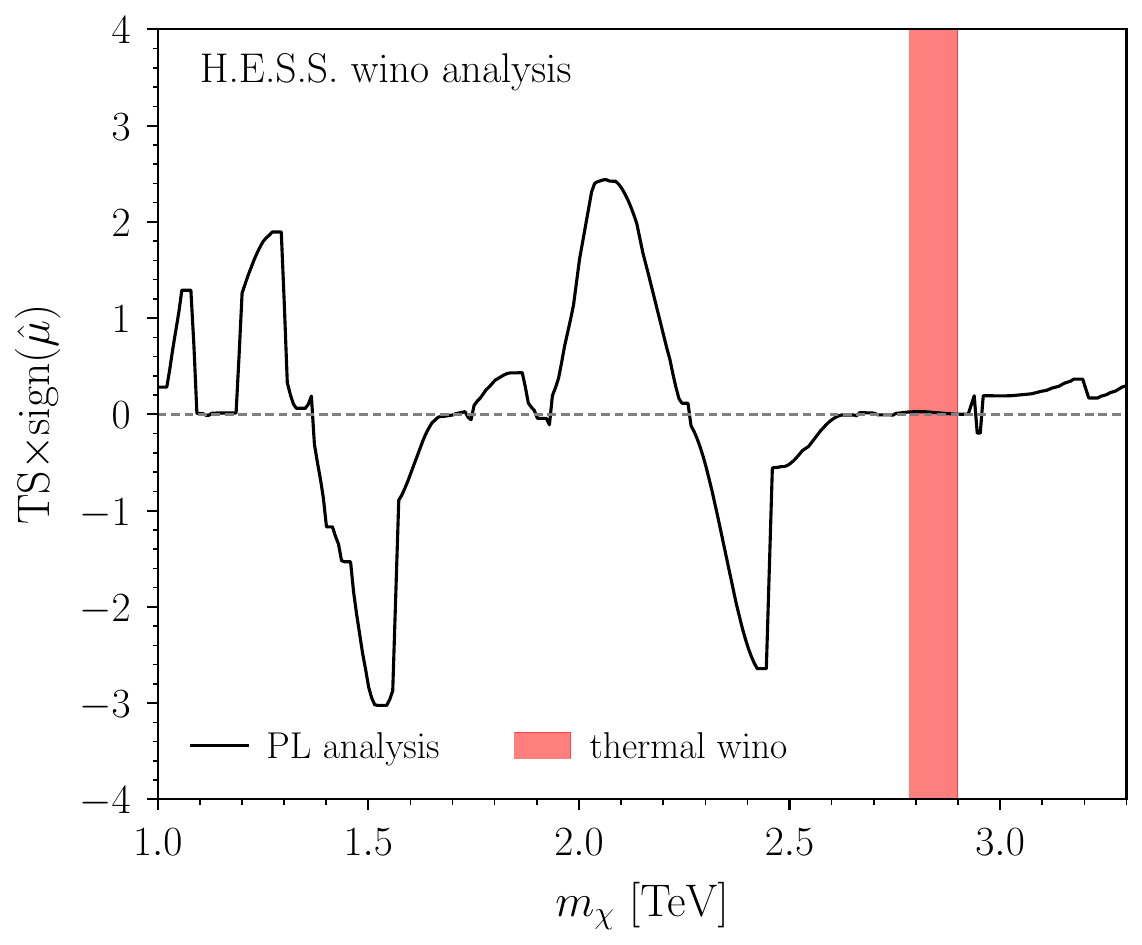}
\end{center}
\vspace{-0.7cm}
\caption{ As in Fig.~\ref{fig:HESS_bf_data} but for the wino model.  We find no significant excesses in favor of the wino model. }
\label{fig:HESS_TS_wino}
\end{figure}

In Fig.~\ref{fig:HESS_TS_wino} we illustrate the discovery TS from the power-law analysis for the wino versus the wino mass, for a mass range in the vicinity of the expected thermal wino mass. 
No significant excesses are found (beyond $\sim$1$\sigma$ in local significance), and thus we find no evidence for a wino-type DM model. 

\begin{figure}[!t]
\begin{center}
\includegraphics[width=0.49\textwidth]{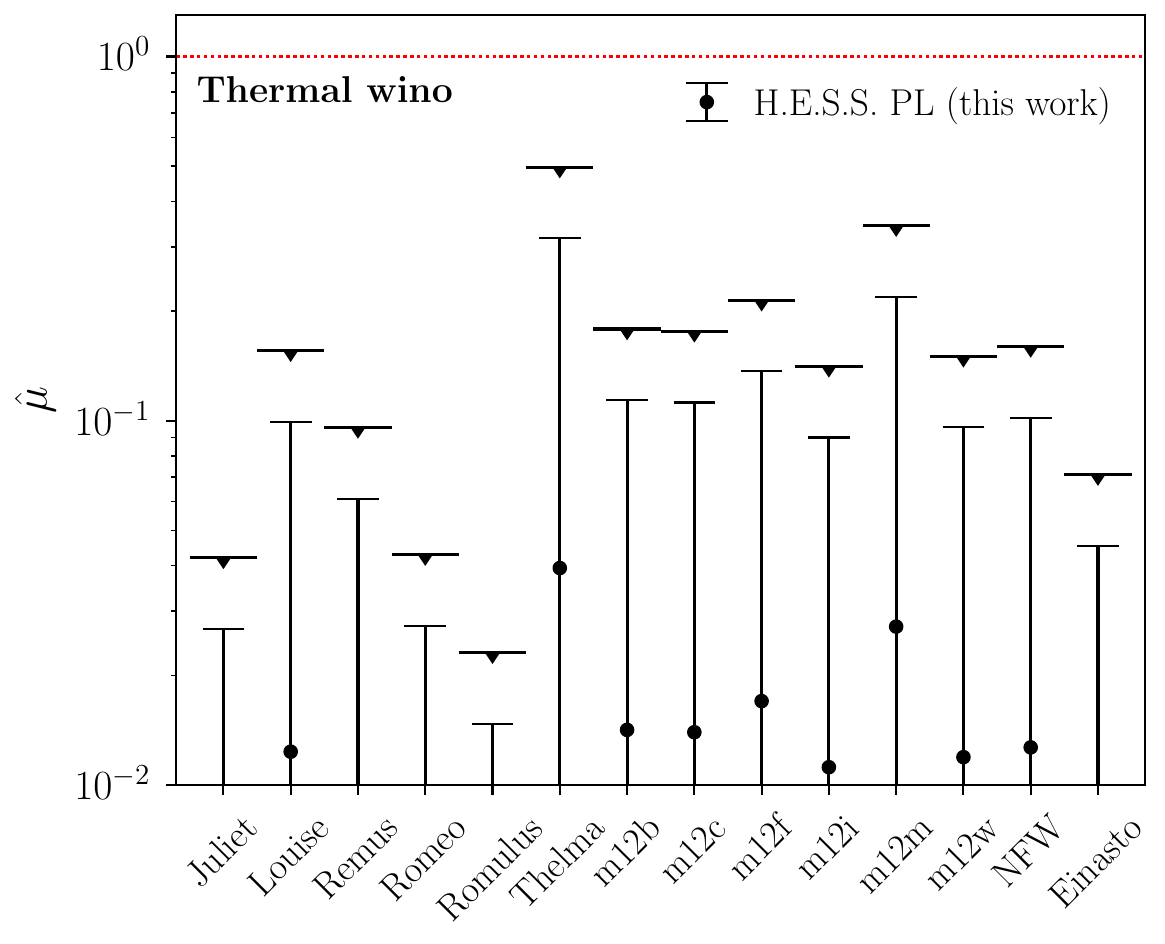}
\end{center}
\vspace{-0.7cm}
\caption{As in Fig.~\ref{fig:HESS_mu_data} but for the wino model, with $m_\chi = 2.84$ TeV fixed to the expected thermal wino mass. We also indicate the 95\% one-sided upper limits on $\mu$ for the different DM profiles. We exclude the thermal wino cross section for all of the DM profiles considered. (We emphasize this conclusion is subject to the various assumptions adopted in our modeling of the H.E.S.S.-II IGS data.)}
\label{fig:HESS_J_wino}
\end{figure}

Fig.~\ref{fig:HESS_bf_wino} illustrates that we exclude the thermal wino model assuming the Romulus DM profile; however, Romulus has the highest central density of all the DM profiles considered (see Fig.~\ref{fig:J}).  A more relevant question is whether we are able to exclude the wino model for all 14 DM profiles (12 FIRE-2 halos plus the NFW and Einasto profiles) that we consider in this work. As shown in Fig.~\ref{fig:HESS_J_wino}, the wino is indeed excluded for all 14 models.
Here, we fix $m_\chi = 2.84$ TeV and show the best-fit cross section (and 1$\sigma$ error bars) for the different DM profiles, as in Fig.~\ref{fig:HESS_mu_data} for the higgsino.  Additionally, we indicate the 95\% one-sided upper limits. The expected wino cross section is excluded by at least a factor of two for all DM profiles considered.
While cautioning the various caveats associated with our reconstruction of the H.E.S.S. instrument response, the combination of H.E.S.S. data with modern DM profile estimates for Milky Way like galaxies supports the thermal wino not constituting the observed DM.

\subsection{H.E.S.S. projection: full data set and no OFF subtraction}

In this subsection we project how the full data set collected in the H.E.S.S.-II IGS may be leveraged to obtain stronger constraints.
For this projected analysis, we consider data collected within the inner 5$^\circ$ of the GC, binned in annuli of $1^\circ$ each.  We imagine analyzing the data in each annulus separately and then constructing a joint likelihood for $\mu$ by combining the appropriate profile likelihoods in the individual annuli, as we do in Sec.~\ref{sec:fermi} using Fermi-LAT data.  We fix the higgsino mass to its thermal value and throughout this subsection we use data between 0.5 TeV and 4 TeV.

\begin{figure}[!t]
\begin{center}
\includegraphics[width=0.49\textwidth]{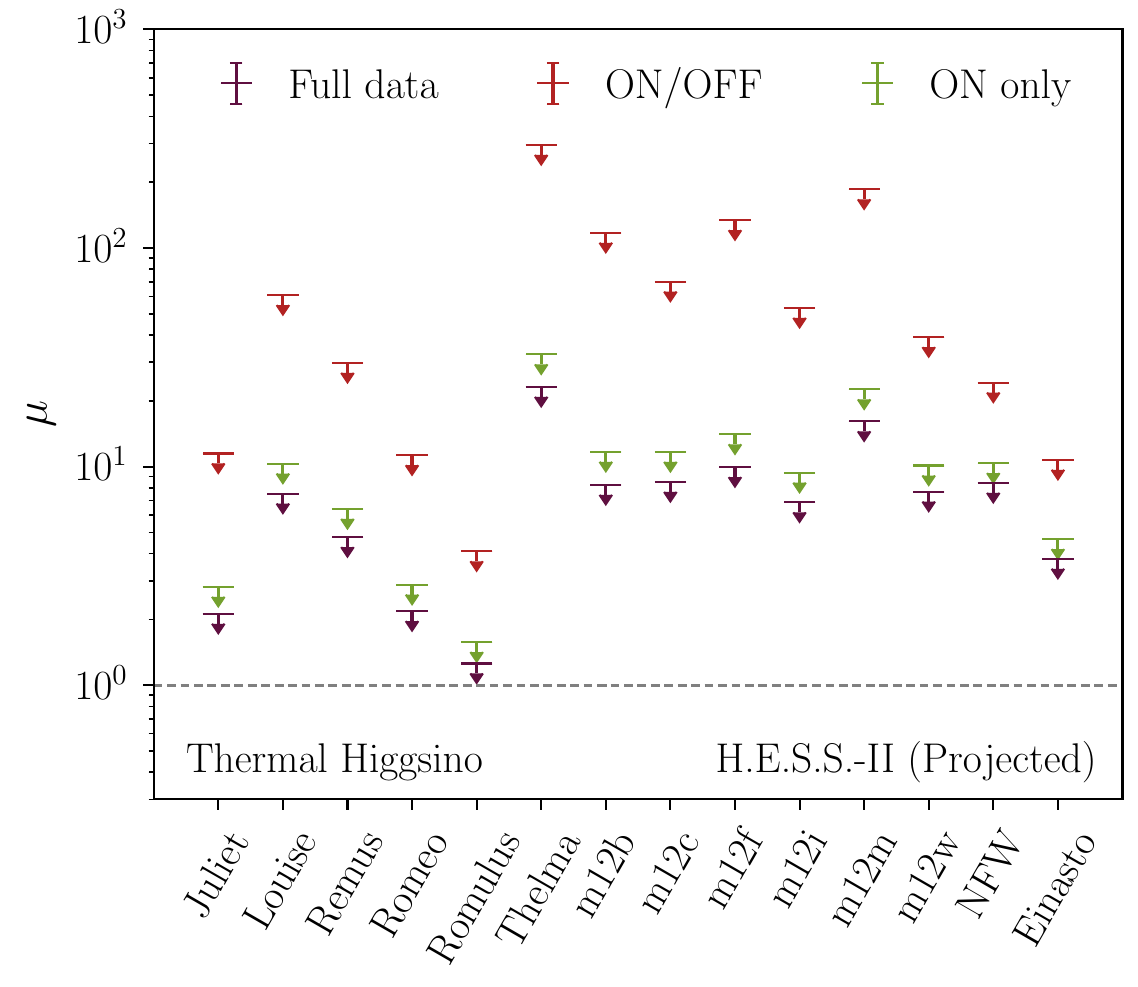}
\end{center}
\vspace{-0.7cm}
\caption{The projected sensitivities of idealized H.E.S.S-II data analyses for the thermal higgsino with fourteen different DM distributions, illustrated through the projected 95\% upper limit (U.L.) on the annihilation cross-section, parameterized through the $\mu$ rescaling parameter, under the null hypothesis.
We assume the 546 hours of data contained within the IGS but include counts collected out to 5$^\circ$ away from the GC, binned in 1$^\circ$ annuli. All projections shown here assume an ideal, fixed-background scenario.  We illustrate the improvement in going from the ON minus OFF procedure to an analysis of the ON-region data only to an analysis of the full data set.  The full-data approach nears the necessary sensitivity to probe the thermal higgsino for the Romulus DM profile, serving as motivation for the H.E.S.S. collaboration to perform a dedicated higgsino analysis without OFF subtraction and using data-driven spectral templates for the cosmic-ray-induced emisison.}
\label{fig:HESS_projection_thermal}
\end{figure}

We begin by considering the ON minus OFF analyses.  As discussed earlier in this section, while OFF subtraction removes the cosmic-ray-induced emission the residual Galactic diffuse emission must be modeled. In Fig.~\ref{fig:HESS_projection_thermal} the projected upper limits labeled `ON/OFF' assume that the diffuse emission is modeled perfectly with no free nuisance parameters; that is, the normalization of the diffuse emission spectral template is fixed to its true value.
The sensitivity of this idealized, projected search is nearly identical to that we achieve in the actual data (in Sec.~\ref{sec:hess_ON_OFF}), as shown in {\it e.g.} Fig.~\ref{fig:HESS_mu_data}.

Recall that in Sec.~\ref{sec:hess_on_only} we analyze the ON data only, fully stacked over the ROI, for evidence of the higgsino model. In Fig.~\ref{fig:HESS_projection_thermal} we show the projected sensitivity from an idealized version of this analysis that both incorporates spatial information through the concentric annuli and also which assumes that the cosmic-ray-induced background emission and the astrophysical diffuse emission are perfectly modeled, with no nuisance parameters. The improvement in sensitivity of the projections in Fig.~\ref{fig:HESS_projection_thermal} relative to those achieved in the real data ({\it e.g.}, Fig.~\ref{fig:HESS_mu_data}) are primarily due to lack of nuisance parameters in the idealized analysis. Fig.~\ref{fig:HESS_projection_thermal} thus suggests that with better spectral templates for the astrophysical diffuse emission, which enable a larger energy window to be used in the analysis and thus less signal/background degeneracy, the H.E.S.S. collaboration would be able to perform an improved search for higgsinos relative to those presented in Sec.~\ref{sec:hess_on_only}.  Moreover, the projected upper limits in Fig.~\ref{fig:HESS_projection_thermal} labeled `Full data' show the further improvement possible by incorporating all of the data within the ROI, apart from the masked regions, and not just the data in the ON ROI.  The additional data volume leads to a modest improvement in sensitivity and pushes the sensitivity of H.E.S.S. towards that necessary to probe the thermal higgsino model for the Romulus DM profile. This estimate strongly motivates H.E.S.S. to perform a dedicated analysis for higgsino DM without OFF subtraction using a spectral template for the cosmic-ray-induced emission. 

\section{CTA Projections for higgsino DM}
\label{sec:CTA_projections}

In this section we project sensitivity of the upcoming CTA to a higgsino DM signal.\footnote{The power of CTA to test DM is widely appreciated, {\it e.g.} Refs.~\cite{CTAConsortium:2012fwj,Pierre:2014tra,Silverwood:2014yza,Lefranc:2015pza,CTA:2015yxo,CTA:2020qlo,Abe:2024cfj}. Our focus here is to extend these discussions to the well motivated thermal higgsino.}
In doing so, we use the CTA detector properties outlined in Sec.~\ref{sec:detector}.  The key characteristics that improve the performance of CTA relative to H.E.S.S. for the higgsino signal are a substantially larger FOV, an increased effective area, and a factor of roughly 2.2 improvement in energy resolution at 1 TeV.  Additionally, while H.E.S.S. is only sensitive to gamma-rays above $\sim$200 GeV, CTA could push this lower threshold down by an order of magnitude.
Incorporating the low-energy data into the analysis may strengthen the sensitivity to the higgsino model because of the continuum signal generated by the $W^+W^-$ and $ZZ$ annihilation channels~\cite{Rinchiuso:2020skh}.
Nevertheless, at such low energies one must contend with substantial astrophysical background and use the subtle differences in spectral shape between the higgsino continuum model and the astrophysical backgrounds to probe the DM model, just as was the case for the Fermi continuum search  in~\cite{Dessert:2022evk}.  Furthermore, unlike for the Fermi-LAT analysis in~\cite{Dessert:2022evk}, an analysis of the low-energy CTA data would also have to deal with the large cosmic-ray-induced background, making it significantly more complicated.
Accordingly, we take a benchmark analysis range for the thermal higgsino signal that extends from 500 GeV to 4 TeV in order to only focus on the cleaner line-like signal. We return later in this section, however, to discuss the impact of extending the analysis down to 10 GeV.

\begin{figure}[!t]
\begin{center}
\includegraphics[width=0.48\textwidth]{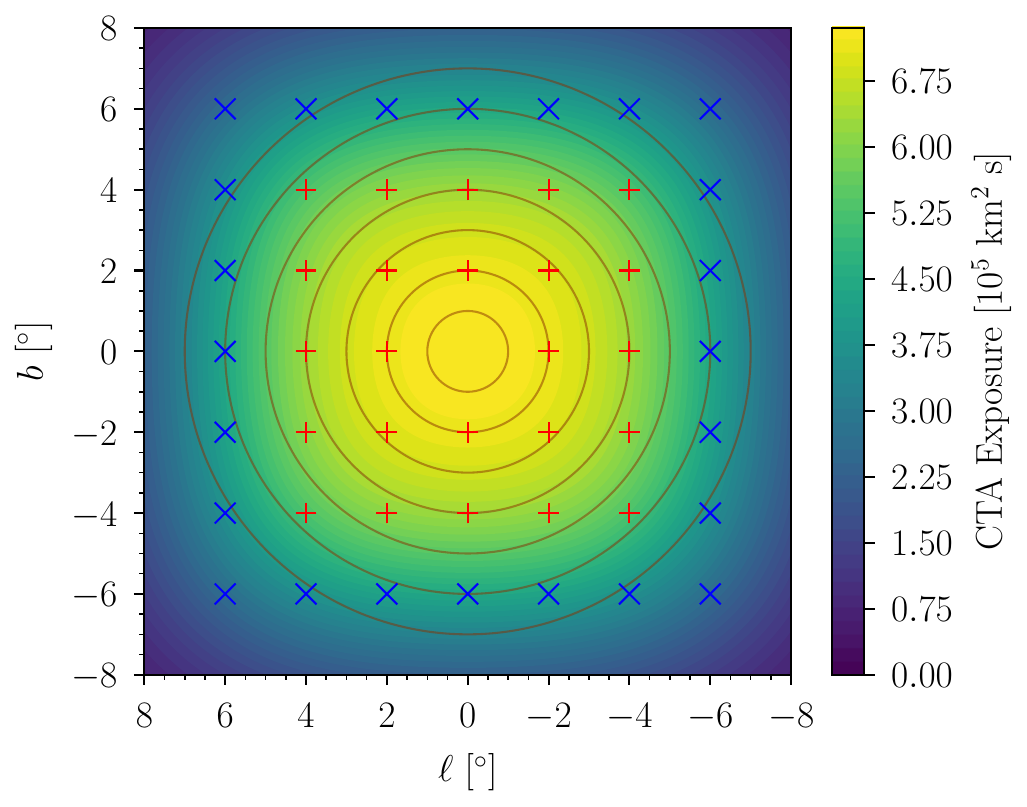}
\end{center}
\vspace{-0.7cm}
\caption{A hypothetical exposure map for CTA observations, on which we base our projections in this section. This map is evaluated at an energy of 1 TeV, with 500 hrs of exposure time distributed evenly over the inner pointings (red $+$) and 300 hrs distributed evenly over the outer pointings (blue $\times$).  In the analysis, we apply a 1$^\circ$ mask to the Galactic plane, removing the region $|b|<0.5^{\circ}$. Also overlaid are concentric rings of 1$^\circ$ each around the GC, which form the spatial bins for our analysis.}
\label{fig:CTA_exposure}
\end{figure}

We project the sensitivity for CTA based on a hypothetical pointing strategy illustrated in Fig.~\ref{fig:CTA_exposure}, assuming 500 hours of total data within the inner 5$^\circ$ of the GC (equally split among the red pointings) and 300 hrs of data between $5^\circ - 8^\circ$ from the GC (equally split among the blue).  We use the CTA instrument response projections described in Sec.~\ref{sec:detector}, assuming for simplicity that the exposure degradation with off-set angle is approximately uniform with energy.  We note that broadly the sensitivity to $\langle \sigma v\rangle$ scales as one over the square root of the data taking time assuming statistics dominated uncertainties, and thus the projections of this section may be easily rescaled to reflect evolving observational expectations.

We focus on projecting sensitivity for the thermal higgsino model.  We perform the analysis in 10 concentric annuli of $1^\circ$ each out from the GC, though ultimately only the inner 7 of these meaningfully contribute, and we mask the Galactic plane within  latitudes $|b| \leq  0.5^\circ$.  We consider a background formed of two components: an isotropic cosmic-ray-induced piece whose spectra is given by the CTA data challenge (Fig.~\ref{fig:edispbkg}) and a diffuse astrophysical component following the Fermi {\tt p8r3} emission model (see Fig.~\ref{fig:fermi_diffuse_map}).  In addition to the ON/OFF procedure described in previous sections, we consider a ``two-template" model analysis that allows us to take advantage of the full set of collected data.  Here, aside from the signal component, the model contains spectral templates for both the cosmic-ray and diffuse contributions.  In practice, a template for the cosmic-ray component might be obtained by taking dedicated OFF data far from the GC but under comparable observational conditions, and the astrophysical background template is given by the Fermi model.
For simplicity, we assume no mismodeling of either of these components to start, assigning a nuisance parameter for the normalization of each. For the ON/OFF subtraction analysis we assume the cosmic-ray background has been cleanly removed, but we include a template to model the residual astrophysical diffuse flux with a nuisance parameter for its overall normalization.  We note that the nuisance parameters slightly degrade the sensitivity to the higgsino signal because of partial degeneracy between the background models and the signal mode (see {\it e.g.} Fig.~\ref{fig:TS_proj_CTASWGO}); the degeneracy may be lessened by enlarging the energy window beyond $0.5$ TeV to 4 TeV or otherwise incorporating priors, while the degeneracy would be increased by narrowing the energy window.

\begin{figure}[!t]
\begin{center}
\includegraphics[width=0.48\textwidth]{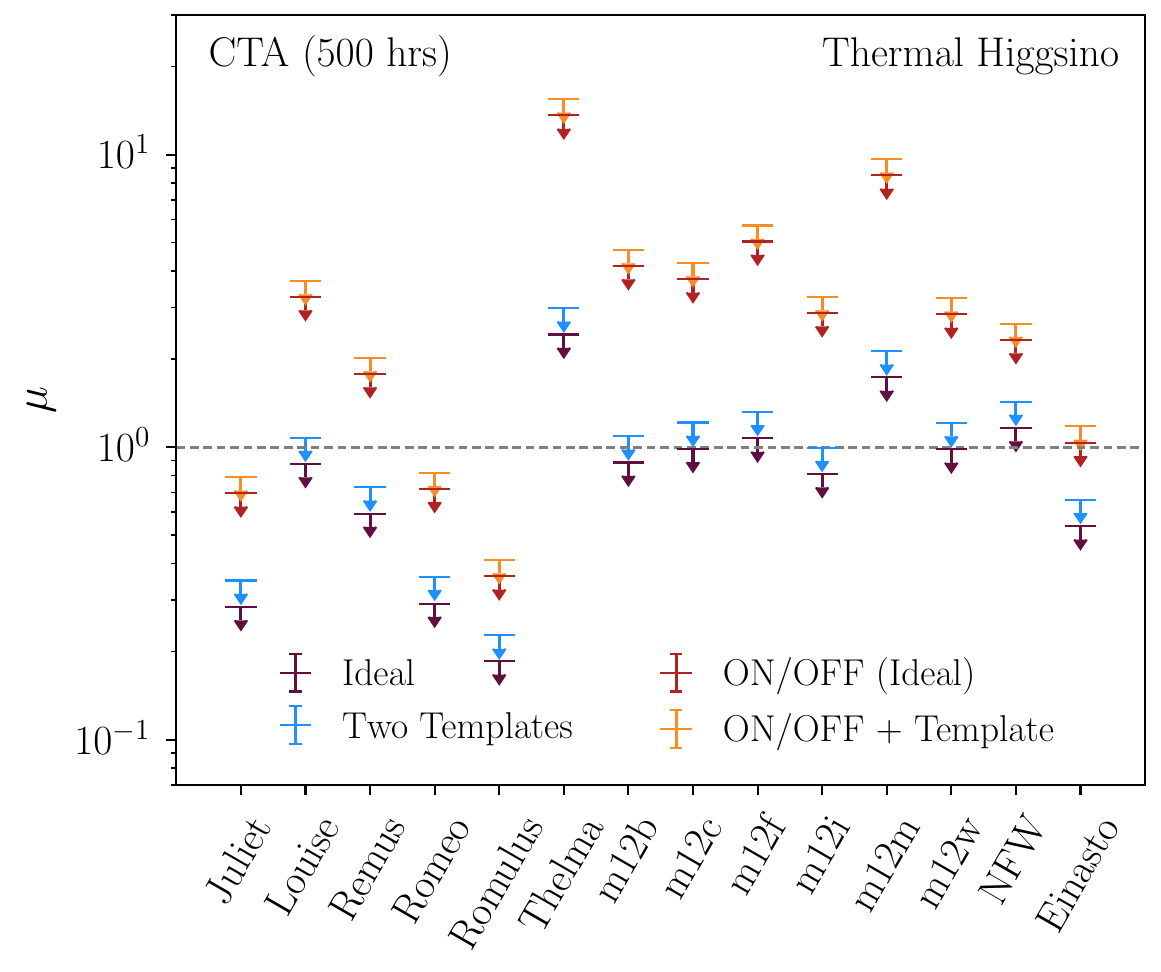}
\end{center}
\vspace{-0.7cm}
\caption{The expected sensitivity of CTA to the thermal Higgsino for various DM distributions. (See Fig.~\ref{fig:CTA_exposure} for the assumed exposure map.)
As illustrated, CTA will be able to discover or otherwise place stringent limits on thermal higgsino DM in most of the DM profiles considered here. However, an ON-OFF approach that discards a large amount of collected data  degrades this sensitivity and significant discovery potential is lost in several profile scenarios.  We project sensitivity both for ideal analyses, for which the normalizations of the spectral templates are fixed to their true values, and more realistic analyses where the template normalizations are floated as nuisance parameters. For these projections we incorporate data between 500 GeV and 4 TeV to focus only on the line-like feature at the endpoint.}
\label{fig:CTA_sensitivity}
\end{figure}

In Fig.~\ref{fig:CTA_sensitivity} we project the expected 95\% upper limit on the annihilation cross section for the thermal higgsino model under the null hypothesis using the analysis frameworks described above.  We also show the expected sensitivity for the optimal analyses of the full and ON-OFF data sets, where the nuisance parameters of the background template(s) are fixed to their true value(s).  We find that analyses of the full data set in optimum conditions, or indeed in the two-template framework, are expected to discover or disfavor the thermal higgsino at considerable significance in the majority of the DM profiles considered.
The corresponding discovery TS in these two cases is shown in Fig.~\ref{fig:TS_proj_CTASWGO}.
The performance of the two-template analysis is $\sim$80\% the sensitivity of ideal under the benchmark energy range.
In the most pessimistic scenario, for the Thelma profile, the upper limit falls short of the thermal higgsino prediction by a factor of 3.0 (2.4) for the two template (ideal) analysis.
Collecting an additional factor of nine in data volume is likely implausible, and so this shortcoming would need to be overcome by other factors such as an optimized pointing strategy or even the deployment of additional telescopes.
(For example, the Omega configuration of CTA could improve the reach to a line-like signal by a factor of 2~\cite{Abe:2024cfj}.)
This point further highlights the importance of better understanding the expected DM density profile in the inner parts of the Milky Way.

In contrast, the ON-OFF subtraction approach would be able to reach the thermal higgsino prediction in roughly a third of the DM profiles considered.
The signal and exposure loss in the ON-OFF approach is severe, particularly for cored profiles such as Thelma, resulting in substantially degraded projected sensitivities relative to those from the un-subtracted analyses.  We note that the performance of the ON-OFF analysis is somewhat sensitive to the particular choice of pointing locations, and we do not make an attempt to particularly optimize our choices for this effect. In general the pointings that optimize the sensitivity depends on the specific DM profile considered.

\begin{figure}[!t]
\begin{center}
\includegraphics[width=0.48\textwidth]{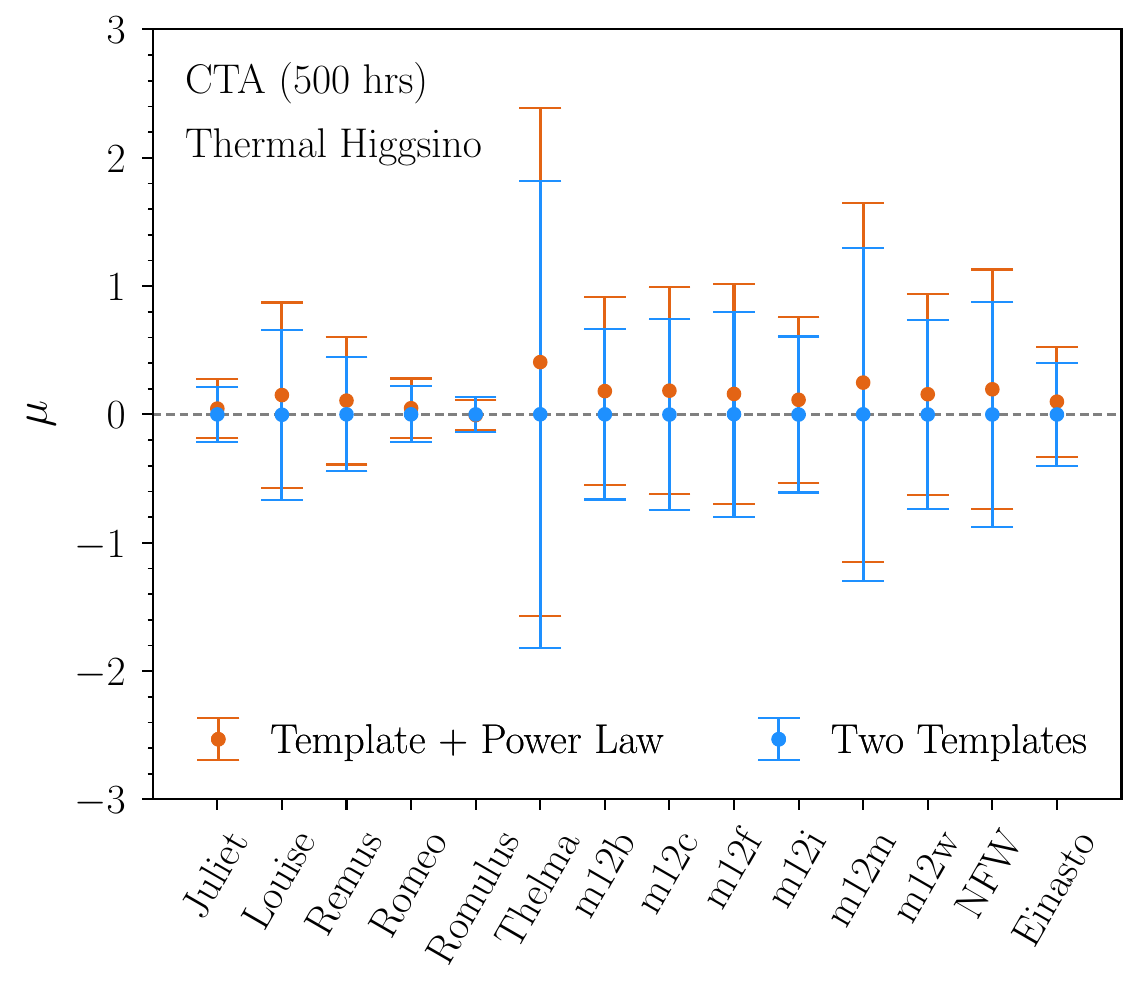}
\end{center}
\vspace{-0.7cm}
\caption{The best-fit and $1\sigma$ uncertainties for a thermal higgsino search on null Asimov data, modeling the astrophysical component with the {\tt p8r3} diffuse model (blue) and with a power law (orange).  The Asimov data itself is generated with the {\tt p8r3} model, and the analysis is conducted over the benchmark energy range $0.5 \, \, {\rm TeV} < E < 4 \, \, {\rm TeV}$.  In both cases, the cosmic ray spectral template is assumed to have no mismodeling. The template plus power-law analysis maintains comparable sensitivity to the more idealized two-template analysis with minimal bias. We stress that the actual spectral templates used in a realistic analysis would almost certainly differ from the simple example here, but the example demonstrates that robust and sensitive analysis frameworks should be possible without using OFF subtraction for CTA.}
\label{fig:CTA_bias}
\end{figure}

\begin{figure}[!t]
\begin{center}
\includegraphics[width=0.48\textwidth]{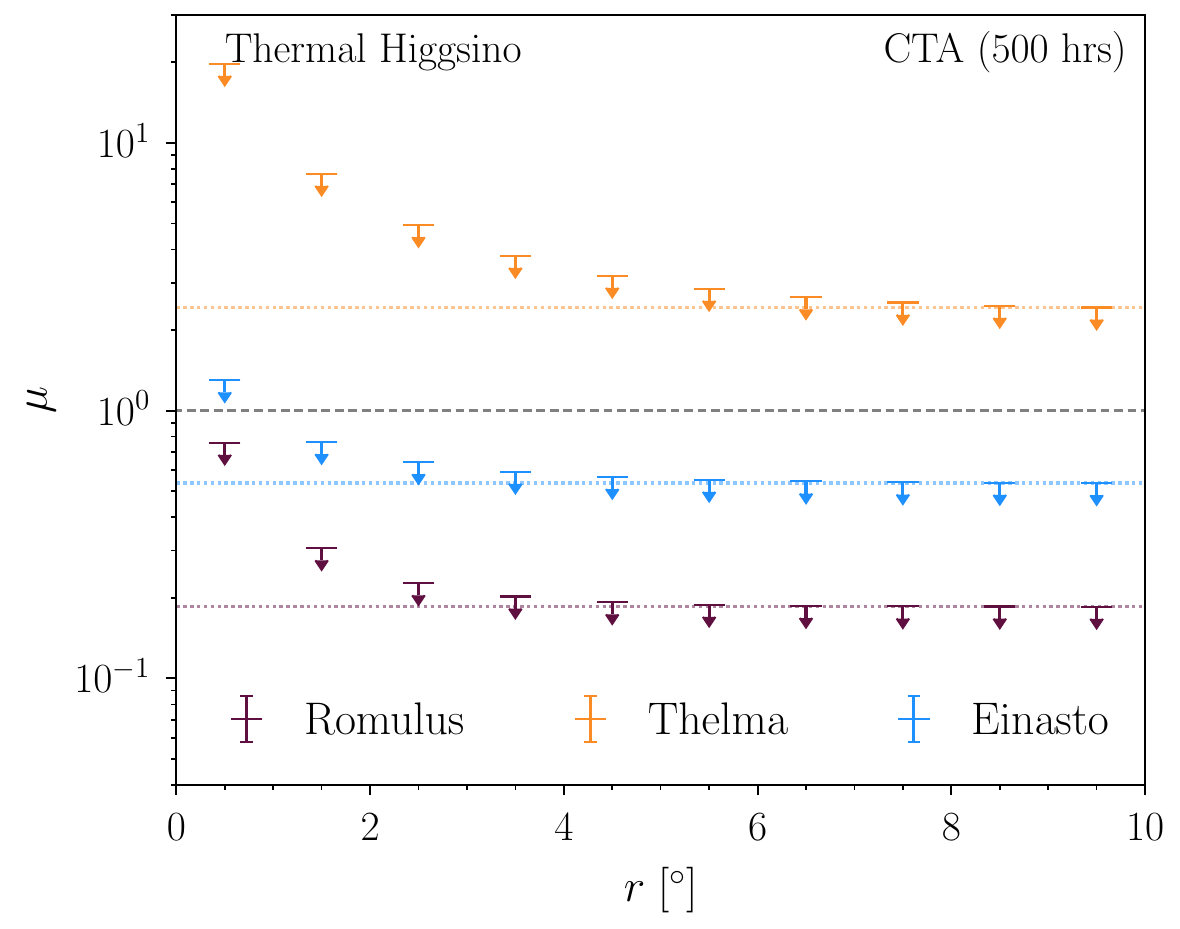}
\end{center}
\vspace{-0.7cm}
\caption{ The expected 95\% upper limit to thermal higgsino annihilation under the null hypothesis from a joint analysis of $1^\circ$ annuli within a given radius $r$, for three different DM profile assumptions: Romulus,  Einasto (particularly cuspy), and Thelma (particularly cored). The dotted horizontal lines show the total sensitivity when all ROIs are joined over. Note that these sensitivities assume the exposure map shown in Fig~\ref{fig:CTA_exposure}, which has its own spatial structure. Despite the raw exposure dropping steeply past $r\sim 5^\circ$,  additional sensitivity of $\sim 30\%$ can still be gleaned from including these outer annuli if the underlying profile is Thelma-like.} 
\label{fig:CTA_v_radius}
\end{figure}

\begin{figure}[!t]
\begin{center}
\includegraphics[width=0.48\textwidth]{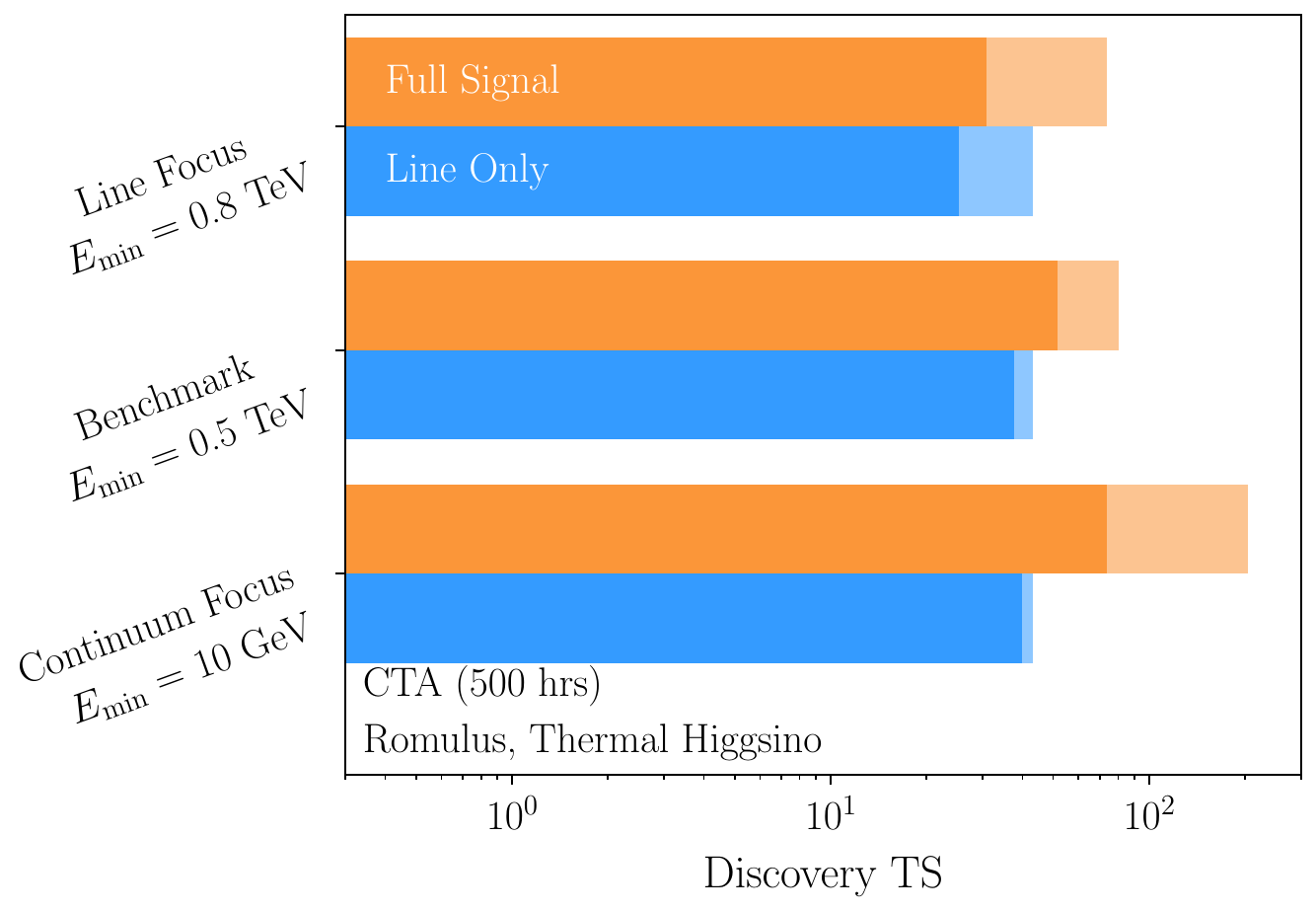}
\end{center}
\vspace{-0.7cm}
\caption{The discovery TS of a thermal higgsino for analysis windows that focus more on the line feature or the continuum spectrum, as compared to the benchmark choice. (Note that all of these analysis windows extend to 4 TeV.) The orange bars correspond to an analysis searching for the full higgsino signal, while the blue corresponds to that searching for a pure line at the thermal higgsino prediction (cf. Fig.~\ref{fig:xsec}). Note that the underlying data assumes a 1.08 TeV higgsino. For each, the two-template analysis is shown in darker color, while the ideal fixed-background case is shown in light.  As illustrated, for the full signal a substantial amount of sensitivity can ostensibly be gained from including the continuum flux in the search, but the optimum gain in signal-to-noise is not achievable in analysis frameworks that require modeling the (much larger and more spectrally degenerate) background. If the signal model is simplified to a pure line, no information is gained by going to lower energies, and the discovery TS is reduced by $\sim 10\%$ in realistic analysis windows. For both signal models, extending the background to lower energies introduces enormous complications in terms of accurately modeling the astrophysical diffuse emission.}
\label{fig:CTA_energy_ranges}
\end{figure}

Next, we take a first step towards assessing the susceptibility of the CTA analyses to mismodeling the spectral data. For simplicity, we only consider mismodeling the astrophysical emission.
To do so we model the astrophysical emission by a power-law, with two nuisance parameters describing the normalization and spectral index.  We additionally include a spectral model for the cosmic-ray-induced emission with its own nuisance parameter for the overall amplitude of that component. The simulated (Asimov) data is, as usual, generated with the Fermi \texttt{p8r3} Galactic diffuse model such that the model used to analyze the data is unable to perfectly describe the background counts.  We use our fiducial $0.5 \, \, {\rm TeV} < E < 4 \, \, {\rm TeV}$ energy range, but emphasize that narrowing the energy range may help mitigate mismodeling at the expense of weakening the sensitivity to a putative signal. In Fig.~\ref{fig:CTA_bias} we show the expected best-fit annihilation cross sections and 1$\sigma$ uncertainties under the null hypothesis for the CTA analysis. (Here, we only consider the analysis of the un-subtracted data, and furthermore we assume no mismodeling of the cosmic-ray-induced emission.) 
Without any bias the expected best-fit cross sections would be zero for all of the DM profiles. However, due to the imperfect power-law model there is a slight bias for each of the assumed DM profiles, though in all cases it is much less than 1$\sigma$ in significance.  For reference, we also show the analogous quantities for the analyses using two spectral templates, which by construction have no mismodeling.

A further question to consider is the impact of the analysis choices we have made on the eventual CTA sensitivity. Firstly, we consider how spatial information impacts the sensitivity, given our assumed exposure map.
In Fig. \ref{fig:CTA_v_radius} we show the reach to the thermal higgsino as we include additional annuli away from the GC, in each case performing a joint analysis over annuli within the specified radius $r$.
We illustrate the results for three different DM profiles: Romulus, which has the largest expected DM signal, Einasto, which is a particular cuspy profile, and Thelma, which is the most cored profile. We find that despite a steep loss of raw exposure at $\sim$$5^\circ$ away from the GC, as much as 30\% more sensitivity can be gained from including data up to $10^\circ$ in cored profiles such as Thelma.
In contrast, the relative gain in information for steeper profiles such as Romulus and Einasto is essentially negligible.
This indicates that a more egalitarian observation strategy, where exposure is more evenly distributed within the inner 10$^\circ$, may particularly benefit the ``challenging" DM profile possibilities -- if the Milky Way happens to look akin to Thelma or m12m -- and bring them closer to detection.  This serves as further motivation for better understanding the uncertainty in the DM density profile for the Milky Way.

It is important to determine the relative information contained within the line and continuum contributions of the signal: in addition to the benchmark analysis that includes signal photons down to 500 GeV discussed so far, we consider one that focuses exclusively on the line signature and truncates at 800 GeV while still extending to 4 TeV.  We also consider an analysis that extends the energy range down to 10 GeV, where the effective area of the instrument becomes functionally zero and the signal model consists overwhelmingly of continuum emission.
The results are demonstrated in Fig.~\ref{fig:CTA_energy_ranges}, which illustrates the discovery TS of the ideal (light) and two-template (dark) full-data analyses for each of these choices. (Recall that the fundamental difference in the two-template analyses is that the spectral templates are assigned nuisance parameters for their normalizations; profiling over these nuisance parameters reduces the sensitivity because of partial degeneracy between these models and the signal mode.)   We find that the theoretical sensitivity to the thermal higgsino may increase by as much as a factor of $\sim$1.6 in the cross section (or $\sim$2.5 in the discovery TS) upon inclusion of the continuum region; this is in rough agreement with the discussion in Ref.~\cite{Rinchiuso:2020skh}, where only an ideal analysis was performed, and the higgsino endpoint contribution was not included. 
However, due to its spectral degeneracy with a much larger background, the sensitivity increase in the template analysis case is more modest, $\sim$30\%.
It is also important to note that an analysis performed over a wider energy range carries a more severe danger of mismodeling.
By way of contrast, restricting the analysis to a narrower window around the line incurs only a minor penalty (relative to the benchmark) in terms of the ideal signal-to-noise.
Thus, while theoretically significant information regarding the signal is contained within the continuum, in realistic analyses of CTA  data it is unclear whether this information can be harnessed efficiently without incurring significant systematic bias. Note that the Fermi analysis in Ref.~\cite{Dessert:2022evk} focused on the low-energy continuum emission but unlike for CTA that work did not also have to contend with significant cosmic-ray-induced background emission. 

It is worth considering how much one gains in detection power for the higgsino by using the higgsino spectral template, including endpoint corrections as illustrated in Fig.~\ref{fig:xsec}, versus only assuming a line-like spectral template. We address this question by creating Asimov data with the full higgsino model, as used for example in the orange bands for Fig.~\ref{fig:CTA_energy_ranges}. However, we then analyze the mock data only with a line spectral template for the DM model, with results shown in blue. We find that using a line spectral template instead of the higgsino spectral template decreases the expected discovery TSs by around 10\% for the `Line Only' analysis. In addition, the imperfect signal model and lack of endpoint contribution for line prediction result in biasing both the best-fit cross-section and mass away from the truth at the 20\% level.

Lastly, the various choices we make regarding the CTA analysis -- such as observational strategy and background modeling -- can be compared to that adopted by the CTA Collaboration in a recent study that appeared as this work was being finalized~\cite{Abe:2024cfj}.
Although that work did not consider the thermal higgsino, it projected the reach to a DM-annihilation-induced line-like signature amongst other possible final states.
A very similar analysis strategy was adopted in that work: the ON data alone was analyzed in a narrow energy window, with a parametric power-law-like model adopted for the background.
The analysis in that work also assumed 500 hrs of GC observations, however with a noticeably different spatial distribution relative to what we adopt (see, {\it e.g.}, Fig.~\ref{fig:CTA_exposure}).
In particular, that work equally divided the time between nine pointing locations defined by all combinations of $\ell \in \{0^\circ,\pm 1^\circ\}$ and $b \in \{0^\circ,\pm 1^\circ\}$.
Correspondingly, the analysis they performed focused on a much smaller region of the sky near the GC.  Still, as shown in App.~\ref{app:CTA_comp}, we find nearly identical projected sensitivity relative to that shown in~\cite{Abe:2024cfj} for the process $\chi \chi \to \gamma\gamma$, at least assuming the Einasto DM profile.  This comparison provides confidence to the higgsino projections in this section.

\section{SWGO Projections for higgsino DM}
\label{sec:SWGO_projections}

\begin{figure}[!t]
\begin{center}
\includegraphics[width=0.49\textwidth]{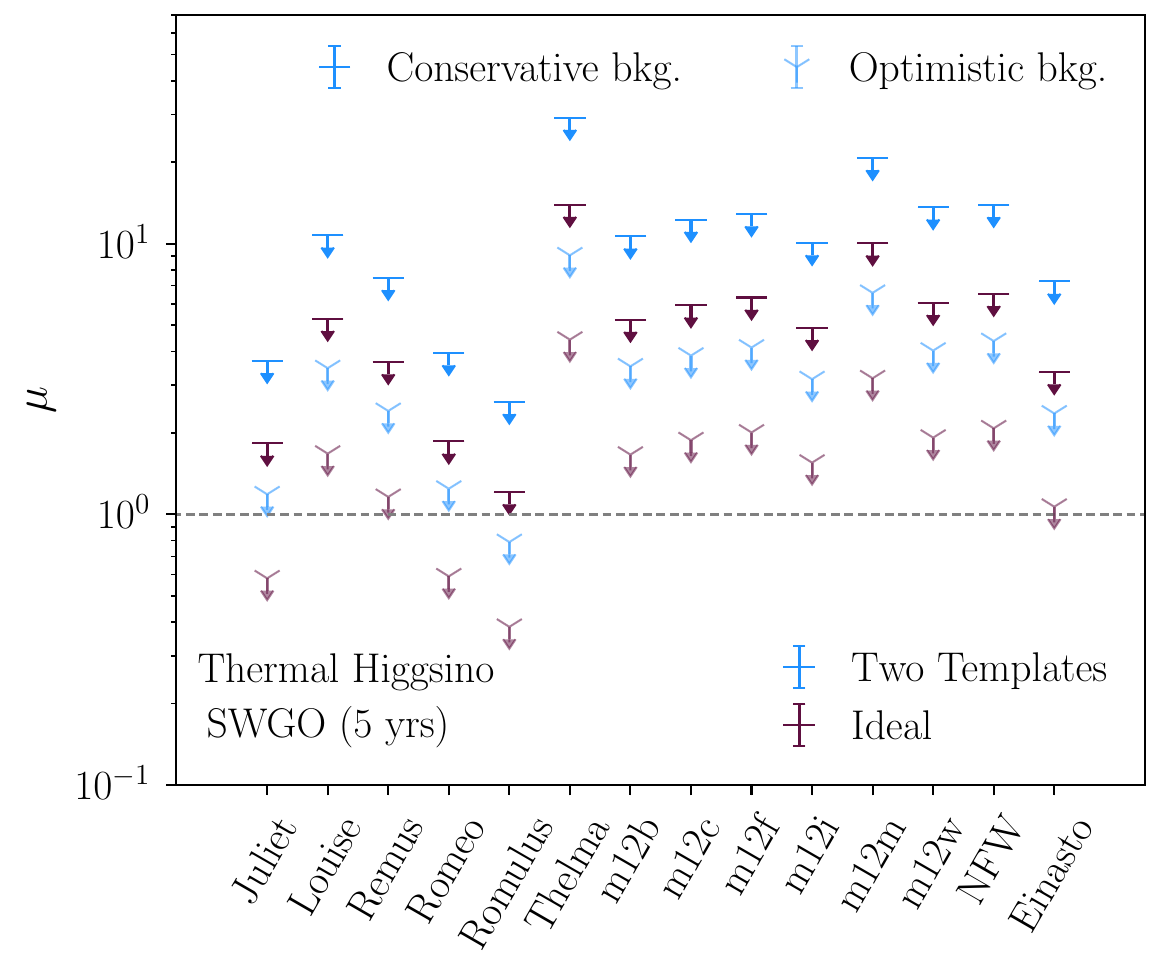}
\end{center}
\vspace{-0.7cm}
\caption{The estimated sensitivity of SWGO to the thermal higgsino, in analogy with Fig.~\ref{fig:HESS_projection_thermal}.
We project 95\% upper limits on annihilating thermal higgsinos with a fiducial SWGO setup, assuming 5 years of observation time  and two different cosmic ray rejection scenarios.
Contrasting with similar figures such as Fig.~\ref{fig:CTA_sensitivity}, the two template analysis strategy represents a substantial decrease in sensitivity relative to the ideal, fixed-background limit, largely due to the coarser energy resolution and consequently a larger signal-background degeneracy in spectral shape relative to that found with {\it e.g.} CTA.}
\label{fig:SWGO_sensitivity}
\end{figure}

We now project the sensitivity of SWGO to the higgsino signal.
Again we adopt the detector characterization from Sec.~\ref{sec:detector}. We reiterate that SWGO, which is based off of the existing HAWC detector, has a substantially different instrument response relative to IACTs such as H.E.S.S. and CTA.
In particular, HAWC and SWGO are characterized by their large FOVs and efficient accumulation of exposure time, although these advantages come at the cost of substantially poorer spectral resolution, degrading the characteristic line-like feature that is a prime indicator of the thermal higgsino. In addition, it is more difficult for SWGO to achieve similar levels of cosmic ray rejection for an equal gamma ray acceptance, compared to H.E.S.S. (for HAWC, see {\it e.g.} \cite{Abeysekara:2017mjj}).

We project the higgsino reach of SWGO assuming a fiducial scenario of 5 yrs of exposure time at a rate of 6 hrs/day, and we analyze a region around the GC consisting of the innermost 10 degrees, divided into 1 degree annuli.  We mask the Galactic plane at $|b| \leq 0.5^\circ$, and we assume a uniform exposure map across the entire region of interest. We convolve both the Fermi diffuse model and the signal maps with a $0.4^\circ$ point-spread function in accordance with the expected angular resolution at a TeV~\cite{Albert:2019afb}. We consider two modes of background rejection, as described in Sec.~\ref{sec:detector}: one scaled directly from the HAWC performance (``conservative"), and one assuming an order of magnitude improvement in rejection from this (``optimistic").  We include energies between 100 GeV and 4 TeV to assess the full sensitivity of the instrument, before any possible analysis cuts related to mismodeling considerations.

With this data set we then focus on the performance of the two template model as compared with the fixed-background, ideal scenario, where the normalizations of the two templates are fixed to their true values. The models for the cosmic-ray-induced background for SWGO are illustrated in Fig.~\ref{fig:edispbkg}. Again, no mismodeling is assumed; for SWGO a faithful model of cosmic-ray-induced flux may be obtained without allocating dedicated observing time, simply by taking as a template the observed flux in blank sky regions of its larger FOV, far away from the GC. As before, the astrophysical component is modeled using the Fermi \texttt{p8r3} spectral template.

In Fig.~\ref{fig:SWGO_sensitivity} we project the expected 95\% upper limit on the annihilation cross-section of thermal higgsinos under the null hypothesis for the optimistic and conservative background rejection levels. 
The associated discovery TS for the optimistic scenario is shown in Fig.~\ref{fig:TS_proj_CTASWGO}.
Comparing with Fig.~\ref{fig:CTA_sensitivity}, we find that a version of SWGO with these specifications is less competitive than CTA at searching for higgsino DM, even in optimistic background rejection scenarios. Ultimately the enhanced observation time of SWGO does not overcome the larger background that results from the weakened cosmic-ray rejection as well as the broader energy resolution ($\sim$20\% as opposed to $\sim$6\% for CTA). Further, the template modeling approach is considerably weaker than the ideal background subtraction, since the discrimination power of the signal is dominated by the line shape, which is smeared out due to the larger energy resolution.
Nonetheless, it is interesting to note that ``low and flat" DM profile realizations  that are challenging to search for, in particular m12m and Thelma, appear to benefit from uniform exposure within a large FOV, since there is still comparable amounts of signal to be gained at $\sim$10$^{\circ}$ from the GC. 
Large FOV observatories such as SWGO may have a unique advantage to discovering DM in more cored profile realizations, if improvements to the current specifications, especially the energy resolution and background rejection, are possible.

\section{Discussion}
\label{sec:discussion}

The higgsino is the canonical WIMP DM candidate. It is minimal, predictive, well-motivated by both UV and infrared considerations, and could emerge as the first hint of a wealth of new physics beyond the SM addressing the electroweak hierarchy problem and gauge unification.  While the higgsino direct detection scattering cross section is likely below the neutrino floor, the thermal higgsino is a major science driver of the next-generation of proposed particle colliders~\cite{Bottaro:2022one,Capdevilla:2021fmj,Accettura:2023ked,Saito:2019rtg,EuropeanStrategyforParticlePhysicsPreparatoryGroup:2019qin,Canepa:2020ntc,Black:2022cth,Fukuda:2023yui,Franceschini:2022sxc,Narain:2022qud,Aime:2022flm,MuonCollider:2022xlm}.

In this paper we show that evidence for the ``smoking gun'' signature of this DM candidate -- the line-like feature at the endpoint of the spectrum -- is well within reach of the upcoming CTA and, to a lesser extent, SWGO.
As a testing ground for the analyses that could be deployed in those instruments, we study in detail the reach of the existing H.E.S.S. instrument, along with the Fermi-LAT, to the line-like feature from the thermal higgsino.
In particular, we considered an analysis of the ON H.E.S.S. data using parametric and separately non-parametric models for the cosmic-ray-induced and astrophysical backgrounds, both of which obtain enhanced sensitivity, without evidence for bias, relative to the canonical ON minus OFF analyses currently performed by H.E.S.S..  Our work strongly motivates a dedicated search for the higgsino signal with H.E.S.S. data by the Collaboration, including accurate spectral templates for the cosmic-ray-induced emission.

We show that the sensitivity of the Fermi-LAT is within a factor of a few of what is needed to probe the line-like feature from the thermal higgsino across a broad range of possible DM profiles.  It is also interesting to consider the reach proposed successors to Fermi, such as the Advanced Particle-astrophysics Telescope (APT)~\cite{Alnussirat:2021tlo,APT:2021lhj}.\footnote{See also~\cite{Xu:2023zyz} for the reach of APT to model-independent continuum final states.} The instrument is proposed to achieve roughly an order of magnitude improvement over Fermi in the effective area although at the cost of a factor of $\sim$2 reduction in the energy resolution. A dedicated projection, beyond the scope of this work, is needed to assess the sensitivity of the APT to higgsino annihilation, but a back-of-the-envelope estimate, using the scaling relations presented in this work, suggests that APT would be sensitive to higgsino annihilation for some of the DM profiles considered here.

In the process of analyzing the H.E.S.S. data from \cite{HESS:2022ygk}, we identify an issue in several DM searches the instrument has performed in the inner Galaxy.
In particular, at the flux levels that work claims to achieve (or equivalently the size of the effective area), the Galactic diffuse emission should have been detected with enormous significance.
The fact that it was not suggests that the detected flux, and therefore the inferred effective area, appear to have been incorrectly estimated by a factor $\sim$8.
Correcting this would weaken the DM limits claimed in that work by this same factor, and if the issue is confirmed it will be important to determine if other DM searches with H.E.S.S. have also been impacted.
This point is discussed further in Apps.~\ref{app:WW},~\ref{app:Line}, and~\ref{app:dwarf}.

In summary, our projections make the science case for CTA South particularly compelling, as it will be able to probe the thermal higgsino model across a wide variety of possible DM profiles for the Milky Way.
In this work we rely on the 12 analogue Milky Way galaxies from the FIRE-2 cosmological simulations. 
As seen throughout this work, for instance in Fig.~\ref{fig:TS_proj_CTASWGO}, the differences between these simulations can be substantial, to the extent that for some profiles ({\it e.g.}, Romulus) CTA would obtain a discovery well over $5\sigma$, whereas for others ({\it e.g.}, Thelma) CTA would see at best a $1\sigma$ hint. It is an urgent task to better understand the physical aspects that go into driving these differences in DM profiles so that more work can be done to shrink the uncertainties on the DM profile in the inner Galaxy of our own Milky Way.  

\section{Acknowledgements}

{\it
We thank Christopher Dessert and Pat Harding for helpful comments, Yujin Park for collaboration during the early stages of this project, and especially Joshua Foster for a number of useful discussions and assistance with the Fermi data.
Lastly, we thank Martin Beneke, Torsten Bringmann, Dan Hooper, Tomohiro Inada, and Tracy Slatyer for comments on a draft version of this work.
BRS was supported in part by the DOE Early Career Grant DESC0019225.
The work of NLR and WLX was supported by the Office of High Energy Physics of the U.S. Department of Energy under contract DE-AC02-05CH11231. 
This research used resources of the National Energy Research Scientific Computing Center (NERSC), a U.S. Department of Energy Office of Science User Facility located at Lawrence Berkeley National Laboratory, operated under Contract No. DE-AC02-05CH11231 using NERSC award HEP-ERCAP0023978.
}

\appendix

\section{H.E.S.S. Inner Galaxy Survey and Detector Response Inference}
\label{app:HESS_detector}

In this appendix we describe our inference of the observed counts and detector response of the H.E.S.S.-II IGS based on the publicly released information provided in~\cite{HESS:2022ygk}.
The main information presented in that work which we use are the morphology of the exposure map, the locations of 14 IGS pointings, the total exposure time (546 hrs) between the IGS and 2014-2015 Sgr A* observations, and the total ON and OFF counts summed over energy bins for each ROI, 25 in total, between $0.5^\circ$ and $3.0^\circ$ away from the GC. The energy bins adopted in this analysis are 67 log-spaced bins whose edges span 0.17 to 66.2 TeV. In addition, numerous figures in the Supplemental Materials of Ref.~\cite{HESS:2022ygk} display observed count rates and fluxes over a subset of ROIs, and offer valuable insight into  the energy-dependence of the detector response and on-sky fluxes.
Although this represents a considerable amount of information, we emphasize that critical information is not provided.
This includes the specific exposure times spent at each pointing location, the precise pointing locations for and total exposure time accumulated by the dedicated Sgr A* observations, the explicit response of the detector and its variation on an observation-by-observation basis, and the distribution of ON and OFF counts over energy bins for each annulus.  

Here we primarily concern ourselves with reconstructing (i) the magnitude of the on-axis effective area $A_{\rm eff}(E)$ and (ii) the distribution of exposure times on a pointing-by-pointing basis. The inference of the latter point also requires the reconstruction of the dedicated 2014-2015 Sgr A* observations.  It is important to emphasize that several of  our results will fundamentally depend on the fidelity of these inferences. Explicitly, our re-analysis of public H.E.S.S.-II data and estimation of astrophysical contamination will depend on both of these reconstructed quantities: different pointing locations will yield different ON-OFF residuals from both diffuse emission and a putative DM signal.  Our inference of the numerical value by which we correct the H.E.S.S. effective area is also dependent on these reconstructions, although we find it encouraging that two independent estimations of this factor -- via geometry of the exposure map and via anchoring the diffuse flux to observed ON-OFF excesses -- yield similar results.  We estimate the sensitivity of the astrophysical and DM ON-minus-OFF residual fluxes to the assignment of exposure times at the end of this section, and show that the those are robust at the $\sim$10\% level.  

A further caveat to stress is that we have no ability to capture observation-by-observation fluctuation in the detector response or observed on-sky background.
We instead work with averaged quantities and assume that the spatial and spectral structure of the effective areas are essentially similar, at least when averaged pointing-by-pointing.  This is an imperfect assumption, but it is difficult to imagine these intrinsic fluctuations driving the large factors of discrepancy suggested from our analysis.

\subsection{Reconstruction of the On-Axis Effective Area}
\label{subsec:Aeff_reconstruction}

\begin{figure*}[!t]
\begin{center}
\includegraphics[width=0.48\textwidth]{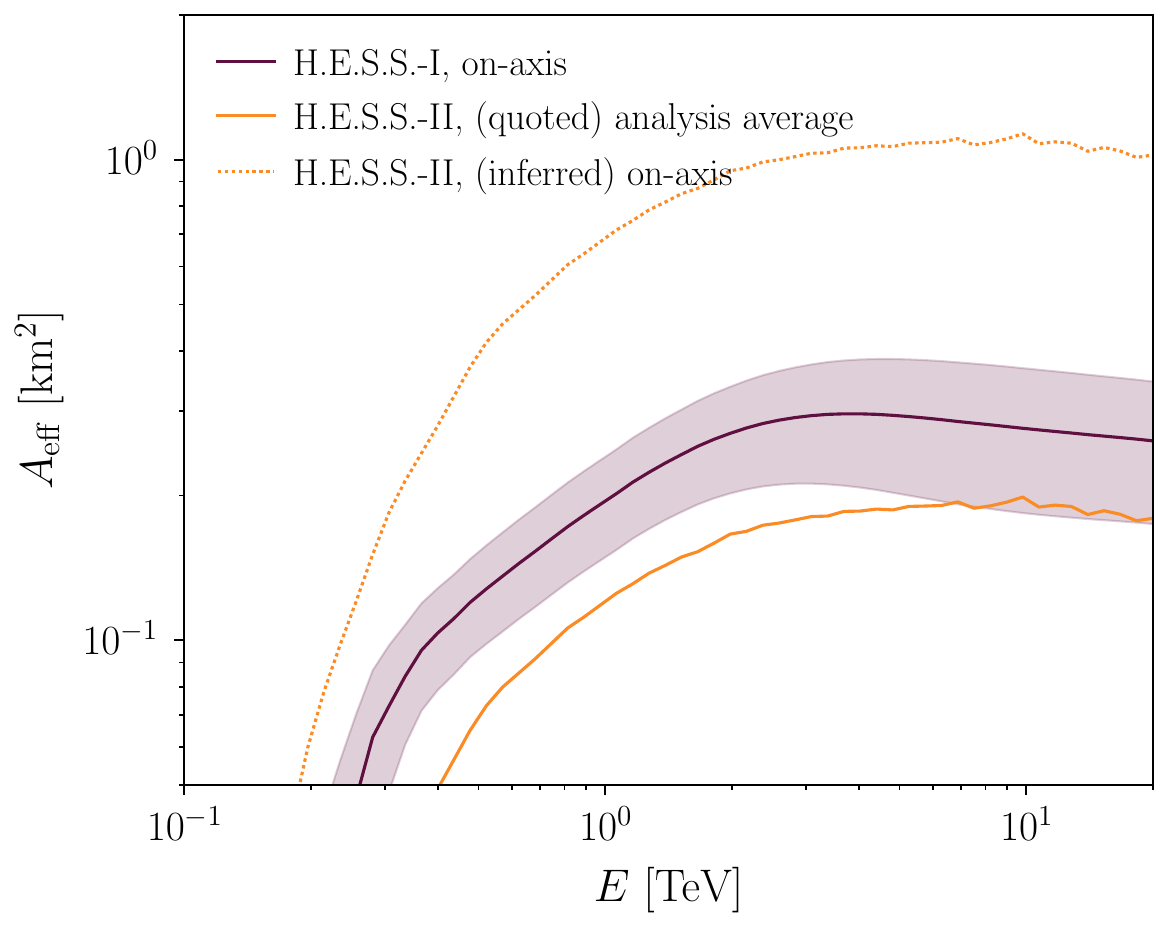}
\includegraphics[width=0.48\textwidth]{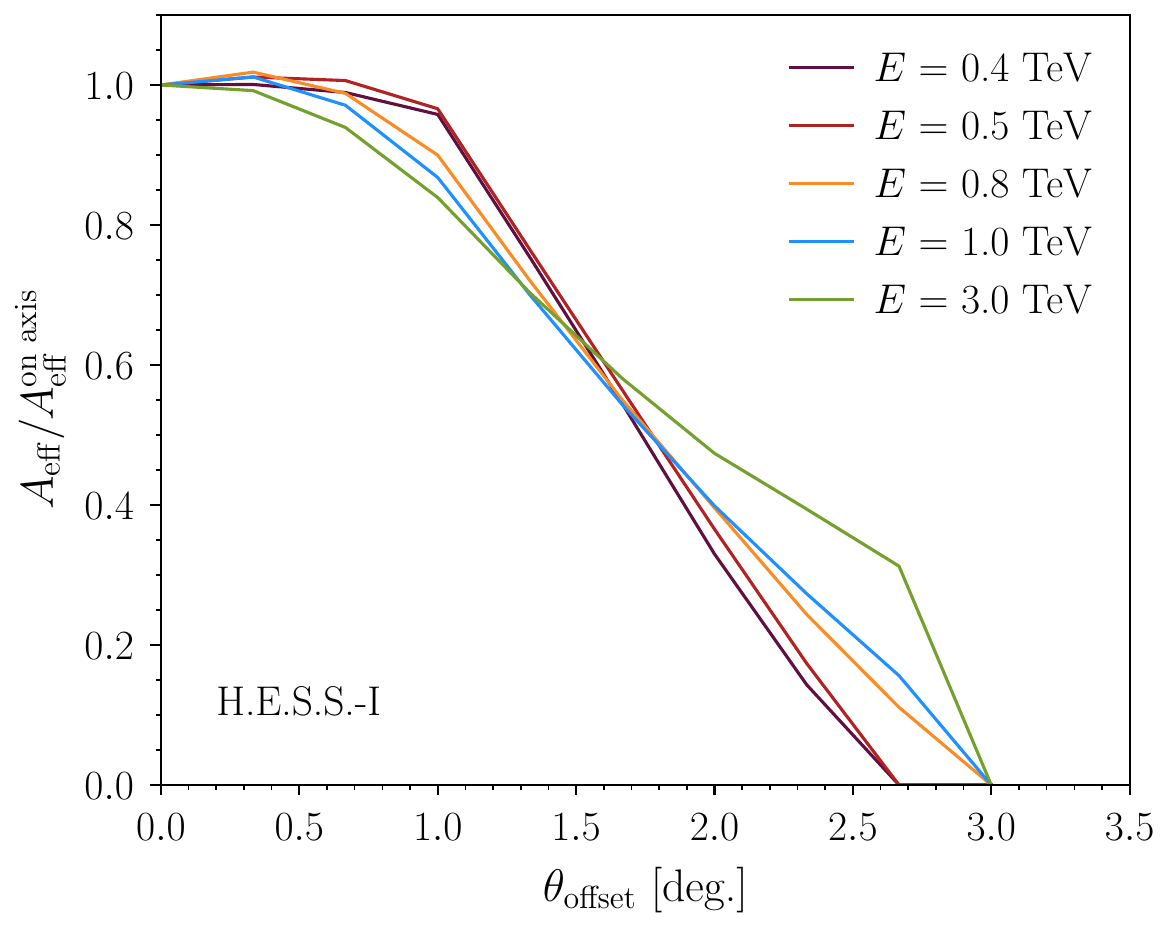}
\end{center}
\vspace{-0.7cm}
\caption{Similar to Fig.~\ref{fig:Aeff} but focusing on the H.E.S.S. instrument response. (Left panel) The expected scale and energy-dependence of the effective area.  The purple line (shaded region) represents the median  (standard deviation) of on-axis effective areas spanned by 38 blank sky observations from H.E.S.S.-I public data release 3~\cite{HESS:2018zix} taken at zenith angle $\leq 40^\circ$.  Also shown in orange is the quoted analysis average effective area of the H.E.S.S.-II DM search (solid), and what the inferred on-axis value would be based on the geometry of the exposure map (dotted). (Right panel) The spatial profile of effective area as a function of offset angle for a range of energies, taken from the same set of H.E.S.S.-I observations. As shown, the spatial responses of the instrument at $\sim$ TeV energies or lower, which accounts for the vast majority of counts observed, are reasonably consistent with one another.}
\label{fig:AppA_Instrument_Response}
\end{figure*}

An averaged effective area $\overline{A_{\rm eff}}(E)$ can be straightforwardly inferred by taking the ratio of Figs.~5 and 6 of the Supplemental Material in Ref.~\cite{HESS:2022ygk}.
This ``analysis" effective area is an average specific to the search conducted in~\cite{HESS:2022ygk}, encapsulating the full amount of data contained in the chosen ROIs, across the full set of observation runs, after symmetric ON and OFF regions have been constructed and masking applied. This is the key quantity that links fluxes -- both signal and background -- to counts: the total number of ON counts observed within energy bin $j$ is related to the stacked flux as, 
\begin{equation}
N^{\rm ON, tot}_j = \frac{d \Phi_j^{\rm ON, tot}}{dE d\Omega} \cdot \delta E_j  \cdot \overline{A_{\rm eff}}(E_j) \cdot t_{\rm exp}^{\rm tot} \cdot \Delta \Omega^{\rm tot},
\label{eq:Nonanalysis}
\end{equation}
where $t_{\rm exp}^{\rm tot} = 546$ hrs and $\Delta \Omega^{\rm tot}=6.4\times 10^{-3}$ sr is the post-masking area within the ON region $r \in [0.5^\circ, 3^\circ]$.
Here we focus on the ON region to illustrate the logic of our inference, but by construction the ON and OFF regions receive equal exposure and cover equal solid angles on the sky.
It is worth stressing again that this is the quantity that links the expected DM flux -- a first-principles prediction -- to expected counts seen in the survey,  and thus ultimately our concern about this effective area will translate into concern about the science results in~\cite{HESS:2022ygk}.   This H.E.S.S.-II analysis effective area as quoted is illustrated in the left panel of Fig.~\ref{fig:AppA_Instrument_Response} and is expected to be smaller than the on-axis response of the instrument simply because the analysis averages over areas larger than the FOV and only a subset of data is ultimately analyzed.  Also plotted is the H.E.S.S.-I on-axis response, the median and standard deviation of 38 observations\footnote{We choose here blank-sky observations observed at zenith angle $\leq 40^\circ$.} contained in the public release.

It is informative to try and estimate what a similar average on-axis effective area for the H.E.S.S.-II data would look like. (Here we mean the average on-axis response over numerous observation runs, as opposed to averaged spatially over the analysis ROI.)  To do this, we note that the total number of counts seen in spectral bin $j$ can also be written 
\begin{equation}
N_j^{\rm ON, tot} =  \sum_{i} N^{{\rm ON}, i}_{j} = \sum_{i}\frac{d \Phi_j^{{\rm ON}, i}}{dE d\Omega} \cdot \delta E_j  \cdot \mathscr{E}^{i} (E_j),
\label{eq:Nonfull}
\end{equation}
set by the flux observed within the ON region at each of the pointings, which are indexed by $i$. The exposure accumulated around a pointing location $i$ is in turn set by integrating the acceptance over the sky-positions of the ON region, 
\begin{equation}
    \mathscr{E}^{i} (E_j) = t_{\rm exp}^i \times \int_{\rm ON} d\vec \Omega \, A^i_{\rm eff}(E_j, \vec \Omega).
\end{equation}

In order to make progress, we assume that the detector response is essentially similar across all observing runs, for both the IGS and Sgr A* data sets.
This allows the energy and spatial dependence of the effective area to factorize, 
\begin{equation}
A^i_{\rm eff}(E_j, \vec\Omega) = A^{\operatorname{on-axis}}_{\rm eff} (E_j)  \times \zeta (|p^i - \vec \Omega|),
\end{equation}
where $\zeta$ models the fall-off of effective area as a function of offset angle $|p^i - \vec \Omega|$ with respect to pointing location $p^i$.  The left panel of Fig.~\ref{fig:AppA_Instrument_Response} shows that the on-axis response might be expected to fluctuate around 50\% about the median.  The right panel, conversely, shows the spatial profile of the H.E.S.S.-I instrument response over a range of energies relevant to this discussion. We model the spatial response around the behavior at energies of 1 TeV, as a Gaussian with
\begin{equation}
\zeta(r) =  \exp\left( - \frac{r^2}{2\sigma^2}\right)\!, \hspace{0.5cm}  \sigma = 1.45^\circ.
\label{eq:Aeff_offset}
\end{equation}
While the spatial falloff is ill-described by a Gaussian at higher energies, we find that at TeV energies or lower (the energies dominating the observed flux) the spatial response is fairly energy-independent and a Gaussian is a reasonable description of the H.E.S.S.-I response. Finally, under this assumption the averaged effective area inherits the same energy dependence as its on-axis counterpart,
\begin{equation}
A^{\operatorname{on-axis}}_{\rm eff} (E) \equiv \alpha \overline{A_{\rm eff}}(E).
\end{equation}

To determine $\alpha$, we match the predicted counts from \eqref{eq:Nonanalysis} and \eqref{eq:Nonfull}, and then assume that the on-sky flux is isotropic, an assumption well-justified if the cosmic-ray background dominates the counts. With this assumption, $\alpha$ is purely geometric and can be computed from the distribution of exposures across pointing locations, 
\begin{equation}
\alpha \simeq \frac{ t_{\rm exp}^{\rm tot} \cdot \Delta \Omega^{\rm tot}}{ \int_{\rm ON} d\vec \Omega \; \sum_i t^i_{\rm exp}  \zeta (|p^i - \vec \Omega|) }.
\end{equation}
For the set of inferred exposure times for each pointing determined in Sec.~\ref{subsec:t_exp_reconstruction}, this factor evaluates to 5.7. 
The largeness of this factor is somewhat inherent to the inefficiencies of the ON-OFF subtraction figure, and the fact that most of the ON region is observed far from on-axis by any of the pointings.
As is apparent in Fig.~\ref{fig:AppA_Instrument_Response}, this would imply an on-axis response of H.E.S.S.-II that peaks at roughly $1\text{ km}^2$, which would be much larger than expected (cf. Fig.~\ref{fig:Aeff}).
Equivalently, the analysis average effective area quoted by~\cite{HESS:2022ygk} is considerably larger than should be expected with a H.E.S.S.-type instrument and the IGS observation strategy.
Given that the claimed $\overline{A_{\rm eff}}(E)$ is not too inconsistent with the H.E.S.S.-I on-axis area, we expect the true analysis quantity -- and thus the true amount of DM counts expected in the search -- to be approximately a factor of 6 lower, which should be compared with the factor of 8 we concluded from a study of the expected diffuse emission in Sec.~\ref{sec:HESS_real}.

\subsection{Inference of Pointing-by-Pointing Exposure Times}
\label{subsec:t_exp_reconstruction}

\begin{figure*}[!tb]
\begin{center}
\includegraphics[width=0.48\textwidth]{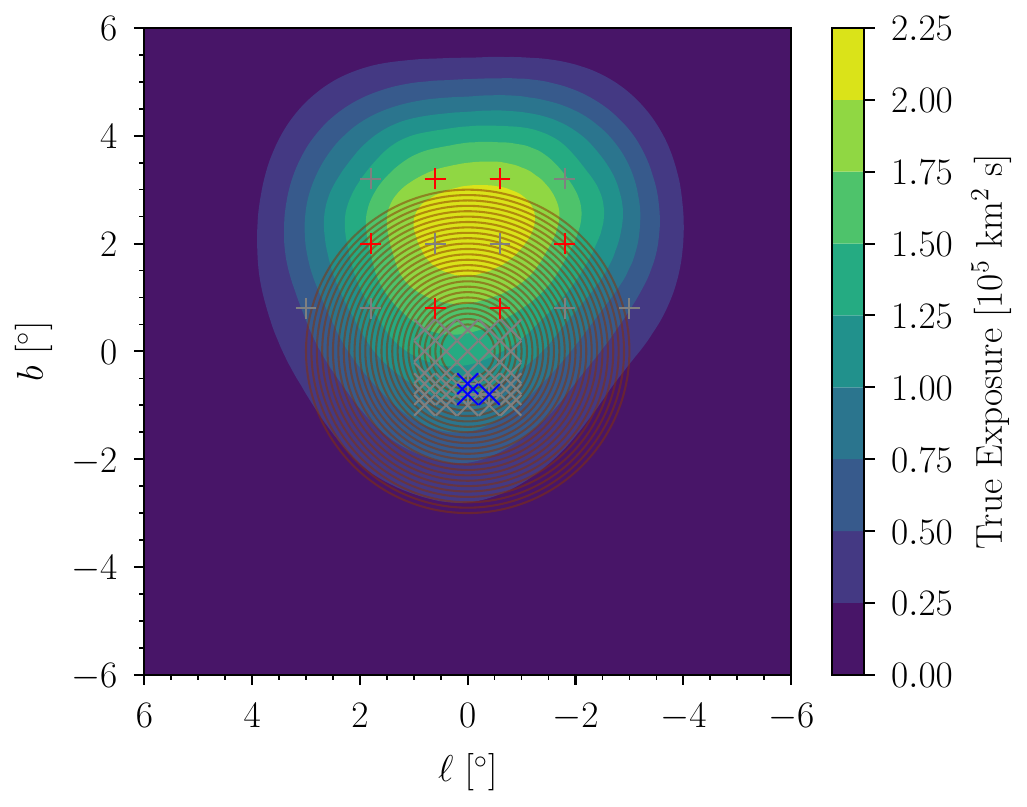}
\includegraphics[width=0.48\textwidth]{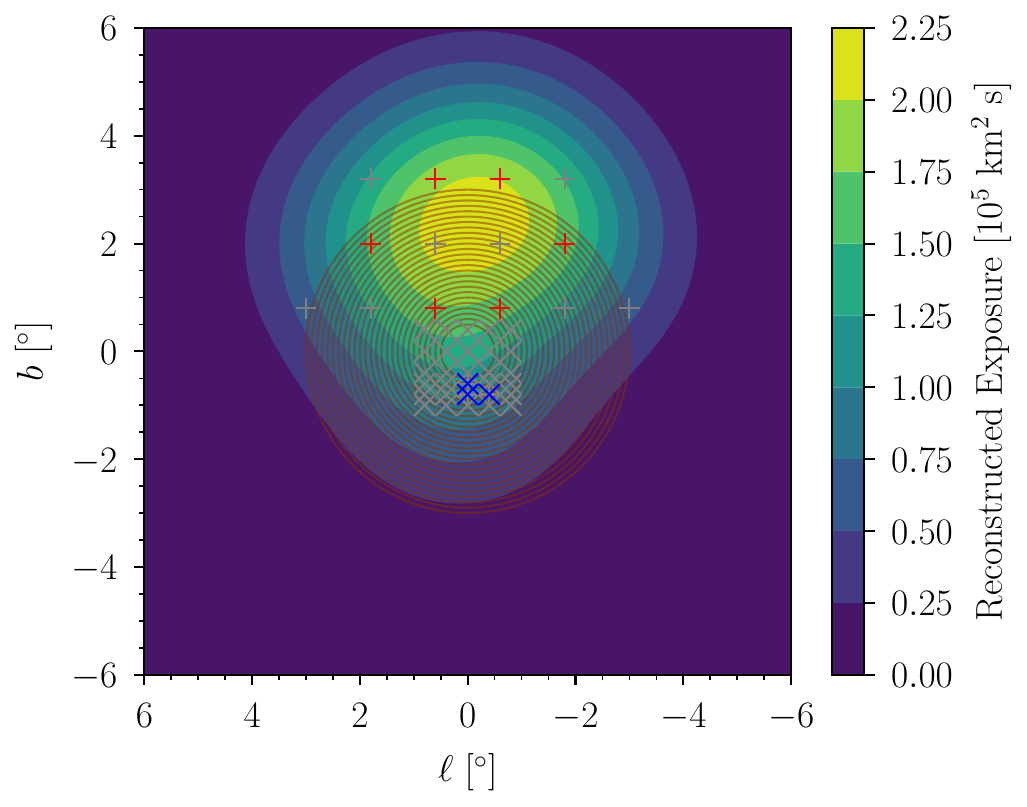}
\includegraphics[width=0.48\textwidth]{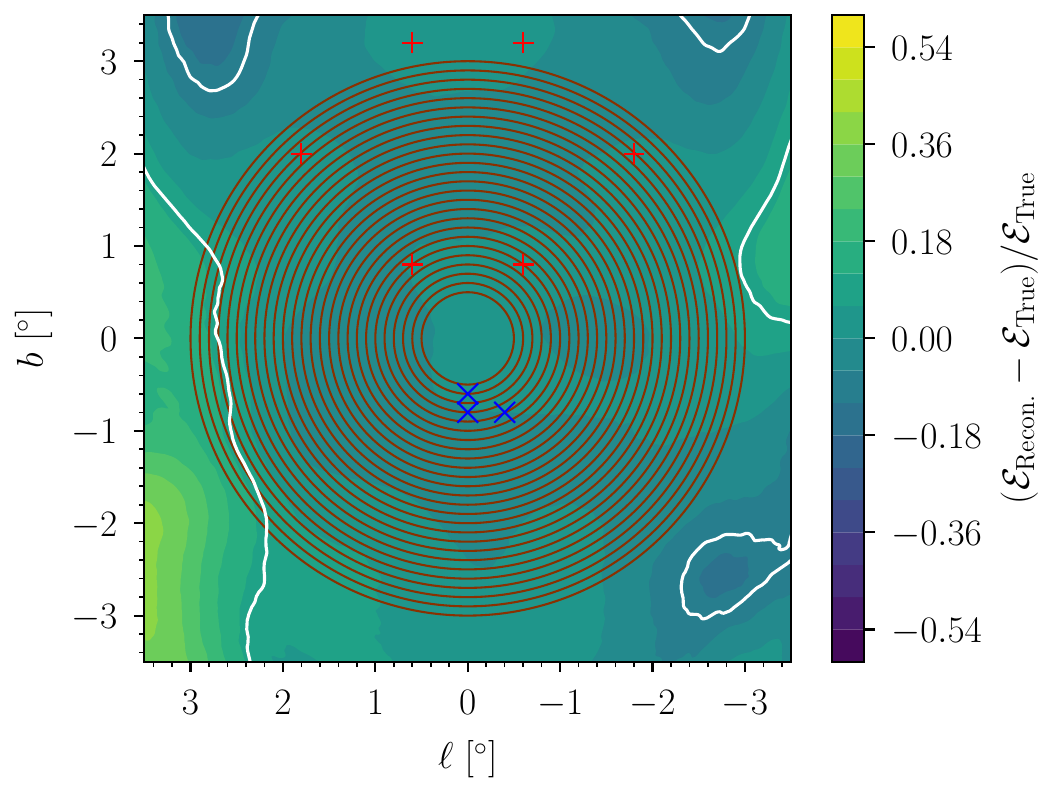}
\end{center}
\vspace{-0.7cm}
\caption{The true (top left) exposure map reproduced from~\cite{HESS:2022ygk} and our best-fit reconstruction (top right), given the set of (candidate) pointing locations shown.  ``$+$'' indicates a pointing location given as part of the inner Galaxy survey, though only the pointings in red return a non-zero best-fit exposure time.  ``x'' indicates a candidate pointing location for the dedicated Sgr A* observations that make up a portion of the H.E.S.S.-II GC data set. The markers in blue indicate locations that return a non-zero best-fit exposure time.  These times, as well as the locations of these pointings, are listed in Tab.~\ref{tab:exposure_times}. (Bottom panel) The residuals of the true minus reconstructed exposure map; the region delineated by the white contour reproduces the structure of the true exposure map at the $\pm 10\%$ level, and covers the vast majority of the analysis ROI.}
\label{fig:reconstruced_exp}
\end{figure*}

\begin{table*}[]
    \centering
\begingroup
\setlength{\tabcolsep}{7pt} 
\renewcommand{\arraystretch}{1.4} 
    \begin{tabular}{c | c c c c c c c c c c c c c c c  }
    \hline
         Pointing &  1-4 & 1-5 & 1-6 & 1-7 & 1-8 & 1-9 & 2-5 & 2-6 & 2-7 & 2-8 & 3-5 & 3-6 & 3-7 & 3-8 \\
         \hline 
         $\ell \, [^\circ]$  & -3.0 & -1.8 &  -0.6 & 0.6 &1.8& 3.0& -1.8& -0.6& 0.6& 1.8& -1.8& -0.6& 0.6& 1.8 \\
         $b \,[^\circ]$  &  0.8 & 0.8 & 0.8 & 0.8 & 0.8 & 0.8 & 2.0 & 2.0 & 2.0 & 2.0 & 3.2 & 3.2 & 3.2 & 3.2 \\ 
         $t_{\rm exp}$ [hrs] &  0 & 0 & 70.8 & 5.2 & 0 & 0 &  96.2 & 0 & 0 & 118.2 & 0 & 76.9 &  116.4 & 0  \\
    \hline
    \end{tabular}
    \begin{tabular}{c| c c c }
        Pointing & S-1 & S-2 & S-3 \\
    \hline
        $\ell \, [^\circ]$  & 0 &  0 & -0.4 \\
        $b \, [^\circ]$  & -0.6 & -0.8 & -0.8 \\
        $t_{\rm exp}$ [hrs]  & 11.3 & 42.8 & 25.7\\
    \hline
    \end{tabular}
\endgroup
\caption{Our best approximations to the pointing locations and exposure times that make up the exposure used for the H.E.S.S. DM annihilation search within the inner Galaxy in~\cite{HESS:2022ygk}. H.E.S.S. provides the pointing locations for the IGS pointings (labeled 1-4 through 3-8) but does not provide pointing locations for the dedicated Sgr A* observations, our inferences for which are labeled S-1 to S-3. We compute these best-fit exposure times $t_{\rm exp}$ per pointing, along with the pointing locations of the Sgr A* observations, by matching a coarse-grained model exposure map to that presented in~\cite{HESS:2022ygk}; see text for details. While several pointing locations return a best-fit exposure time of 0 hours, we recover a total observing time in good agreement with the 546 hrs reported by the collaboration.}
\label{tab:exposure_times}
\end{table*}

Next we consider how the 546 hrs of data collected for the IGS survey is distributed amongst the individual pointings. The full exposure map, reproduced in Fig.~\ref{fig:HESS_II_exposure}, represents the combined observations taken over 2014-2020 as part of the IGS and dedicated Sgr A* observations by H.E.S.S.. As discussed above, we assume that around each pointing the fall-off of exposure with beam angle is consistent with that summarized in Fig.~\ref{fig:Aeff} and well-modeled by~\eqref{eq:Aeff_offset}.  The remaining unknowns for reproducing the full map are then the pointing locations where the 2014-2015 Sgr A* data were taken and the time spent at each of these locations.
To evaluate this, we down-bin the exposure map to 40$\times$40 pixels, indexed by $k$, between $\ell, b \in [-5^\circ, 5^\circ]$, and compare model to data with the statistic
\begin{equation}
\Delta \chi^2 = \sum_k \left( \sum_i t^i_{\rm exp}\zeta(\theta_{ik}) - \tilde{\mathcal{E}}_k \right)^2\!,
\end{equation}
where the exposure $\tilde{\mathcal{E}}_k$ is normalized such that 
\begin{equation}
\sum_k \tilde{\mathcal{E}}_k =  t^{\rm tot}_{\rm exp} \times \sum_k \zeta (\theta_k),
\end{equation}
and $t^{\rm tot}_{\rm exp} = 546$ hrs. 

For the positions of the Sgr A* observations, we consider a grid of 30 possible positions close to the GC, between $\ell \in [-0.8^\circ, 0.8^\circ]$ and $b \in [-1.0^\circ, 0.4^\circ]$. We find the maximum likelihood configuration of the exposure time distribution over the 14 IGS pointings and the 30 potential Sgr A* pointings; our bestfit reconstruction is given in Table~\ref{tab:exposure_times}.  We find, surprisingly, that several of the provided H.E.S.S.-II IGS pointing locations were consistently best fit to zero exposure time, and that only three of the proposed Sgr A* pointings were preferred by the model. At the end, we find a reasonably faithful reconstruction of the exposure map in Fig.~\ref{fig:HESS_II_exposure} with only 6 of the 14 listed IGS pointings. Also, we note that our best-fit reconstruction yields a total exposure time slightly larger than stated, at 560 hrs, which we attribute to some variance in the on-axis acceptance on a run-by-run basis.  In Fig.~\ref{fig:reconstruced_exp} we show the pointing locations we consider, our best-fit reconstructed exposure map, and the relative residual between true and modeled maps.  As shown, the reconstruction is evidently imperfect, but agreement between the two everywhere within the total analysis ON region is within 10\%, and we expect our re-analysis of the H.E.S.S.-II data to be faithful to approximately this order.

\subsection{Estimation of Diffuse and Signal Counts}

In the previous two subsections we estimate the detector response and the observational strategy used in Ref.~\cite{HESS:2022ygk}.
We now use these to estimate the expected astrophysical and DM counts that should be seen in both the ON and OFF regions.  

A key observation in the main text is that the Galactic diffuse emission seen by Fermi-LAT should have produced a much larger asymmetry in the observed ON and OFF counts than reported by Ref.~\cite{HESS:2022ygk} -- see Fig.~\ref{fig:HESSonoffDiff} and surrounding text for a discussion.  This points to a fundamental mismatch in the TeV gamma-ray sky seen by Fermi and that seen by H.E.S.S., if the background flux and analysis acceptance of the latter are taken as quoted.  This discrepancy, in conjunction with the surprisingly small flux reported in the right of Fig.~\ref{fig:edispbkg}, which would require an extremely efficient cosmic-ray rejection, and the overly large on-axis effective area computed in App.~\ref{subsec:Aeff_reconstruction}, lead us to conclude that the on-sky flux and effective area must be under/over-reported by a common factor, such that the total observed counts remains as stated.  Anchoring the ON-OFF residual to the expected anisotropic flux predicted by the Fermi diffuse model produces a relative factor of $\sim$$8.1$. 

Of course, we emphasize that this claim is reliant on our inference of experimental properties not explicitly given, and so it is important to study how sensitive our statements are to variations in these inferences. We study this here. Specifically, we consider whether the exposure time could plausibly be divided between pointings such that the flux and effective areas quoted in \cite{HESS:2022ygk} could be consistent with the diffuse emission not appearing significantly in the ON minus OFF data.

\begin{figure}[!t]
\begin{center}
\includegraphics[width=0.485\textwidth]{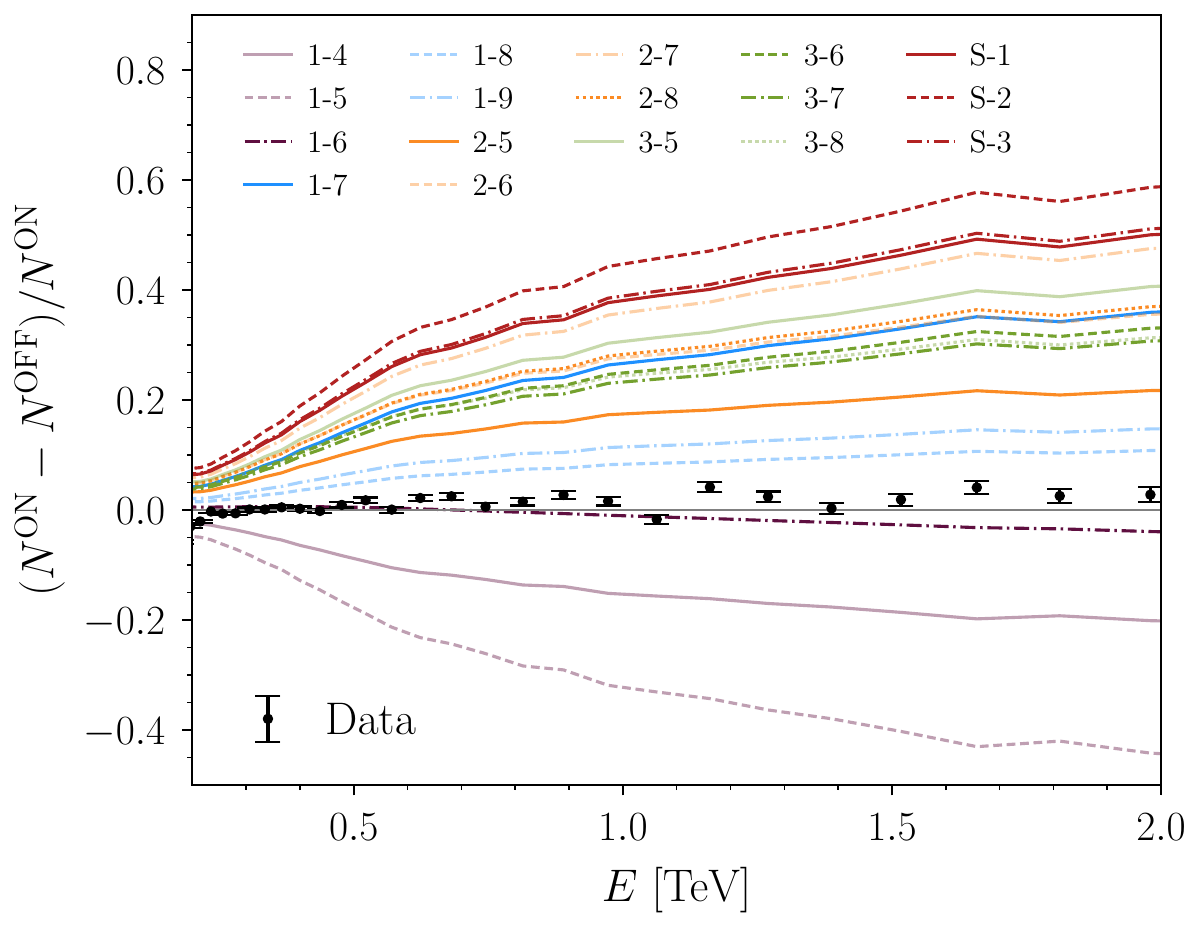}
\end{center}
\vspace{-0.7cm}
\caption{The fractional ON-OFF asymmetry generated by diffuse emission in the full analysis ROI around each pointing.
The pointings that in our fit prefer a non-zero exposure time are shown with higher opacity.
The near-zero asymmetry observed in the data cannot be accounted for by a realistic reweighting of the different pointings.
The results shown here follow from a direct extraction from~\cite{HESS:2022ygk}.
If we apply a correction factor that reduces the predicted diffuse counts by a factor of $\sim$8 (see Sec.~\ref{sec:HESS_real}), the data and prediction would be in good agreement.}
\label{fig:Diffuse_Signal_Difference}
\end{figure}

In Fig.~\ref{fig:Diffuse_Signal_Difference} we show the expected ON-OFF asymmetry (assuming no DM contribution) per ON photon, compared to the asymmetry observed in the data.
The expected asymmetry is shown for each of the 14 provided IGS pointings and 3 potential Sgr A* pointings.
By construction of the ON and OFF regions, the approximately isotropic cosmic-ray contribution should cancel in the difference, and therefore the asymmetry should solely originate from the diffuse emission.
In constructing the expected emission for each curve here, we assume the cosmic-ray flux is at the level of the observed flux presented in \cite{HESS:2022ygk} (again, see Fig.~\ref{fig:edispbkg}).

Figure~\ref{fig:Diffuse_Signal_Difference} demonstrates that almost all pointings would generate a large positive ON-OFF asymmetry from astrophysical emission and one that is considerably larger than what is observed in the data.
The notable exceptions to this are pointings 1-4 and 1-5, where the lower observing latitude allows a large amount of diffuse flux to enter the OFF region, and the flux in the ON region is significantly suppressed by the shape of the applied mask. Note that this negative asymmetry is not present in pointings 1-8 and 1-9 since the masking (and diffuse flux) is not left-right symmetric.  In this case, if the background flux is truly as low as reported, in order to achieve the near-zero ON-OFF asymmetry seen in the data roughly half of the photon counts would have to be collected around 1-4 or 1-5. In our reconstruction, we find a best-fit of zero exposure time in both of these locations, but even visually in Fig.~\ref{fig:HESS_II_exposure} the low-latitude IGS pointings (1-4 through 1-9, which encompass the less-positive ON-OFF predicted rates, in addition to the overtly negative ones) obviously do not dominate the collected exposure. If at least half of data collection is done around pointings with $b \geq 2$, a clear underestimate from Fig.~\ref{fig:HESS_II_exposure}, then the remaining half must be split exclusively between pointings 1-4 and 1-5 to cancel out the resulting $\sim 30\%$ expected asymmetry. This is clearly simply not the case, indicating that this discrepancy between H.E.S.S. and Fermi observations of the GC gamma-ray sky goes beyond the specifics of our exposure time reconstruction.
If, however, the cosmic-ray flux were a factor of 8.1 larger, then the predicted asymmetries in the figure would all be decreased by roughly the same amount, and it would not be challenging to choose a breakdown of pointings that matches the observed data.

Although the exact asymmetry can vary noticeably between pointings for the diffuse emission, if we consider instead the signal DM would generate in the ON-OFF data, we find that for all pointings there is more counts in the ON rather than OFF region. Of course, this is exactly as would be expected given the purpose of the ON and OFF regions.
More to the point, the asymmetry does not vary considerably between the specific pointings, suggesting that errors in our exact reconstruction of the exposure map are unlikely to significantly impact searches for DM in the ON minus OFF data, certainly no larger than the 10\% errors in our reconstructed exposure map.
Further, if we work with the ON data set alone as we advocate in the main text, the number of DM counts is considerably more stable than in the ON-OFF data set.
This demonstrates that the results of our H.E.S.S. re-analysis should likewise be robust against the details of the reconstructions discussed here.
(This is further validated by the results in App.~\ref{app:WW} where we show we can almost exactly reproduce the key results of \cite{HESS:2022ygk}.)

\section{H.E.S.S. analysis in annuli}
\label{app:annuli}

In Sec.~\ref{sec:HESS_real} we analyze the H.E.S.S data, reconstructed in App.~\ref{app:HESS_detector}, stacked over the full ROI, thereby removing all spatial information. In this appendix we present results of analyses performed within individual annuli, with the caveat -- as emphasized in App.~\ref{app:HESS_detector} -- that our reconstruction of the data in the annuli is less reliable than for the full ROI. Nevertheless, our focus is primarily to show that the results when including spatial information are broadly similar to those we presented in the main text.

The H.E.S.S. analysis in \cite{HESS:2022ygk} divided the sky from 0.5$^\circ$ to $3.0^\circ$ from the GC into 25 annuli each of width $0.1^\circ$. Here we combine these into eight larger regions that amount to combining the following set of annuli: 1-6, 7-11, 12-15, 16-18, 19-20, 21-22, 23-24, and 25.  We analyze the stacked data in each of these sub-regions independently, with independent nuisance parameters, and then we construct the joint likelihood for the $\mu$ parameter at fixed $m_\chi$ by taking the product of the profile likelihoods in the individual sub-regions. For simplicity, we only present results here for the power-law analysis procedure, with the same sliding-window-size as in the main text for the analysis of the fully-stacked ROI.

\begin{figure}[!t]
\begin{center}
\includegraphics[width=0.49\textwidth]{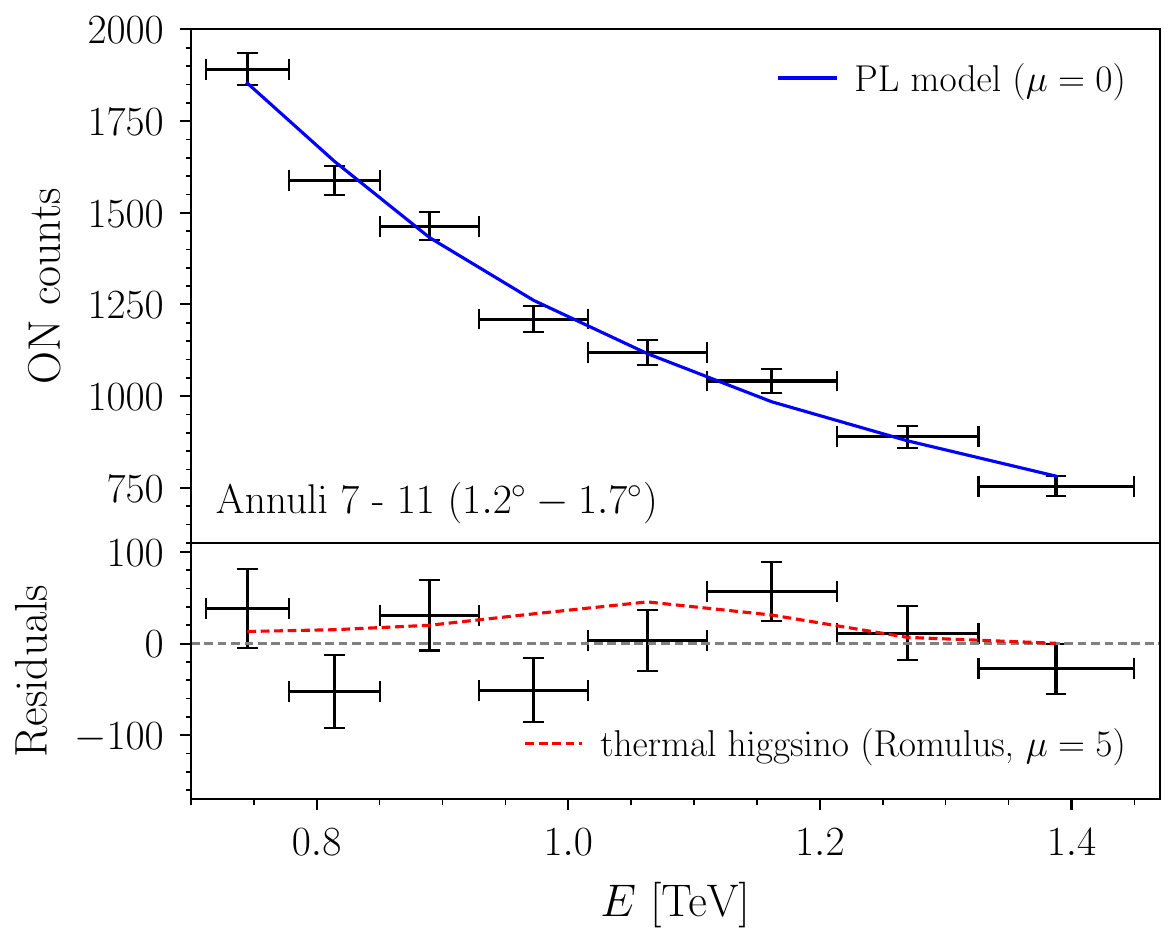}
\end{center}
\vspace{-0.7cm}
\caption{As in Fig.~\ref{fig:HESS_null_signal_fits} (left panel) but for the analysis of the data in the sub-region given by annuli 7 through 11 (in the notation of Ref.~\cite{HESS:2022ygk}).  }
\label{fig:annulus_1}
\end{figure}

\begin{figure}[!t]
\begin{center}
\includegraphics[width=0.49\textwidth]{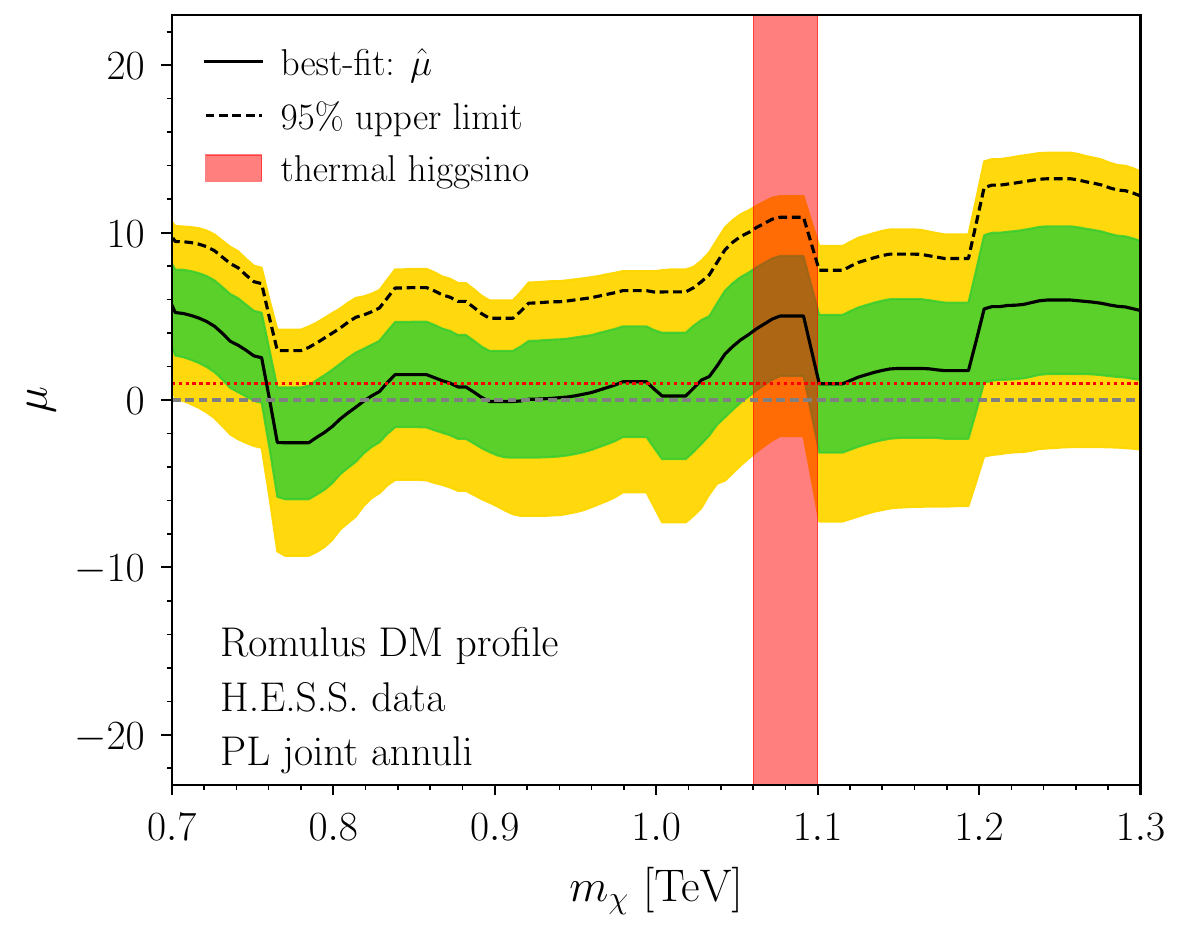}
\end{center}
\vspace{-0.7cm}
\caption{As in Fig.~\ref{fig:HESS_bf_data}, but for the analysis joint over sub-ROI's as described in App.~\ref{app:annuli}.}
\label{fig:joint_limit}
\end{figure}

In Fig.~\ref{fig:annulus_1} we illustrate an example fit of the power-law model to the data in the sub-region given by annuli 7 through 11, stacked. As in the case of {\it e.g.} Fig.~\ref{fig:HESS_null_signal_fits} for the full-ROI-stacked data, we find no visual signs of mis-modeling.
Again adopting the Romulus profile, the joint limit as a function of $m_\chi$ is shown in Fig.~\ref{fig:joint_limit}.
Comparing to the left panel of Fig.~\ref{fig:HESS_bf_data}, we see that the joint analysis over annuli leads to nearly identical results in this case to that found in the stacked analysis, at least for this DM profile.

\section{A cross-check on the H.E.S.S. $\chi \chi \to W W$ inner Galaxy sensitivity estimate}
\label{app:WW}

\begin{figure*}[!tb]
\begin{center}
\includegraphics[width=0.49\textwidth]{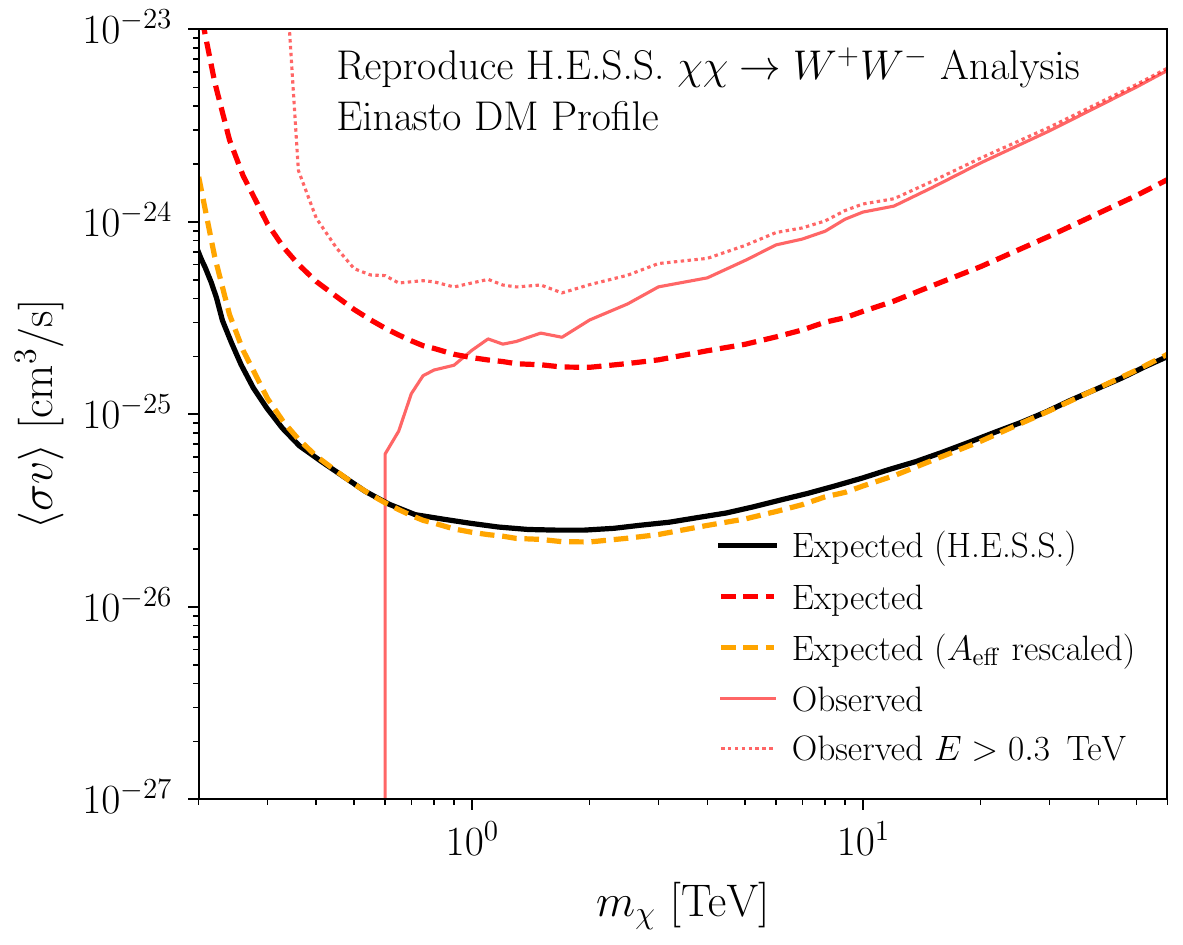}
\includegraphics[width=0.48\textwidth]{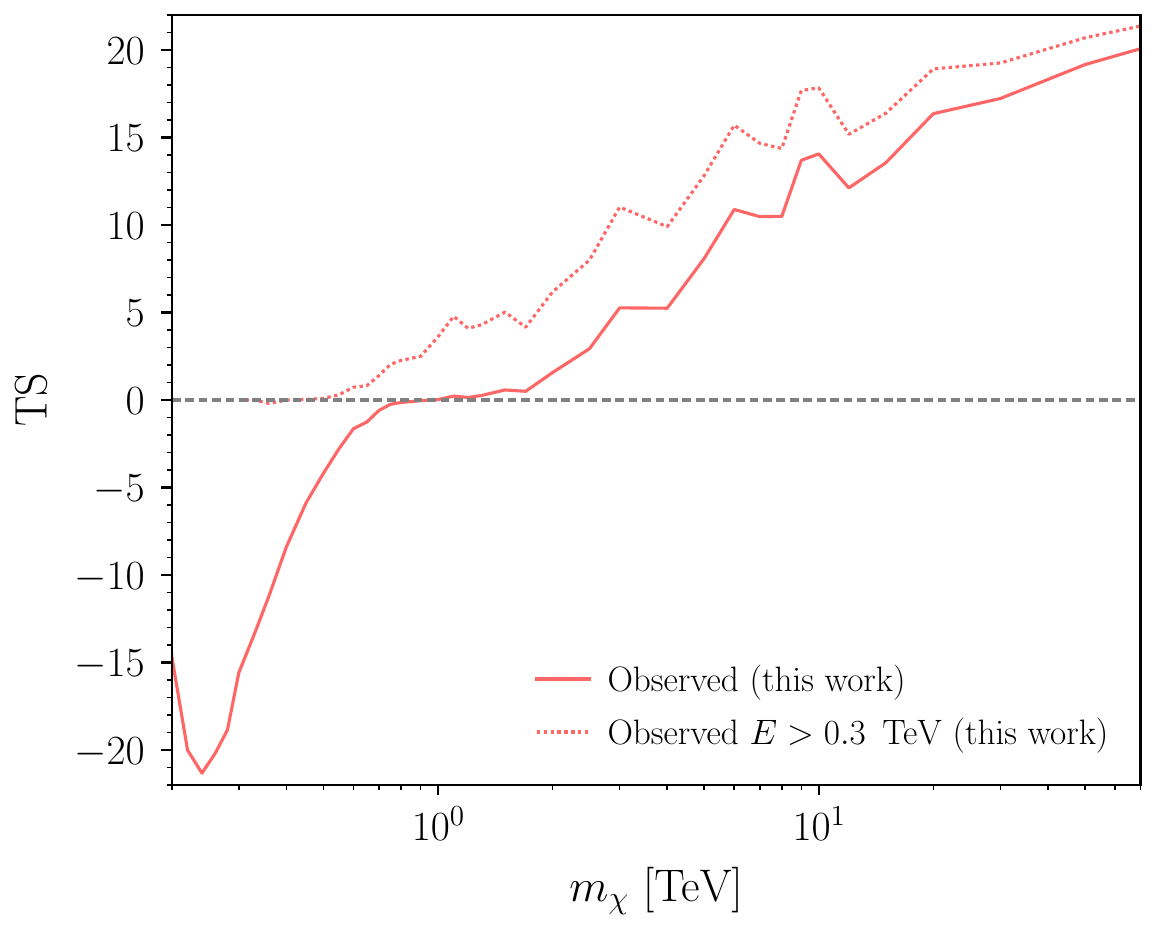}
\end{center}
\vspace{-0.7cm}
\caption{(Left panel) A reproduction of the H.E.S.S. analysis $\chi \chi \to W W$ search performed in \cite{HESS:2022ygk}, used to provide a cross check on our estimates for the H.E.S.S. instrument performance.
As can be seen, if we rescale the effective area up by a factor of $\sim$$8$ from what we believe it should be, we obtain the results in dashed orange that are in good agreement with the H.E.S.S. published limits (in black) from~\cite{HESS:2022ygk}.
Without this, our effective limits in dashed red are correspondingly weaker.
The actual limits are heavily impacted by the first few energy bins (with $E < 0.3\,$TeV) where the OFF counts far exceed the ON counts, leading to a preferred negative cross section, therefore we also show the inferred limit with those bins removed.
(Right panel) The observed TS associated with the two observed limits in our reproduced analysis (when the best fit cross section is negative we define the TS as negative also.)
As can be seen there is a strong preference for a signal in the data, although as argued in this work this is likely arising for residual astrophysical diffuse emission.}
\label{fig:CrossCheck}
\end{figure*}

As detailed in the main body, our procedure for determining the H.E.S.S.-II instrumental response made a number of assumptions that may differ from the exact instrumental response in practice.
To provide a cross check on our approach, in this appendix we reproduce the primary analysis in the H.E.S.S.-II DM search of \cite{HESS:2022ygk}, in particular their search for DM annihilation through the $W W$ channel.

In our analysis, we divide the equivalent ROI used in \cite{HESS:2022ygk} into the eight rings described in App.~\ref{app:annuli}, rather than the twenty-five that work used, although this distinction should only have a minor impact.
Spectrally, we use the same energy bins as in the H.E.S.S. analysis.
To repeat the steps taken to infer the H.E.S.S.-II response, \cite{HESS:2022ygk} provides information on both the total flux and the total flux multiplied by the effective area, each as a function of energy.
The ratio of these provides the instrumental effective area as a function of energy, which when combined with knowledge of the energy binning, ROI, total observation time (546 hours), and a model for the energy resolution, provides us with the full instrumental response in \eqref{eq:dN/dE} required to forward model a DM signal. Additionally, the flux multiplied by the effective area is provided for the ON and OFF region, which when combined with the energy binning and observation time allows us to infer the total number of counts in each energy bin for the ON and OFF region.
As discussed in the main text, we are concerned that the effective area we extract from the above procedure is a factor of $\sim$8 too large, and so by default we reduce the effective area by this value.  However, here we show results with and without that change.

This information is sufficient to construct a simple Gaussian likelihood analysis of the data.
We take Poisson errors on the ON and OFF counts, and then form the ON-OFF data sets, which in the model adopted by \cite{HESS:2022ygk} should be background free. (In this section, importantly, for the purpose of direct comparison we follow~\cite{HESS:2022ygk} and do not model the astrophysical diffuse emission.) 
We can then determine the limit we obtain on a $\chi \chi \to W W$ DM signal.
The relative $J$-factors follow from the reconstruction of the exposure map and consequent ON \& OFF regions detailed in App.~\ref{app:HESS_detector}.

The limit we observe, and the corresponding TS in favor of a positive (or negative) signal are shown in the left and right panels of Fig.~\ref{fig:CrossCheck}.
We display the TS as negative if the best fit cross section is less than zero.
As seen, there is strong preference for a negative signal due to a large excess of OFF over ON events in the bins with $E < 0.3\,$TeV;  therefore, we also show results with those bins removed. In both cases, the preference for a positive spectral excess is clear, although as discussed in \cite{HESS:2022ygk} this is more likely due to unmodeled diffuse emission.

More important than the specific limit we obtain is to compare how our expected sensitivity matches that of H.E.S.S.-II.  This provides a cross check on the instrumental response and exposure map we have reconstructed and is largely separate from any fluctuations or excesses in the data. To obtain the expected limit, we set the central value of the ON-OFF data to zero, consistent with the no background hypothesis, but keep the error bars fixed.
Analyzing these null Asimov data provides the expected limit we show in the left of Fig.~\ref{fig:CrossCheck}. 
We show two versions of the expected limit, and each are compared to the no systematics expected limit published in \cite{HESS:2022ygk}.
The first of our results, in dashed red, uses the corrected effective area, and is roughly a factor of 8 weaker than what H.E.S.S. claimed. This is similar in magnitude to the correction factor we applied to the effective area, and indeed if we scale $A_{\rm eff}$ up by that factor, we obtain the results in dashed orange.
As can be seen, these are in good agreement with the H.E.S.S. expected limit.

Our conclusions from the fact we can apparently accurately reconstruct the H.E.S.S. results are twofold.
Firstly, this provides a partial validation on our extraction of the H.E.S.S. instrument response from \cite{HESS:2022ygk}.
(Of course, the validation for our purposes is not perfect in that we are validating a search for a broad DM spectrum of photons as opposed to the line signature of interest.)
Secondly, these results suggest that if the H.E.S.S. effective area used in \cite{HESS:2022ygk} has indeed been overestimated, then it appears that the associated DM sensitivity could have been also.

\section{H.E.S.S. sensitivity to $\chi \chi \to \gamma \gamma$}
\label{app:Line}

\begin{figure}[!tb]
\begin{center}
\includegraphics[width=0.49\textwidth]{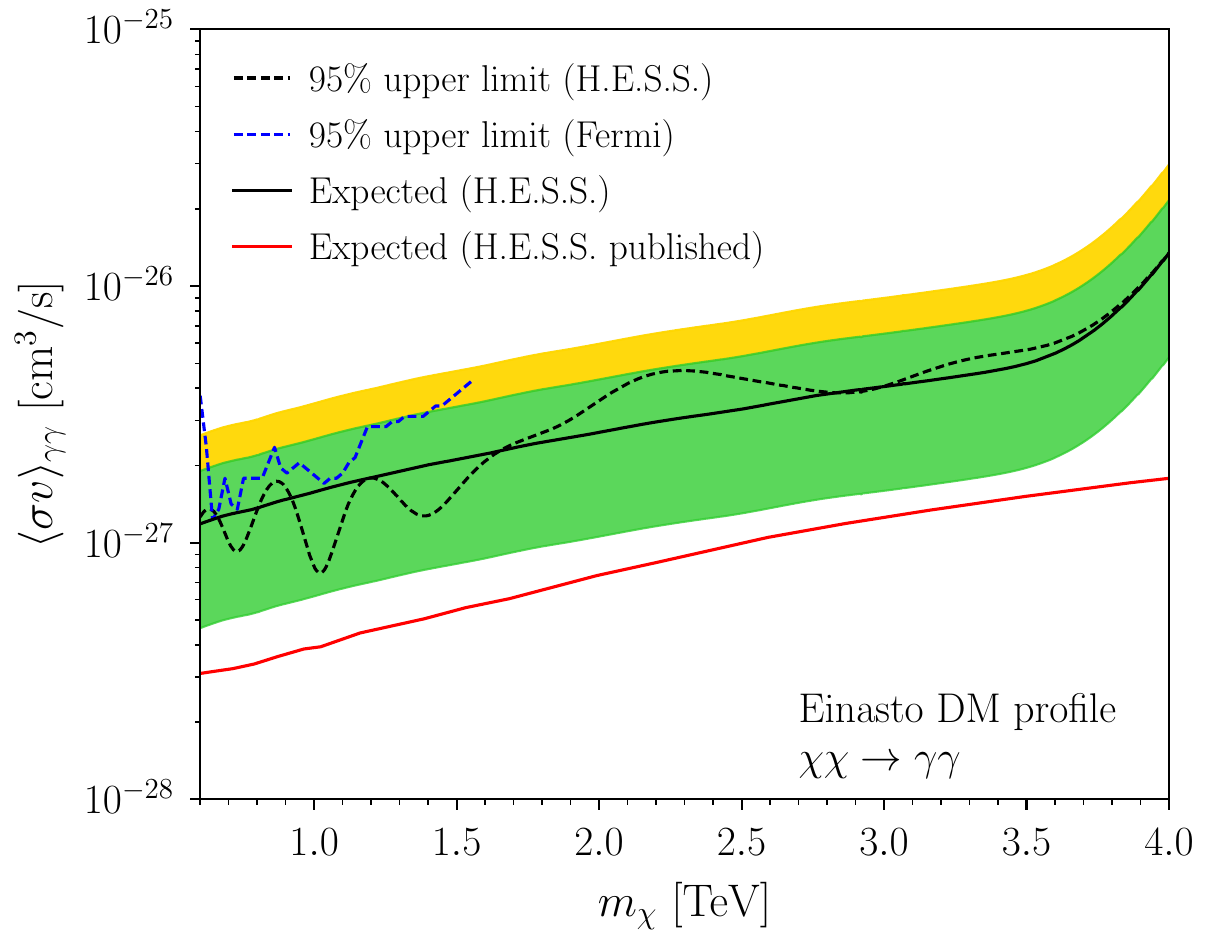}
\end{center}
\vspace{-0.7cm}
\caption{H.E.S.S. and Fermi sensitivity to a DM line signal, $\chi \chi \to \gamma \gamma$, at TeV masses for the Einasto DM profile.
We show the expected and observed limit of a H.E.S.S. analysis performed joint over annuli (as in App.~\ref{app:annuli}) with a PL background model.
This should be contrasted with the published sensitivity by the H.E.S.S. collaboration in \cite{HESS:2018cbt}, which also used an analysis of the inner Galaxy, but with almost a factor of two less data.
That the H.E.S.S. limit is considerably stronger raises the possibility that there may be similar issue in the sensitivities determined in that work to those we suspect impact \cite{HESS:2022ygk}, see the text for further discussion.
Lastly, for comparison we show the sensitivity to the same signal obtained with Fermi.}
\label{fig:CrossCheck-Line}
\end{figure}

The primary focus of this work is a search for the specific signature predicted by the annihilation of higgsino DM, which predicts a different spectrum of photons than were DM to simply annihilate to a line through $\chi \chi \to \gamma \gamma$.
This point is exemplified in Fig.~\ref{fig:xsec}.
Nonetheless, searches for a direct line from DM are common, and it is a simple modification of the higgsino searches performed in the main text to repeat these for the line signal.  Moreover, computing the upper limit on $\chi \chi \to \gamma \gamma$ allows us to directly compare to the previous H.E.S.S. analysis in~\cite{HESS:2018cbt}.

With this motivation, in Fig.~\ref{fig:CrossCheck-Line} we show the expected and observed H.E.S.S. limit on $\chi \chi \to \gamma \gamma$, as well as the observed limit for Fermi (both computed in this work).
For H.E.S.S. we perform a joint analysis over eight annuli as in App.~\ref{app:annuli}, using a PL model for the background, effectively as in Fig.~\ref{fig:joint_limit}.
For Fermi our analysis is identical to that in Sec.~\ref{sec:fermi} other than for the changed signal shape.
(Note that the Fermi limits may in principle be extended up to 2 TeV, as in~\cite{Foster:2022nva}, though we do not do so here since we are primarily interested in energies near 1 TeV.)
In both cases, as opposed to the Romulus DM profile generally adopted in this work we instead show the upper limits assuming the Einasto profile to facilitate a comparison with the existing H.E.S.S. results presented in \cite{HESS:2018cbt}.

The most important comparison here is between the H.E.S.S. results we present and those published by the collaboration in \cite{HESS:2018cbt}.
The H.E.S.S. results are also determined from observations of the inner Galaxy, exploiting in total 254 hrs of data.
The limits we derive follow our extraction of the data from the IGS analysis of~\cite{HESS:2022ygk} (with a correction factor of $\sim $8 applied to the effective area), which focused on continuum rather than line-like signatures.
Nevertheless, the later H.E.S.S. data in \cite{HESS:2022ygk} contained 546 hrs of observations, more than a factor of two greater than in \cite{HESS:2018cbt}.
While in practice it is important exactly where the additional observations were taken compared to the GC, we would expect our sensitivity should be marginally stronger than that obtained in \cite{HESS:2018cbt}.
Figure~\ref{fig:CrossCheck-Line} shows the opposite to be true.
Unfortunately Ref.~\cite{HESS:2018cbt} provides far less information than \cite{HESS:2022ygk}, so it is not possible to extract their analysis effective area, or other instrumental properties that would be needed to perform a more detailed reproduction of their analysis.
Nevertheless, we take our results as suggestive that if the issue we suggest has impacted \cite{HESS:2022ygk} were confirmed, it may also impact the DM line-like limits claimed in \cite{HESS:2018cbt}.

Let us also comment on the difference between the H.E.S.S. and Fermi sensitivity.
As Fig.~\ref{fig:CrossCheck-Line} shows, H.E.S.S. achieves slightly stronger sensitivity that Fermi over the mass ranges considered.
On the other hand, the difference is marginal.
As Fig.~\ref{fig:HESS_mu_data} demonstrates, for the slightly different higgsino spectrum Fermi is marginally more sensitive, so we are operating in the regime where small differences in the DM spectral shape, and spatial shape through the assumed DM profile, can have an impact.

\section{Dwarf galaxy sensitivity estimate}
\label{app:dwarf}

\begin{figure}[!t]
\begin{center}
\includegraphics[width=0.49\textwidth]{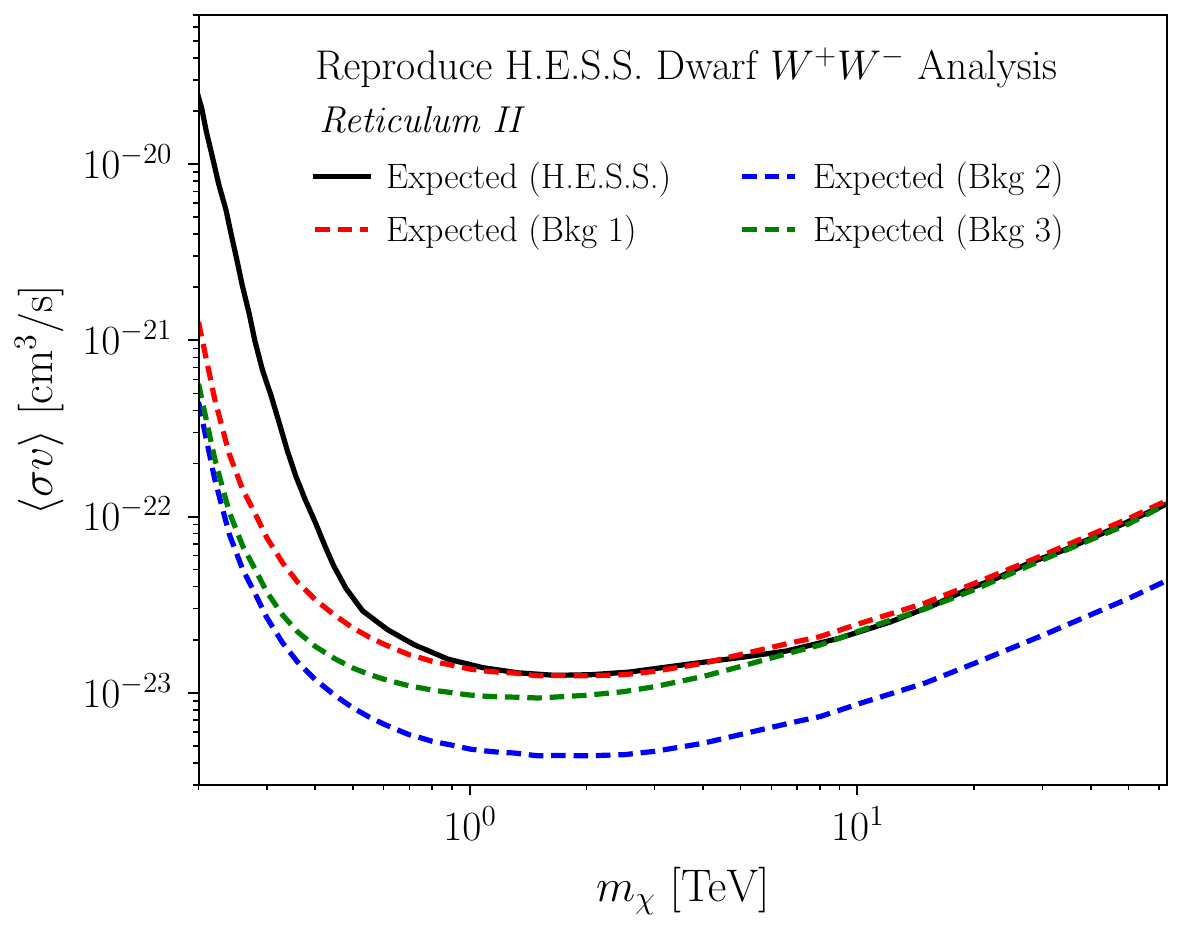}
\end{center}
\vspace{-0.7cm}
\caption{An attempt to reproduce the expected $WW$ sensitivity obtained from the 18.3 hour observation of the Reticulum II Dwarf galaxy in \cite{HESS:2020zwn}.
Full details are provided in the text, but we compare the published results from H.E.S.S. in \cite{HESS:2020zwn} (black), to results we derive using three different background models.
For the first (red, Bkg 1), we took the background as the observed flux in \cite{HESS:2022ygk}, but rescaled by $\sim$$8.1$ to a larger value to correct effective area discussed in the main text.
Our second background model (blue, Bkg 2) is the result if we instead use the uncorrected effective area.
Finally, the third model (green, Bkg 3) is determined by fitting the total hadronic (proton + Helium) cosmic ray spectrum to the observed number of OFF events quoted in \cite{HESS:2020zwn}.
We emphasize a number of assumptions were introduced to obtain these results -- most importantly we used the public H.E.S.S. I instrument response rather than the equivalent for H.E.S.S. II -- we refer to the text for a full discussion of these.
}
\label{fig:DwarfCheck}
\end{figure}

One of the conclusions of this work is that there are reasons to be concerned that the H.E.S.S. results in \cite{HESS:2022ygk} may have overestimated their sensitivity to DM by a factor $\sim$$8$.
If such a rescaling is required, one could worry whether it has also impacted other H.E.S.S. analyses.
Unfortunately, recent H.E.S.S. DM searches often provide far less public information than was given in \cite{HESS:2022ygk}, making them challenging to reproduce.
Nevertheless, with several assumptions we can reproduce the search for DM annihilation in Milky Way dwarf galaxies presented in \cite{HESS:2020zwn}, and we outline our procedure for doing so and the results that follow in this appendix.

The analysis in \cite{HESS:2020zwn} focused on several recently discovered ultra-faint dwarf spheroidal galaxies. Results for the $WW$ annihilation channel are provided for each of the objects considered, and therefore we focus on their analysis of Reticulum II.
We match the energy binning of the H.E.S.S. analysis, use their observation time of 18.3 hours, and that the ON region was a circle of 0.2$^\circ$ radius, with the OFF region eight such circles.
In the official analysis, the ON region was divided into two concentric sub-regions, each of width 0.1$^\circ$, by contrast here we consider a single spatial ROI, and we do not expect a significant impact from this choice.
To determine the signal flux in the ON region we need an estimate for the $J$-factor of Reticulum II, and we follow the official analysis and take $J_{\rm ON} = 10^{19.2 \pm 0.6}~{\rm GeV}^2/{\rm cm}^5\cdot{\rm sr}$. 
(The OFF region is taken sufficiently far from the dwarf that we assume there is no DM flux from that region.)
To account for the uncertainty in the $J$-factor, the full results in \cite{HESS:2020zwn} treat this contribution as a nuisance parameter with a log-normal distribution.
In the Asimov procedure we adopt, the nuisance parameter can be profiled over analytically, and we find that it roughly weakens the limit as if one had used a $J$-factor 1$\sigma$ lower for the signal prediction.
Note, however, that in the results \cite{HESS:2020zwn} provide for individual dwarfs, the $J$-factor is fixed to its central value (no profiling is performed), and therefore we adopt the same procedure in our results.

For the instrument response, assumptions are required.
We use the public H.E.S.S.-I instrument response shown in Fig.~\ref{fig:Aeff}. This is clearly not correct, as \cite{HESS:2020zwn} was a H.E.S.S.-II analysis, and in particular we expect the effective area to be underestimated below 1 TeV, although the limits on DM masses well above this should be less impacted.
Additionally, the observation of Reticulum II was taken in wobble mode, with the central axis of the telescope varied between 0.5-0.7$^\circ$ from the center of the dwarf.
To compute the effective area for the ON and OFF regions, we assume the instrument pointed exactly 0.6$^\circ$ away.

The final ingredient we need to estimate the DM sensitivity is the expected background rate for these observations.
There are many ways this could be chosen, and we consider three possibilities, which we label ``Bkg 1-3''.
For Bkg 1 we take the observed flux in the H.E.S.S. inner Galaxy analysis of \cite{HESS:2022ygk}, but scaled up by the $\sim$$8.1$ factor that we believe may have been overestimated in the effective area.
For Bkg 2, we take the observed flux from the inner Galaxy study directly, with no correction factor applied.
Finally, for Bkg 3 we consider a completely different approach to inferring the background.
We take the total hadronic (proton and Helium) cosmic ray flux observed by AMS-02~\cite{AMS:2015tnn} to provide the background spectrum and fix the normalization such that the number of counts the flux generates in the OFF region matches the number observed in the Reticulum II observation. As the flux steeply rises at low energies, this procedure is undoubtedly biased by our use of the H.E.S.S.-I effective area.

Our expected limits for each background model, together with the published H.E.S.S. result, are shown in Fig.~\ref{fig:DwarfCheck}.
These results should be interpreted with caution given the many assumptions underpinning their derivation. Nevertheless, given the reasonable agreement at high masses (where our use of the H.E.S.S.-I effective area is less of a concern) for Bkg 1 and 3, we see no immediate reason to suspect a similar factor of $\sim$$8$ is impacting these results. If that were the case, we might expect the results to be in closer agreement with Bkg 2, when in fact we find that gives rise to a stronger limit.

\section{Comparison to CTA Collaboration projections for $\chi \chi \to \gamma\gamma$}
\label{app:CTA_comp}

In the final stages of preparation of this work Ref.~\cite{Abe:2024cfj} appeared, which projects sensitivity of the upcoming CTA South to DM annihilation, including $\chi\chi \to \gamma\gamma$, though that work does not consider the higgsino model. Reference~\cite{Abe:2024cfj} uses a similar un-subtracted analysis framework to that advocated for in this work. It is thus useful to directly compare the sensitivity projected in~\cite{Abe:2024cfj} to that we find with our analysis procedure. As discussed in the main article, while  Ref.~\cite{Abe:2024cfj} also assumes 500 hrs of data in the GC region, the data in~\cite{Abe:2024cfj} is assumed to be more concentrated towards the GC than in our work.

\begin{figure}[!t]
\begin{center}
\includegraphics[width=0.49\textwidth]{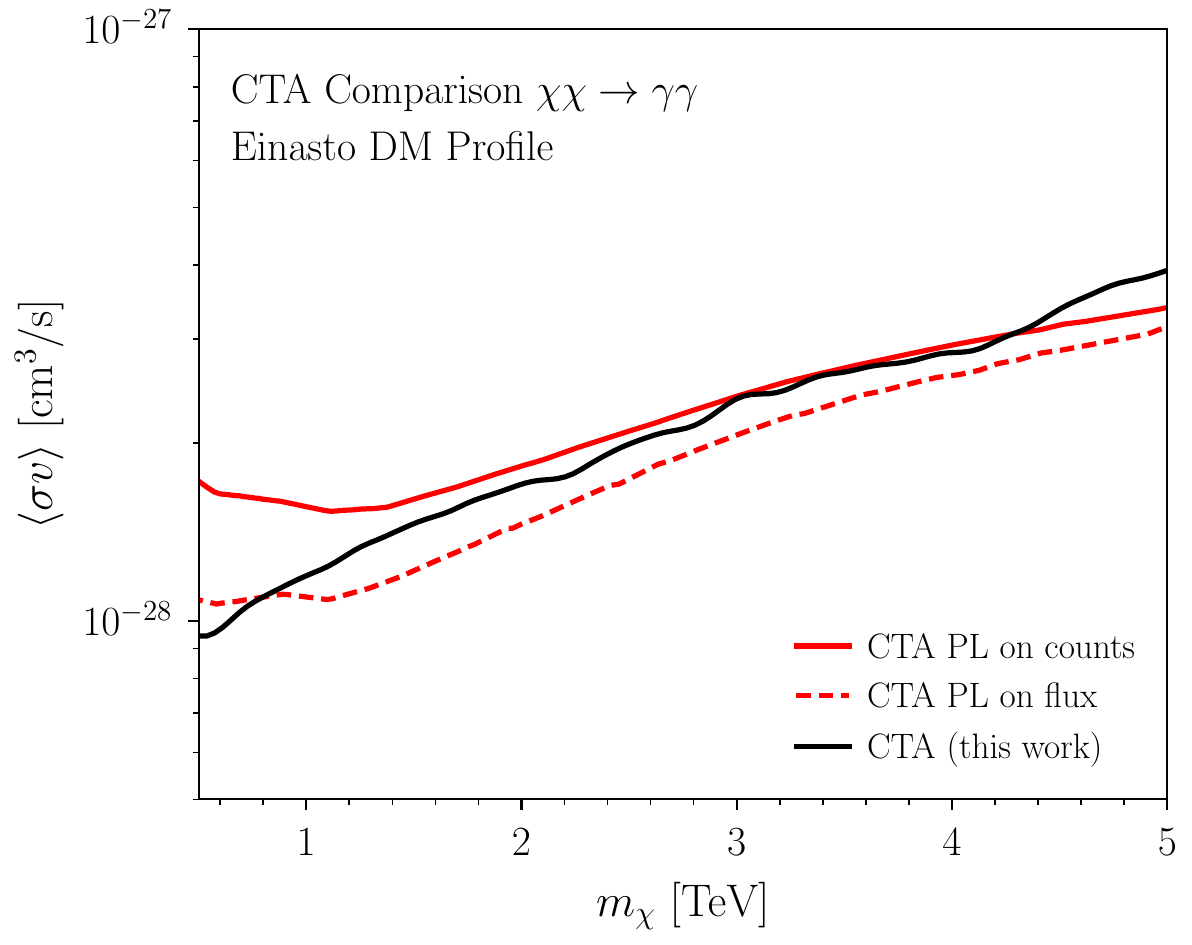}
\end{center}
\vspace{-0.7cm}
\caption{The projected sensitivity to the process $\chi \chi \to \gamma \gamma$ from CTA South assuming 500 hrs of data in the GC region from this work (labeled `CTA (this work)') and for two different analysis proposals in Ref.~\cite{Abe:2024cfj} (see text for details). The agreement between our projected 95\% upper limits gives confidence to our CTA higgsino projections in the main article.
}
\label{fig:CTA_comparison}
\end{figure}

In Fig.~\ref{fig:CTA_comparison} we compare the projected 95\% upper limits under the null hypothesis on the annihilation cross-section for the process $\chi \chi \to \gamma\gamma$ relative to the upper limits projected in~\cite{Abe:2024cfj}. We assume the Einasto profile to make contact with~\cite{Abe:2024cfj}.  We compute our projected upper limits (`CTA (this work)') through a procedure analogous to that we perform for the higgsino signal, as described in Sec.~\ref{sec:CTA_projections}. In particular, we use our two-template approach, with spectral templates for the cosmic-ray-induced gamma-ray background and astrophysical gamma-ray background, with overall normalization parameters treated as nuisance parameters. We assign independent nuisance parameters to each annulus, with the annuli as in Sec.~\ref{sec:CTA_projections}, and then we combine the results between annuli to compute the joint profile likelihood for $\langle \sigma v \rangle$. 

There are two minor differences between our procedure in Sec.~\ref{sec:CTA_projections} and that performed here. First, we search for a pure line final state, $\chi \chi \to \gamma\gamma$, instead of the higgsino model. Second, since we show results across a range of $m_\chi$ we use a floating energy window encompassing energies in $(m_\chi/2,4 \cdot m_\chi)$, which roughly matches the energy we use for the thermal higgsino in Sec.~\ref{sec:CTA_projections}.

We compare our projected upper limit with those from~\cite{Abe:2024cfj} in Fig.~\ref{fig:CTA_comparison}.  That work presented two different analysis approaches, both of which used sliding energy windows (of width around 8 times the energy resolution of the detector) and both of which are similar to the proposed strategies in this work. The first, labeled `CTA PL on count' models the background emission simply as a power-law over the narrow energy range. The second, labeled `CTA PL on flux' assumes that the cosmic-ray-induced emission is perfectly modeled, with no nuisance parameters, and models the residual astrophysical emission by a folded power-law. Our two-template approach is slightly different than both of these methods but has a projected sensitivity roughly between the two across most of the mass range considered.  This comparison helps give confidence to our higgsino projections, considering the similarities between the higgsino signal and that from the $\chi \chi \to \gamma\gamma$ process.

\bibliography{refs}

\end{document}